\documentclass[fleqn,aps,prb,twocolumn,showpacs,floatfix,longbibliography]{revtex4-2}
\usepackage{amsmath,amssymb,amsfonts,float,graphics,epsfig,epstopdf,color,verbatim,tabularx,bm,multirow,appendix,hyperref}
\usepackage{amsmath}
\usepackage{amssymb}
\usepackage{mathtools}
\usepackage{graphicx}
\usepackage{lipsum}
\usepackage{bm}
\usepackage{hyperref}
\usepackage{url}
\usepackage[utf8]{inputenc}
\usepackage{subfigure}
\usepackage{slashed,bbm}
\usepackage{graphics,psfrag,epsfig}
\usepackage{dsfont}
\usepackage{setspace}
\usepackage{wasysym}
\usepackage{slashed}
\usepackage{lipsum}
\usepackage{braket}
\usepackage{physics}
\usepackage{enumerate}%

\usepackage{cancel}
\usepackage{hyperref}
\usepackage{xcolor}

\definecolor{dark-red}{rgb}{0.4,0.15,0.15}
\definecolor{dark-blue}{rgb}{0.15,0.15,0.4}
\definecolor{medium-blue}{rgb}{0,0,0.5}
\hypersetup{
colorlinks, linkcolor={dark-blue},
citecolor={dark-blue}, urlcolor={medium-blue}
}
\usepackage{ulem}
\newcommand{\be}{\begin{equation}}
\newcommand{\ee}{\end{equation}}
\newcommand{\bea}{\begin{eqnarray}}
\newcommand{\eea}{\end{eqnarray}}

\renewcommand{\i}{\text{i}}

\newcommand{\e}{e}

\begin{document}

\title{Symmetry-enforced many-body separability transitions}

\author{Yu-Hsueh Chen}
\affiliation{Department of Physics, University of California at San Diego, La Jolla, California 92093, USA}
\author{Tarun Grover}
\affiliation{Department of Physics, University of California at San Diego, La Jolla, California 92093, USA}

\begin{abstract}
\noindent
We study quantum many-body mixed states with a symmetry from the perspective of \textit{separability}, i.e., whether a mixed state can be expressed as an ensemble of short-range entangled (SRE) symmetric pure states. We provide  evidence for `symmetry-enforced separability transitions' in a variety of states, where in one regime the mixed state is expressible as a convex sum of symmetric SRE pure states, while in the other regime, such a representation is not feasible. We first discuss Gibbs state of Hamiltonians that exhibit spontaneous  breaking of a discrete symmetry, and argue that the associated thermal phase transition can be thought of as a symmetry-enforced separability transition. Next, we study cluster states in various dimensions subjected to local decoherence, and identify several distinct mixed-state phases and associated separability phase transitions, which also provides an alternate perspective on recently discussed `average SPT order'. We also study decohered $p+ip$ superconductors, and find that if the decoherence breaks the fermion parity explicitly, then the resulting mixed state  can be expressed as a convex sum of non-chiral states, while a fermion-parity preserving decoherence results in a phase transition at a non-zero threshold that corresponds to spontaneous breaking of fermion parity. Finally, we briefly discuss systems that satisfy NLTS (no low-energy trivial state) property, such as the recently discovered good LDPC codes, and argue that the Gibbs state of such systems  exhibits a temperature-tuned separability transition.

\end{abstract}

\maketitle
\tableofcontents
\section{Introduction} \label{sec:intro}

Suppose one has the ability to apply unitary gates that act in a geometrically local fashion on a many-body system. Starting from a product state, a specific  circuit composed of such gates results in a specific pure state, and an ensemble of such circuits can therefore be associated with the mixed state $\rho = \sum_i p_i |\psi_i\rangle \langle \psi_i|$ where the pure state $|\psi_i\rangle$ is prepared with probability $p_i$. If one is limited to only constant depth unitary circuits, then the corresponding mixed state can be regarded as `short-range entangled' or `trivial' \cite{werner1989, hastings2011topological}, which generalizes the notion of short-range entangled pure state \cite{verstraete2005renormalization, chen2010local, hastings2005quasiadiabatic, bravyi2006lieb, chen2011classification,zeng2015gapped}. In parallel with the notion of symmetry protected topological phases for pure states \cite{gu2009tensor,pollmann2012symmetry,chen2013symmetry,schuch2011classifying}, it is then natural to define a trivial/short-ranged entangled \textit{symmetric} mixed state (a `sym-SRE' state) as one that can be obtained from an ensemble of pure states, where each element of the ensemble is prepared with only a constant depth circuit consisting of local, symmetric gates under some given symmetry. Motivated from experimental progress in controllable quantum devices where both unitary quantum dynamics and decoherence play an important role \cite{bruzewicz2019trapped,henriet2020quantum,kjaergaard2020superconducting,preskill2018quantum}, in this paper we will explore mixed state phase diagrams where in one regime a mixed state is sym-SRE, and in the other regime, it is not. We will call such phase transitions `symmetry-enforced separability transitions', since a sym-SRE state is essentially separable \cite{werner1989} (i.e. a convex sum of unentangled states) upto short-distance correlations generated by constant-depth unitaries. In the absence of any symmetry constraint, analogs of such transitions were recently studied in Ref.\cite{chen2023separability} in the context of decohered topologically ordered mixed states  \cite{lee2023quantum, fan2023diagnostics,bao2023mixed,lu2023mixed,su2023conformal, wang2023intrinsic, sang2023mixed}. To make progress, we will try to leverage our understanding of the complexity of preparing pure many-body states using unitaries.  Some of the questions that will motivate our discussion are: Do there exist separability phase transitions when pure-state symmetry protected topological (SPT) phases are subjected to decoherence, and if yes, what is the universality class of such transition? When a 2d chiral pure state (e.g. the ground state of an integer quantum Hall phase) is subjected to local decoherence, can the resulting density matrix be expressed as a convex sum of non-chiral states? Can the conventional, finite temperature phase transitions corresponding to the spontaneous breaking of a global symmetry be also thought of as separability transitions?

As an example, consider the transverse field Ising model on a square lattice. We provide an argument (Sec.\ref{sec:SSB}) that the Gibbs state for this model can be prepared using an ensemble of finite-depth local unitary circuit at all temperatures, including at $T \leq T_c$, where $T_c$ is the critical temperature for spontaneous symmetry breaking. It is crucial here that one is not imposing any symmetry constraint on the unitaries. This is consistent with previous works \cite{lu2020structure,wu2020entanglement,wald2020entanglement} where evidence was provided that the mixed-state entanglement corresponding to a Gibbs state that exhibits spontaneous symmetry breaking remains short-ranged at all non-zero temperatures, including at the finite temperature critical point (assuming absence of any co-existing finite-temperature topological order). However, if one only allows access to an ensemble of short-depth unitary circuits composed of \textit{Ising symmetric} local gates, then using results of Ref.\cite{lu2023mixed}, we provide a rigorous argument that the Gibbs state can not be prepared for any $T \leq T_c$. We expect similar results to hold  for other symmetry broken Gibbs states as well. Therefore, the conventional, finite-temperature symmetry breaking phase transition in a transverse-field Ising model can be thought of as a symmetry-enforced separability transition. This statement is true even when the transverse field is zero (i.e. for a classical Ising model) - the quantum mechanics still plays a role since the imposition of symmetry implies that one is forced to work with `cat' (GHZ) states, which are long-range entangled.

In the context of pure states, a well-known example of symmetry enforced complexity is an SPT phase whose ground state can not be prepared using a finite depth circuit composed of symmetric local gates \cite{gu2009tensor,pollmann2012symmetry,chen2013symmetry,schuch2011classifying}. Recent works have provided a detailed classification of SPT phases protected by zero-form symmetries that are being subjected to decoherence using spectral sequences and obstruction to a short-ranged entangled (SRE) purification \cite{ma2022average,ma2023topological}.  Progress has also been made in understanding non-trivial decohered SPTs using string operators \cite{de2022symmetry} and  `strange correlators' \cite{lee2022symmetry,zhang2022strange}, concepts that were originally introduced to characterize pure SPT states \cite{den1989preroughening, pollmann2012symmetry, you2014wave}. Here we will be interested in understanding decohered SPT states from the viewpoint of separability, which, as we discuss in Sec.\ref{sec:separability}, is a different notion of entanglement of mixed states than that based on SRE purification considered in Ref.\cite{ma2022average,ma2023topological}. As hinted above, we define a sym-LRE (symmetric, long-range entangled) state as one which does not admit a decomposition as a convex sum of pure states which are all preparable via a finite-depth circuit made of symmetric local gates. If so, it is interesting to ask if there exist separability transitions  between sym-LRE and sym-SRE states as a function of the decoherence rate, analogous to the phase transitions in mixed states with instrinsic topological order \cite{chen2023separability}. We will not consider a general SPT state, and focus primarily on cluster states in various dimensions to illustrate the broad idea. A key step in our analysis will be the following result that was also briefly mentioned in Ref.\cite{chen2023separability} and which we discuss in detail in Sec.\ref{sec:trivial}: for a large class of SPTs, including the cluster states in various dimensions, Kitaev chain in 1d, and several 2d topological phases protected by zero-form $Z_2$ symmetry, one can find local, finite-depth channels that map the pure state to a Gibbs state. We will discuss decoherence induced separability transitions due to such channels in Sec.\ref{sec:trivial}.

When trying to understand complexity of mixed SPT states, we will often find the following  line of inquiry helpful. One first asks: Does assuming that a mixed state is trivial (i.e. decomposable as a convex sum of SRE pure states) lead to an obvious contradiction? If the answer to this question is `yes', then we already know that the mixed state is necessarily non-trivial. In this case, there may still exist interesting transitions between two different kinds of non-trivial mixed states, and we will consider a couple of such examples as well. On the other hand, if the answer to this question is `no', we will attempt to find an explicit decomposition of the mixed state as a convex sum of SRE states. The aforementioned relation between local and thermal decoherence will again be instrumental in making analytical progress.

As an example, consider the ground state of the 2d cluster state Hamiltonian $H$ subjected to a local channel that locally anticommutes with the terms in the Hamiltonian. One can show that resulting decohered state $\rho_d$ takes the Gibbs form: $\rho \propto e^{- \beta H}$ where $\tanh(\beta) = 1-2p$ and $p$ is the decoherence rate. In this example, $H$ has both a zero-form and a one-form Ising symmetry. We will provide arguments that this system undergoes a separability transition as a function of $p$: for $0 < p < p_c$, $\rho$ cannot be decomposed as a sum of pure states that respect the aforementioned two symmetries, while for $p > p_c$, such a decomposition is feasible. Moreover, for $ p > p_c$ we will express $\rho_d$ explicitly as $\sum_m p_m |\psi_m\rangle \langle \psi_m|$, where $|\psi_m\rangle$ are pure, symmetric states that are statistically SRE. More precisely, one can define an ensemble averaged string-correlation, $[\langle S_C \rangle^2] \equiv \sum_m p_m \langle S_C \rangle_m^2$, where $\langle S_C \rangle_m = \langle \psi_m |S_C |\psi_m\rangle/\langle \psi_m|\psi_m\rangle$, and $S_C$ is a string-operator whose non-zero expectation value implies long-range entanglement. We will show that  $[\langle S_C \rangle^2]$ precisely corresponds to a disorder-averaged correlation function in the 2d random-bond Ising model along the Nishimori line \cite{nishimori1981internal}. Therefore, in this example, the separability transition maps to the ferromagnetic transition in the random bond Ising model. For the 3d cluster state, we will find an analogous relation between separability and 3d random-plaquette Ising gauge theory. We note that similar order parameters and  connections to statistical mechanics models also appear in the setting of measurement protocols to prepare long-range entangled SPT states \cite{lee2022measurement,zhu2022nishimori}. We briefly discuss connection to these works.

As another byproduct of the relation between local decoherence and Gibbs states, we also study recently introduced non-trivial class of mixed states which are protected by a tensor product of `exact' and `average' symmetries \cite{de2022symmetry,lee2022symmetry, ma2022average,ma2023topological}. One says that a density matrix $\rho$ has an   `exact symmetry' if  {$U_E \rho \propto \rho$} for some unitary $U_E$, while it has an `average symmetry' if $U_A^\dagger \rho U_A = \rho$ for some unitary $U_A$.  Refs. \cite{de2022symmetry,lee2022symmetry, ma2022average,ma2023topological} have provided several non-trivial examples of such mixed-state SPTs by showing that they possess non-trivial correlation functions, and/or cannot be purified to a short-ranged entangled (SRE) pure state. Here we will focus on examples of such states that are based on cluster states in various dimensions, and using locality/Lieb-Robinson bound \cite{lieb1972finite,hastings2010locality,huang2015quantum}, show that the corresponding mixed states cannot be written as a convex sum of symmetric, pure states. For 1d cluster state, we also provide an alternative proof of non-separability by using the result from Ref.\cite{levin2020constraints} that in one-dimension if a state has an average $Z_2$ symmetry, and its connected correlation functions are short-ranged, then the corresponding `order parameter' and the `disorder parameter' can't be both zero or non-zero at the same time.

Next, in Sec.\ref{sec:fermions} we consider fermionic chiral states subjected to local decoherence. We primarily focus on the ground state of a 2d $p_x+ip_y$ superconductor ($p+ip$ SC) as our initial state (we expect integer quantum Hall states to have qualitative similar behavior). We first consider subjecting this pure state to a finite-depth channel with Kraus operators that are linear in fermion creation/annihilation operators, so that the decoherence breaks the fermion parity symmetry. In the pure state classification of topological superconductors, fermion parity is precisely the symmetry responsible for the non-trivial topological character of the $p+ip$ SC \cite{wen2012symmetry,ryu2010topological}. Therefore, it is natural to wonder about the fate of the mixed state obtained by breaking this symmetry from exact down to average. One potential path to make progress on this problem is to map the mixed state to a pure state in the doubled Hilbert-space using the Choi-Jamiolkowski (C-J) map \cite{jamiolkowski1972linear, choi1975completely} (we will call such a state the `double state', similar to the nomenclature in Ref.\cite{bao2023mixed}). There are interesting subtleties in applying the C-J map to fermionic Kraus operators that we clarify. Following the ideas in Refs.\cite{lee2022symmetry,lee2023quantum,bao2023mixed,su2023conformal}, one may then map the double state to a 1+1-D theory of counter-propagating free CFTs coupled via a  fermion bilinear term, which is clearly relevant and gaps out the edge states in these doubled picture. However, a short-depth channel cannot qualitatively change the expectation value of state-independent operators (i.e. $\tr(\rho O)$ where $O$ is independent of $\rho$) \cite{fan2023diagnostics,lee2023quantum}, and it is not obvious what does the gapping of edge modes imply for the actual mixed state. We conjecture that the physical implication of the gapping of the edge states in the doubled formulation is that the actual mixed state can now be expressed as a convex sum of SRE states with zero Chern number, which is equivalent to the statement that they can be obtained as a Slater determinant of Wannier states, unlike the pure $p+ip$ state where such a representation is not possible \cite{thouless1984wannier,read2000paired,schindler2020pairing}. Therefore, the transition from the pure state to the mixed state can be thought of as a `Wannierizability transition'. We consider an explicit ansatz of such a decomposition, and provide numerical support of our conjecture by calculating the entanglement spectrum and modular commutator of the pure states whose convex sum corresponds to the decohered density matrix. 

A more interesting channel that acts on the 2d $p+ip$ SC corresponds to Kraus operators that are \textit{bilinear} in fermion creation/annihilation operators. To make progress on this problem, we use the C-J map to obtain a field theoretic description for this problem in terms of two counter-propagating chiral Majorana CFTs interacting via a four-fermion interaction, where the strength of the interaction is related to the strength of the interacting decohering channel. This theory admits a phase transition at a critical interaction strength in the supersymmeteric tricritical Ising universality class, which can be thought of as corresponding to spontaneous breaking of the fermion parity. Although we don't have an understanding of this transition directly in terms of the mixed state in the non-doubled (i.e. original) Hilbert space, it seems reasonable to conjecture that at weak decoherence, the density matrix can not be expressed as a convex sum of area-law entangled non-chiral states, while at strong decoherence, it is most naturally expressible as a convex sum of states with GHZ like character that originates from the aforementioned spontaneous breaking of the fermion parity.

Incidentally, the kind of arguments we consider to rule out sym-SRE mixed states in the context of symmetry broken phases or SPT phases also find an application in an exotic separability transition where symmetry plays no role. In particular, we consider separability aspects of Gibbs state of Hamiltonians that satisfy the `no low-energy trivial state' (NLTS) condition introduced by Freedman and Hastings in Ref.\cite{freedman2013quantum}. Colloquially, if a Hamiltonian satisfies the NLTS condition, then any pure state with energy density less than a critical non-zero threshold cannot be prepared by a constant depth circuit. Recently, Ref.\cite{anshu2022nlts} showed that the `good LDPC code' constructed in Ref.\cite{leverrier2022quantum} satisfies the NLTS condition (we note that `Good LDPC codes' \cite{panteleev2022asymptotically,leverrier2022quantum,dinur2023good} have the remarkable property that both the code distance and the number of logical qubits scale linearly with the number of physical qubits). Ref.\cite{anshu2022nlts} already showed that the NLTS condition holds also for mixed states, if one defines the circuit depth of a mixed state as the minimum depth of unitary needed to prepare it by acting on system$\otimes$ancillae, both initially in a product state, where the ancillae are traced out afterwards \cite{anshu2020circuit}. Under such a definition of a non-trivial mixed state (namely, a mixed state that can not be prepared by a constant depth circuit under the aforementioned protocol), even mixed states with long-range \textit{classical correlations} (e.g. the Gibbs state of 3d classical Ising model)  would be considered non-trivial. In contrast, under our definition of a non-trivial mixed state, such classical states will be trivial since they can be written as a convex sum of SRE states. Therefore we ask: assuming that one defines a trivial (non-trivial) mixed state as one which can (can't) be expressed as a convex sum of SRE states, is the Gibbs state of a Hamiltonian that satisfies the NLTS property non-trivial at a low but non-zero temperature? Under reasonable assumptions, in Sec.\ref{sec:ldpc} we provide a short argument that this is indeed the case. This implies that one should expect a non-zero temperature separability transition in such Gibbs states.

In Sec.\ref{sec:connections} we briefly discuss connections between separability criteria and other measures of the complexity of a mixed state such as the ability to purify a mixed state to an SRE pure state, entanglement of the doubled state using C-J map, and strange correlators.

Finally, in Sec.\ref{sec:summary} we summarize our results and discuss a few open questions.

For convenience, below we list some of the acronyms that appear frequently in this work:

\begin{itemize}
	\item SRE: Short-range entangled.
	\item LRE: Long-range entangled.
	\item CDA: Convex decomposition ansatz (for a density matrix).
	\item RBIM: Random bond Ising model.
	\item SPT: Symmetry protected topological.
\end{itemize}

\section{Separability criteria with and without symmetry}
\label{sec:separability}
Motivated from Werner and Hastings \cite{werner1989, hastings2011topological}, we call a mixed state $\rho$ short-range entangled (SRE) if and only if it can be decomposed as a convex sum of pure states
\begin{equation}
	\rho = \sum_m p_m |\psi_m\rangle \langle \psi_m|, \label{eq:SREmixeddef1}
\end{equation}
where each $|\psi_m\rangle$ is short-range entangled (SRE), i.e., it can be prepared by applying a constant-depth local unitary circuit to some product state. The physical motivation for this definition is rather transparent: if a mixed state can be expressed as Eq.\eqref{eq:SREmixeddef1}, only then it can be prepared using an ensemble of unitary-circuits (acting on the Hilbert space of $\rho$) whose depth does not scale with the system size. We note that this definition of an SRE mixed state has been employed to understand phase transitions in systems with intrinsic topological order subjected to thermal or local decoherence \cite{lu2020detecting,chen2023separability}.

One can generalize the notion of an SRE mixed state in the presence of a symmetry. Specifically, we say that a mixed state $\rho$ satisfying $U(g) \rho U^\dagger(g) = \rho,\ \forall g \in G$ is a `symmetric SRE' (sym-SRE in short) if and only if one can decompose it as a convex sum of pure states, where each of these pure states can be prepared by applying a finite-depth quantum circuit made of local gates that all commute with $U$, to a symmetric product state.

Several comments follow.

\begin{enumerate}
	
	\item The `only if' clause in our definition for a (sym-) SRE state is a bit subtle. For example, consider a density matrix where there exists no decomposition that satisfies Eq.\eqref{eq:SREmixeddef1}, but there exists a decomposition $\rho = \sum\limits_{m, |\psi_m\rangle \in \textrm{SRE}} p_m |\psi_m\rangle \langle \psi_m| +  \sum\limits_{m, |\phi_m\rangle \notin \textrm{SRE}} q_m |\phi_m\rangle \langle \phi_m|$ such that the relative weight of the non-SRE states is zero in thermodynamic limit (i.e. $ \sum_m q_m/(\sum_m \left( p_m + q_m \right) \rightarrow 0$ in the thermodynamic limit). In this case, it might seem reasonable to regard $\rho$ as SRE. 
	One may also define an average circuit complexity of a density matrix as $\langle \mathcal{C} \rangle = \textrm{inf} \{ \sum_m p_m \mathcal{C}(\psi_m) \}$, where $\mathcal{C}(|\psi_m\rangle)$ is the minimum depth of a circuit composed of local gates to prepare the state $|\psi_m\rangle$ and the infimum is taken over all possible decompositions of the mixed state $\rho$. One may then consider calling a mixed state $\rho$ as SRE if and only if $\langle \mathcal{C} \rangle$ does not scale with the system size. But even then, there may be special cases where the average behavior is not representative of a typical behavior. We will not dwell on this subtlety further at this point, and use physical intuition to quantify the separability of a density matrix, were we to encounter such a situation.
	
	\item Ref.\cite{hastings2011topological} also introduced a seemingly different definition of an SRE mixed state: Consider a `classical' state $\rho_{cl} \propto e^{-H_{cl}}$, where $H_{cl}$ is a Hamiltonian composed of terms which are all diagonal in a product basis, and which acts on an enlarged Hilbert space $a \otimes s$ where $s$ denotes the system of interest and $a$ denotes ancillae. Then a mixed state $\rho$ may be regarded as SRE if it can be obtained from $\rho_{cl}$ by applying a finite-depth unitary on $s \otimes a$, followed by tracing out $a$. That is, one may consider $\rho$ as SRE if 
	
	\be  
	\hspace{1cm} \rho = \tr_{a} \left(U^{\dagger}e^{- H_{cl}} U/Z\right) \label{eq:SREmixeddef2}
	\ee
	
	where $U$ is a finite-depth circuit and $Z = \tr\left(e^{-H_{cl}}\right)$. We are unable to show that the definition in Eq.\eqref{eq:SREmixeddef1} is equivalent to Eq.\eqref{eq:SREmixeddef2}. Although we will primarily use the former definition (Eq.\eqref{eq:SREmixeddef1}), in Sec.\ref{sec:connections} we will briefly discuss potential connections between the two definitions, and also relation with other diagnostics of mixed-state entanglement.
	
	\item  The symmetry {($U(g)\rho U^\dagger (g)=\rho, \forall g\in G$)} we consider is called weak symmetry  (average symmetry) in Ref.\cite{de2022symmetry} (Ref.\cite{ma2022average}), which highlights its difference with the stronger symmetry  $U(g)\rho = \rho U(g) = e^{i \theta(g)}\rho, \forall g \in G$, termed  strong symmetry (exact symmetry) in Ref.\cite{de2022symmetry} (Ref.\cite{ma2022average}).
	Physically, exact symmetry enforces the constraint that the density matrix \textit{must} be written as an incoherent sum of pure states, where each of them is an eigenstate of $U(g)$ with the same eigenvalue $e^{i \theta(g)}$.
	On the other hand, while the mixed state $\rho$ with only average symmetry can be written as a convex sum of symmetric pure states having different charge under $G$, one may as well express $\rho$ as a convex sum of non-symmetric pure states.
	Therefore, our requirement that each of the pure states respects the symmetry puts a further constraint on a mixed state with only average symmetry.
	
	On that note, Ref. \cite{roberts2017symmetry} defined a symmetric-SRE state for a symmetry $U$ as one which satisfies Eq.\eqref{eq:SREmixeddef2} where  $e^{-H_{cl}}$ is replaced by $P_{\theta(g)} e^{-H_{cl}}$ where $P_{\theta(g)}$ is a projector onto a given symmetry charge $\theta(g)$. Therefore, in this definition one is always working with a density matrix that has an \textit{exact} symmetry. As already mentioned, we will instead only impose the average symmetry in our definition of a sym-SRE state (of course, there may be special quantum channels that happen to preserve an exact symmetry).

	\item An alternative definition of an SRE mixed state was considered in Refs.\cite{anshu2020circuit, ma2022average,ma2023topological}  whereby a mixed density matrix is considered SRE if it can be obtained from a \textit{pure} product state in a system$\otimes$ancillae Hilbert space via a finite-depth unitary followed by tracing out ancillae. In contrast, as already mentioned above in comment $\#2$, Ref.\cite{hastings2011topological}  defines a mixed density matrix as SRE if it can be obtained from the `classical mixed state' $\rho_{cl} \propto e^{-H_{cl}}$ of system$\otimes$ancillae via a finite-depth local quantum channel. Therefore, a mixed state can be trivial/SRE using the definition of Ref.\cite{hastings2011topological} while remaining non-trivial/LRE using the definition of  Refs.\cite{anshu2020circuit, ma2022average,ma2023topological}.  The physical distinction between these two definitions is most apparent when one considers a mixed state for qubits of the form $\rho = \frac{1}{2} \left(|\uparrow \rangle \langle \uparrow| + |\downarrow \rangle \langle \downarrow|\right)$ where $|\uparrow \rangle = \prod_i |\uparrow\rangle_i$ and $|\downarrow \rangle = \prod_i |\downarrow\rangle_i$. This state is clearly separable (unentangled). However, any short-depth purification of this state must be long-ranged entangled. This is because $\tr(\rho Z_i Z_j) - \tr(\rho Z_i) \tr(\rho Z_j)$ is non-zero and the purified state can't change this correlation function due to the Lieb-Robinson bound \cite{lieb1972finite,hastings2010locality} (this is also related to the fact the entanglement of purification \cite{terhal2002entanglement} is sensitive to both quantum and classical correlations, and therefore is not a good mixed-state entanglement measure). Thus the aforementioned $\rho$ will be SRE using definition of Ref.\cite{hastings2011topological}, and LRE using the definition of Ref.\cite{anshu2020circuit,ma2022average,ma2023topological}. Of course, it will also be SRE via Eq.\eqref{eq:SREmixeddef1}, which is the definition we will use throughout this paper.

\end{enumerate}

\section{An illustrative example: Separability transition in the Gibbs state of the 2d quantum Ising model} \label{sec:SSB}
Let us consider an example to illustrate the difference between an SRE mixed state and a sym-SRE mixed state, that will also provide one of the simplest examples of a separability transition. Consider the density matrix $\rho$ for qubits (i.e. objects transforming in the spin-1/2 representation of $SU(2)$) given by $\rho(\beta) = e^{- \beta H}/Z$ where $H$ is a local Hamiltonian that satisfies $U^{\dagger} H U = H$ with  $U = \prod_i X_i$ being the generator of the Ising symmetry, and $Z =\tr  e^{- \beta H}$ is the partition function. Let us further assume that $\rho(\beta)$ exhibits spontaneous symmetry breaking (SSB) for $\beta > \beta_c$ where $ 0 < \beta_c < \infty$ (for a range of other parameters that specify the Hamiltonian). For concreteness, one may choose $H$ as the nearest neighbor transverse-field Ising model on the square lattice, i.e., $H = -\sum_{\langle i, j\rangle }Z_i Z_j - h \sum_i X_i $ although the only aspect that will matter in the following discussion is that $H$ is local with a zero-form Ising symmetry, and the order parameter in the symmetry breaking phase is a real scalar (e.g. one may as well consider a transverse-field Ising model on a cubic lattice). Therefore, for a range of the transverse-field $h$ and $\beta > \beta_c$ (where $\beta_c$ depends on $h$), the two-point correlation function $\tr \left(\rho Z_i Z_j\right) $ is non-zero for $|i - j| \to \infty$. We will argue that $\rho$ is SRE for all non-zero temperatures, while it is sym-SRE only for $\beta < \beta_c$. Partial support for $\rho$ being an SRE at all non-zero temperatures was provided in Refs.\cite{lu2020structure,wu2020entanglement,wald2020entanglement} and we will argue below for an explicit decomposition of $\rho$ in terms of SRE states.

The statement that $\rho$ is not sym-SRE for $\beta \geq \beta_c$ was also hinted in \cite{roberts2017symmetry}, and intuitively follows from the fact that for $\beta > \beta_c$, SSB implies that if one decomposes $\rho$ as a convex sum of symmetric, pure states, those pure states must have GHZ-like entanglement. Let us first consider a rigorous argument for this statement which, upto small modifications, essentially follow the one in Ref.\cite{lu2023mixed} for a closely related problem of non-triviality of a density matrix with an exact symmetry and long-range order.

To show that for $\beta > \beta_c$, $\rho$ can't be a sym-SRE state, let us first decompose $\rho$ as $\rho = \rho_{+} + \rho_{-}$ where $\rho_{\pm} = (\frac{1\pm U}{2})\rho$ are the projections of  $\rho$ onto even and odd charge of the Ising symmetry.
$\rho_{+}$ and $\rho_{-}$ are valid density matrices with an exact Ising symmetry, that is, they satisfy, $U \rho_{\pm} = \pm \rho_{\pm}$. Now let us make the assumption that for $\beta > \beta_c$, $\rho$ is a sym-SRE state. We will show that this assumption leads to a contradiction. Therefore, we write $\rho_{\pm} = \sum_{\alpha} p_{\alpha,\pm} |\psi_{\alpha, \pm}\rangle \langle \psi_{\alpha,\pm}|$ where $p_{\alpha, \pm}$ are positive numbers, and $|\psi_{\alpha,\pm}\rangle$ are SRE states $\forall \alpha$ that satisfy $U |\psi_{\alpha,\pm}\rangle = \pm |\psi_{\alpha,\pm}\rangle$. Since $U$ anti-commutes with $Z_i$, $ \langle \psi_{\alpha, \pm}| Z_i |\psi_{\alpha, \pm}\rangle = 0 $. Further, since $|\psi_{\alpha, \pm}\rangle $ are all SRE states, correlation functions of all local operators  decay exponentially (notably, we assume that the associated correlation length is bounded by a \textit{system-size independent} constant for all $|\psi_{\alpha, \pm}\rangle $), and therefore,   $ \langle \psi_{\alpha, \pm}| Z_j Z_k |\psi_{\alpha, \pm}\rangle -  \langle \psi_{\alpha, \pm}| Z_j  |\psi_{\alpha, \pm}\rangle  \langle \psi_{\alpha, \pm}| Z_k |\psi_{\alpha, \pm}\rangle =  \langle \psi_{\alpha, \pm}| Z_j Z_k |\psi_{\alpha, \pm}\rangle $ vanishes as $|j-k| \rightarrow \infty$. However, this leads to a contradiction, because this implies that $\tr \left(\rho Z_j Z_k\right) = \sum_{\pm} \sum_{\alpha} p_{\alpha,\pm}  \langle \psi_{\alpha, \pm}| Z_j Z_k |\psi_{\alpha, \pm}\rangle $ itself vanishes, which we know can't be true since as mentioned above, for $\beta > \beta_c$, the system is in an SSB phase with long-range order. Therefore, our assumption that $\rho$ is a sym-SRE state for $\beta > \beta_c$ must be incorrect. The same conclusion also holds for $\beta = \beta_c$ since the correlations at the critical point decay as a power-law.

As mentioned in the introduction, our general approach would be to first look for general constraints that lead to a mixed state being necessarily non-trivial. If we are unable to find such a constraint, we will attempt to find an explicit decomposition of the density matrix as a convex sum of SRE states. For example, above, we noted that $\rho$ cannot be a sym-SRE state for $\beta \geq \beta_c$, and we also claimed that $\rho$ is an SRE state for all non-zero temperatures. Let us therefore try to find an explicit decomposition of $\rho$ as a convex sum of SRE pure states for any non-zero temperature, and as a convex sum of symmetric, pure SRE states for $\beta < \beta_c$. The key player in our argument will be a particular convex decomposition ansatz (CDA in short)  that is motivated from ``minimally entangled typical thermal states'' (METTS) construction introduced in Ref.\cite{white2009minimally}, and which was employed in Ref.\cite{lu2020detecting} to show that the Gibbs state of  2d and 3d toric code is SRE for all non-zero temperatures. Note that despite the nomenclature, METTS construction as introduced in \cite{white2009minimally} does not involve minimization of entanglement over all possible decompositions, and is simply an ansatz that is physically motivated (which is why we prefer the nomenclature CDA over METTS for our discussion).

First, let us specialize to zero transverse field. In this case, $\rho$ is clearly an SRE state at any temperature since $\rho \propto \sum_m e^{- \beta E_m} |z_m \rangle \langle z_m|$ where $|z_m\rangle$ denotes a product state in the $Z$-basis and $E_m = \langle z_m | H | z_m\rangle$. To obtain a symmetric convex decomposition, we write:

\be 
\label{Eq:METTS}
\rho = \frac{\sum_{m} e^{-\beta H/2}|x_m\rangle \langle x_{m}| e^{-\beta H/2}}{Z} =\sum_m  |\psi_m\rangle \langle \psi_m|,
\ee 

where the set $\{|x_m\rangle\}$ corresponds to the complete set of states in the $X$ basis, {$|\psi_m\rangle = e^{-\beta H/2}|x_m\rangle/\sqrt{{Z}}$ is the unnormalized wave function}. The states $|\psi_m\rangle$ are clearly symmetric under the Ising symmetry, and their symmetry charge ($= \pm 1$) is determined by the parity of the number of sites in the product state $|x_m\rangle$ where spins point along the negative-$x$ direction. We will now argue that the states $|\psi_m\rangle$ are SRE for $\beta < \beta_c$ and LRE for $\beta \geq \beta_c$. To see this, we first consider the ``partition function with respect to $|\psi_m\rangle$'' defined as $\mathcal{Z}_m = \langle \psi_m |\psi_m\rangle$ and study its analyticity as a function of $\beta$. In this specific example, since transverse field is set to zero, one finds that for all $m$, $\mathcal{Z}_m$ is simply proportional to the partition function of the 2d classical Ising model at inverse temperature $\beta$, and therefore is non-analytic across the phase transition. Similarly, the two-point correlation function $\langle \psi_m |Z_i Z_j| \psi_m\rangle/\langle \psi_m | \psi_m\rangle$ is just the two-point spin-spin correlation function in the 2d classical Ising model, which is long-ranged for $\beta \geq \beta_c$ and exponentially decaying for $\beta < \beta_c$. These observations strongly indicate that $|\psi_m\rangle$ is SRE (and correspondingly, $\rho$ sym-SRE) if and only if $\beta < \beta_c$. Note that the states $|\psi_m\rangle$ are expected to be area-law entangled for all $\beta$. This is because one may represent the imaginary time evolution $e^{-\beta H}|m\rangle$ as a tensor network of depth $\beta$ acting on $|m\rangle$ (which is a product state), which can only generate an area-law worth of entanglement. Further, even the state at $\beta = \infty$ is area-law entangled (= the ground state of $H$). Therefore, short-range correlations are strongly suggestive of short-range entanglement.

Now, let's consider non-zero transverse field. To argue that $\rho$ is SRE for any non-zero temperature, {we again decompose it as $\rho = \sum_m |\psi_m\rangle \langle \psi_m|$ where $|\psi_m\rangle = e^{-\beta H/2}|z_m\rangle/\sqrt{Z}$}. The corresponding  $\mathcal{Z}_m = \langle \psi_m |\psi_m\rangle$ can now be expressed in the continuum limit as an imaginary-time path integral $\mathcal{Z}_m \sim \int_{\phi(\tau = 0) = \phi({\tau = \beta}) = \phi_0} D\phi \,\,e^{- S}$ where $S = \sum_n \int_{k_x,k_y} |\phi(k_x, k_y, n)|^2 (k^2_x + k^2_y + \omega^2_n) + \int_{ \tau = 0}^{\beta} \int_{x,y}\left( r |\phi|^2 + u |\phi|^4 \right)$, $\omega_n = 2 \pi n/\beta$ are the Matsubara frequencies, and, {crucially}, the Dirichlet boundary conditions $\phi(x,y,\tau = 0) = \phi({x,y,\tau = \beta}) = \phi_0(x,y)$ are imposed by the `initial' state $z_m \sim \phi_0(x,y)$. 
Since $\beta \neq \infty$, the discrete sum over the Matsubara frequencies  will be dominated by $\omega_n = 0$,
{which corresponds to space-time configurations that are translationally invariant along the imaginary-time-direction.} 
{
Furthermore, the Dirichlet boundary conditions imply that there is just one such configuration, namely, $\phi(x,y,\tau) = \phi_0(x,y)$ such that $\mathcal{Z}_m \sim e^{S[\phi_0(x,y)]}$, and thus the fluctuations of $\phi$ will be completely suppressed at all non-zero temperatures (including at the finite temperature critical point which corresponds to renormalized $r = 0$)
}
Therefore, we expect that $\mathcal{Z}_m$ will not exhibit singularity across the finite temperature critical point, which indicates that the states $|\phi_m\rangle$ are SRE.

To argue that $\rho$ is sym-SRE for $\beta < \beta_c$, {we now decompose $\rho$ as $\rho = \sum_m  |\psi_m\rangle \langle \psi_m|$ where $|\psi_m\rangle = e^{-\beta H/2}|x_m\rangle/\sqrt{Z}$}. The corresponding  $\mathcal{Z}_m = \langle \psi_m |\psi_m\rangle$ can again be expressed in the continuum limit as an imaginary-time path integral $\mathcal{Z}_m \sim \int D\phi \,\,e^{- S}$ where $S = \sum_n \int_{k_x,k_y} |\phi(k_x, k_y, n)|^2 (k^2_x + k^2_y + \omega^2_n) + \int_{ \tau = 0}^{\beta} \int_{x,y}\left( r |\phi|^2 + u |\phi|^4 \right)$. 
{
Crucially, since the initial state is now a product state in the $X$ basis,  the fields at the two boundaries $\tau = 0,\beta$ are being integrated over all possible configurations
}
Again, the path integral will be dominated by $\omega_n = 0$ which only implies that the dominant contribution comes from configurations $\phi(\tau, x,y) = \phi(x,y)$. 
Therefore, unlike the aforementioned case when the CDA states corresponded to  $e^{-\beta H/2}|z_m\rangle /\sqrt{Z}$, here dominant contribution to $\mathcal{Z}_m$ precisely corresponds to the partition function of the 2d classical Ising model, which is in the paramagnetic phase for {$\beta < \beta_c$}. The correspondence with 2d classical Ising model makes physical sense since the universality class of the phase transition at any non-zero temperature is indeed that of the 2d classical Ising model. 
Therefore, we expect that the states   $|\psi_m\rangle = e^{-\beta H/2}|x_m\rangle/\sqrt{Z}$ are SRE for $\beta < \beta_c$ and LRE for $\beta \geq \beta_c$. Correspondingly, we expect that the Gibbs state is sym-SRE (sym-LRE) for $\beta < \beta_c$ ($\beta > \beta_c$).

To summarize, we provided arguments that the Gibbs state of a transverse field Ising model is an SRE state at any non-zero temperature, and a sym-SRE state only for $\beta < \beta_c$. Therefore, we expect that it undergoes a separability transition as a function of temperature if one is only allowed to expand the density matrix as a convex sum of symmetric states. We expect similar statements for other  models that exhibit a finite temperature zero-form symmetry breaking phase transition.
In the following sections, we will employ broadly similar logic as in this example, with primary focus on topological phases of matter subjected to local decoherence.
Specifically, we write $\rho = \Gamma \Gamma^\dagger$ and employ the following CDA:
\begin{equation}
	\label{Eq:metts}
	\rho = \sum_m \Gamma |m\rangle \langle m| \Gamma^\dagger = \sum_m |\psi_m\rangle \langle \psi_m| = \sum_m p_m |\tilde{\psi}_m\rangle \langle \tilde{\psi}_m|,
\end{equation}
where $|\psi_m\rangle = \Gamma|m\rangle$, $p_m = \langle m| \Gamma^\dagger \Gamma|m\rangle =  \langle \psi_m  |\psi_m\rangle$, and $|\tilde{\psi}_m\rangle = |\psi_m\rangle/\sqrt{  \langle \psi_m  |\psi_m\rangle}$ are normalized versions of $|\psi_m\rangle$.
We note that here $\Gamma $ is not unique (note that $\Gamma$ is not restricted to be a square matrix, see e.g. Ref.\cite{chen2023separability}), and CDA in Eq.\eqref{Eq:METTS} corresponds to choosing $\Gamma = \rho^{1/2}$ for the Gibbs state $\rho$. We will sometimes call states $\{|\psi_m\rangle\}$ that enter a particular CDA as `CDA states'. {We further note that in general we do not know how to find the matrix $\Gamma$ that is `optimal' i.e. a matrix $\Gamma$ that guarantees that the states $\Gamma |m\rangle$ are SRE whenever $\rho$ is SRE. However, as we will see in the rest of the paper, for a large class of problems, after a judicious choice of the basis $\{|m\rangle\}$, the decomposition in the form of Eq.\ref{Eq:metts} turns out to be optimal.}

\section{Separability transitions in  SPT States} 
\label{sec:trivial}

The fundamental property of a non-trivial SPT phase is that it cannot be prepared using a short-depth  circuit consisting of local, symmetric, unitary gates \cite{gu2009tensor,pollmann2012symmetry,chen2013symmetry,schuch2011classifying}. Therefore, it is natural to ask: if an SPT phase is subjected to local decoherence, is the resulting mixed state sym-SRE, i.e., can it be expressed as a convex sum of symmetric, SRE pure states? This is clearly a very challenging question for many-body mixed states since to our knowledge, there does not exist an easily calculable measure of mixed-state entanglement that is non-zero if and only if the mixed state is unentangled \cite{horodecki2009quantum} (if such a measure did exist, then it would be useful to study its universal, long-distance component, similar to topological part of negativity \cite{lu2020detecting,lu2023characterizing,fan2023diagnostics}). As already hinted in the introduction, our general scheme will be to first seek sufficient conditions that make a given mixed state sym-LRE (i.e. not sym-SRE). We will do this by decomposing the decohered state into its distinct symmetry sectors as $\rho = \sum_Q \rho_Q$, with $\rho_Q$ the projection of the density matrix onto symmetry charge $Q$, and then examining whether the assumption of each $\rho_Q$ being an SRE leads to a contradiction. If we are unable to find an obvious contradiction, we will then attempt to use the decomposition outlined in Eq.\eqref{Eq:metts} to express $\rho$ as a convex sum of sym-SRE states. In either of these steps, we will exploit the connection between local and thermal decoherence for cluster states that was briefly mentioned in Ref.\cite{chen2023separability}, and which is described in the next subsection in detail.

\subsection*{A relation between local and thermal decoherence}
Systems with intrinsic topological order typically behave rather differently when they are coupled to a thermal bath, compared to when they are subjected to decoherence induced by a short-depth quantum channel. For example, when 2d and 3d toric codes are embedded in a thermal bath, so that the mixed state is described by a Gibbs state, the topological order is lost at any non-zero temperature \cite{dennis2002,hastings2011topological,yoshida2011,lu2020detecting}.  In contrast, when 2d or 3d toric codes are subjected to local decoherence, then the error-threshold theorems \cite{shor1996fault,aharonov1997fault,kitaev2003fault,knill1998resilient,preskill1998reliable,terhal2015quantum} imply that the mixed-state topological order is stable upto a non-zero decoherence rate \cite{dennis2002,wang2003confinement, fan2023diagnostics, lee2023quantum, bao2023mixed,chen2023separability}. Given this, it is interesting to ask if there exist situations where a local short-depth channel maps a ground state to a Gibbs state. Here we show that this is indeed the case if the corresponding Hamiltonian satisfies the following properties:
\\
(1) It can be written as a sum of local commuting terms where each of them squares to identity:
\begin{equation}
	\label{Eq:ham_spt}
	H = \sum_{j} h_j,\ [h_j,h_k] = 0,\ h_j^2 = I,\ \forall j,k.
\end{equation}
\\
(2) There exists a local unitary $O_j$ which anticommutes (commutes) with $h_k$ if $j = k\, (j \neq k)$:
\begin{equation}
	\begin{aligned}
		\label{Eq:Oj_kraus}
		O_j h_j  O^\dagger_j& = -h_j, \\
		O_j h_k  O^\dagger_j& = h_k \ (j \neq k).
	\end{aligned}
\end{equation}
Specifically, denoting the total system size as $N$, the channel $\mathcal{E} = \mathcal{E}_1 \circ \cdots \mathcal{E}_N$ with
\begin{equation}
	\label{Eq:local_channel}
	\mathcal{E}_j[\rho] = (1-p)\rho + p O_j \rho  O^\dagger_j
\end{equation}
maps the ground state density matrix $\rho_0 $ to a Gibbs state for $H$.
\\
To verify the claim, we first note that 
Eq.\eqref{Eq:ham_spt} implies that $\rho_0$ can be written as the product of the projectors on all sites $\rho_0 = \frac{1}{2^N} \prod_j (I-h_j)$. Now, using Eq.\eqref{Eq:Oj_kraus}, it is straightforward to show that $\mathcal{E}_j[\rho_0] = \frac{1}{2^N} [I-(1-2p)h_j]\prod_{k\neq j} (I-h_k)$. It then follows that the composition of $\mathcal{E}_j$ on all sites gives
\begin{equation}
	\mathcal{E}[\rho_0] = \frac{1}{2^N} \prod_j [I-(1-2p)h_j]. \label{eq:pre-exprho}
\end{equation}
Since $h_j^2 = I$, which implies $e^{-\beta h_j} = \cosh(\beta) I - \sinh(\beta) h_j$, one may now exponentiate Eq.\eqref{eq:pre-exprho} to obtain $\mathcal{E}[\rho_0] = \frac{1}{\mathcal{Z}} e^{-\beta H}$ where $\tanh\beta =(1-2p)$, $\mathcal{Z} = \tr(e^{-\beta H})$. In Sec.\ref{sec:summary}, we also discuss a $\mathbb{Z}_N$ generalization of this construction. For the rest of the paper, the aforementioned $\mathbb{Z}_2$ version will suffice. In the following we will exploit the connection between local and thermal decoherence to study decoherence-induced separability transitions for the cluster states in various dimensions (Secs.\ref{sec:1dcluster},\ref{sec:2dcluster},\ref{sec:3dcluster}). We will also briefly discuss a couple examples where the pure state is protected by a single zero-form symmetry (Sec.\ref{sec:zeroformSPT}).

\subsection{1d cluster state} \label{sec:1dcluster}

\begin{figure*}
	\centering
	\includegraphics[width=0.8\linewidth]{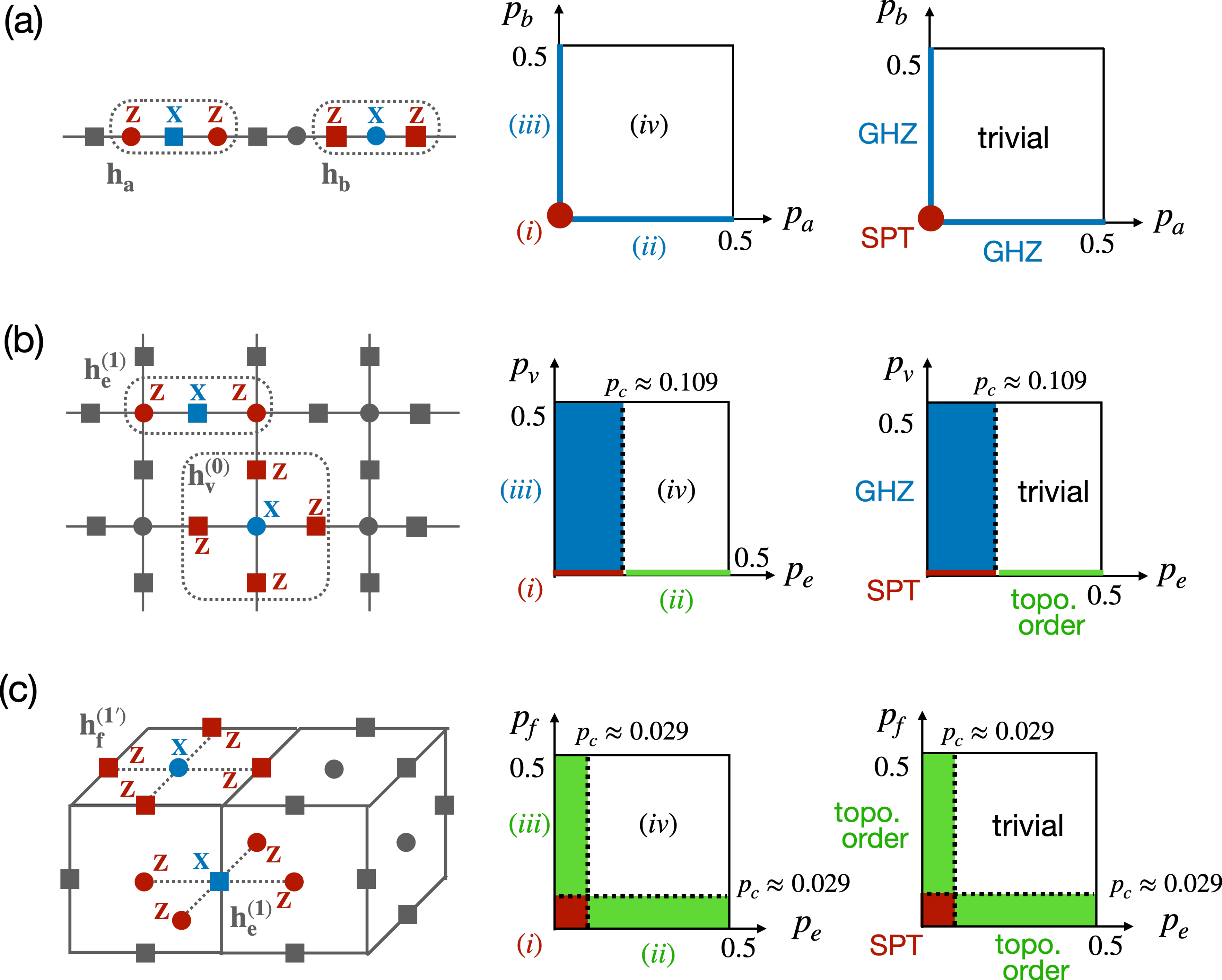}
	\caption{Cluster states under decoherence in (a) 1d, (b) 2d, and (c) 3d.
		The first column depicts the Hamiltonian of cluster states. 
		The second column divides the decohered mixed state as a function of error rates into several regimes that have qualitatively different behaviors. The white regions (region (iv)) in the three phase diagrams denote phases where the mixed state is `sym-SRE' (`trivial'), i.e., it is expressible as a convex sum of symmetric, short-ranged entangled pure states. In contrast, the colored regions or lines (regions (i), (ii), (iii)) denote phases where such a decomposition is not possible (`sym-LRE'). There can be phase transitions from one kind of sym-LRE phase to a different kind of sym-LRE phase as depicted by different colors. The phase diagram is obtained by calculating objects of the form $[\langle O \rangle^2 ]  =   \sum_{Q} P(Q) \left(\langle O \rangle_{Q} \right)^2$ where $O$ corresponds to an appropriate observable that characterizes symmetry-enforced long-range entanglement, and $P(Q)$ is the probability for obtaining the symmetry charge $q$. The $p_c \approx 0.109$ in the second row corresponds to the ferromagnetic to paramagnetic phase transition in the 2d random-bond Ising model along the Nishimori line, while $p_c \approx 0.029$ in third row corresponds to the critical point in the 3d random plaquette gauge model along the Nishimori line. The third column shows the phase diagram obtained by expressing $\rho$ as a convex sum of symmetric states, where each symmetric state $|\psi_m\rangle = \rho^{1/2} |m\rangle$ with  $|m\rangle$ the product state in Pauli-$X$ basis. See main text for more details.
	} 
	\label{Fig:cluster_main}
\end{figure*}

The Hamiltonian for  the 1d cluster state is 
\begin{equation}
	\label{Eq:cluster_ham}
	\begin{aligned}
		H
		& =-\sum_{j=1}^{N} (Z_{b,j-1} X_{a,j}  Z_{b,j} + Z_{a,j} X_{b,j}  Z_{a,j+1})\\ 
		& 
		= \sum_{j=1}^{N} h_{a,j} + h_{b,j}
	\end{aligned}
\end{equation}
where $a$ and $b$ denote the two sublattices of the 1d chain, see Fig.\ref{Fig:cluster_main}(a). $H$ has a global $\mathbb{Z}_2 \times \mathbb{Z}_2$ symmetry generated by
\begin{equation}
	U_a = \prod_j X_{a,j},\ U_b = \prod_j X_{b,j}.
\end{equation}
We assume periodic boundary conditions, so that their is a unique, symmetric, ground state of $H$ which is separated from the rest of the spectrum with a finite gap. It is obvious that $H$ satisfies Eq.\eqref{Eq:ham_spt}. To satisfy Eq. \eqref{Eq:Oj_kraus}, we choose Kraus operators $O_{a/b,j} = Z_{a/b,j}$.
Therefore, under the composition of the channel $\mathcal{E}_{a/b,j}[\rho] = (1- p_{a/b})\rho + p_{a/b}  Z_{a/b,j} \rho Z_{a/b,j}$ on all sites, the pure state density matrix becomes
\begin{equation}
	\label{Eq:1d_cluster_rho}
	\begin{aligned}
		\rho(p_a, p_b) & =  \Big( \frac{1}{\mathcal{Z}_a} e^{-\beta_a \sum_j h_{a,j} } \Big) \Big(\frac{1}{\mathcal{Z}_b} e^{- \beta_b \sum_j h_{b,j}} \Big)  \\
		& = \rho_a(p_a) \rho_b (p_b),
	\end{aligned}
\end{equation}
with $\tanh\beta_{a/b} = (1-2p_{a/b})$ and $\mathcal{Z}_{a/b} = \tr(e^{-\beta_{a/b} \sum_j h_{a/b,j}})$.
In the following, we will  suppress the arguments $p_a, p_b$ in $\rho_a(p_a), \rho_b (p_b)$ if there is no ambiguity. Note that $\rho_a$ and $\rho_b$ commute with each other.
To decompose $\rho$ as a convex sum of symmetric states, we write $\rho = \sum_{Q_a, Q_b}\rho_{Q_a, Q_b}$, where each $\rho_{Q_a, Q_b}$ is an unnormalized density matrix that carries exact symmetry: $U_{a} \, \rho_{Q_a, Q_b} = (-1)^{Q_{a}} \rho_{Q_a, Q_b}$, $U_{b}\, \rho_{Q_a, Q_b} = (-1)^{Q_{b}}\rho_{Q_a, Q_b} $, with $Q_a = 0, 1$ and $Q_b = 0, 1$, so that the sum over $Q_a, Q_b$ contains four terms. The explicit expression for  $\rho_{Q_a, Q_b}$ is given as: $\rho_{Q_a, Q_b}  = \rho_{Q_a} \rho_{Q_b}$, where $\rho_{Q_a} = \rho_a P_{Q_a}$ and $\rho_{Q_b} = \rho_b P_{Q_b}$, and  $P_{Q_{a/b}}  = (I + (-1)^{Q_{a/b}} U_{a/b})/2 $ are projectors.  Note that the probability for a given sector $(Q_a, Q_b)$ is given by $ \tr\left( \rho_{Q_a, Q_b}\right)$ which can be used to obtain the normalized density matrix $\tilde{\rho}_{Q_a,Q_b}$ for a sector $(Q_a,Q_b)$ as $\tilde{\rho}_{Q_a,Q_b} = \rho_{Q_a, Q_b}/ \tr\left( \rho_{Q_a, Q_b}\right)$.

To discuss whether the decohered mixed state $\rho$ is trivial based on our definition of sym-SRE mixed state, we start from considering the special case $p_a > 0,\ p_b = 0$, i.e., the mixed state obtained by applying the aforementioned quantum channel only on sublattice $a$. This case was studied in detail in Ref.\cite{ma2023topological} from a different perspective and is an example of an `average-SPT' phase \cite{lee2022symmetry,zhang2022strange, ma2022average, ma2023topological}. In particular, it was shown in Ref.\cite{ma2023topological} that this mixed state can not be purified to an SRE pure state using a finite-depth local quantum channel. As discussed in Sec.\ref{sec:separability}, our definition of SRE mixed state is a bit different (namely, whether a mixed state can be written as a convex sum of SRE pure states), and therefore, it is worth examining whether this state continues to remain an LRE mixed state using our definition.

When $p_a > 0,\ p_b = 0$, only the sector corresponding to $Q_b = 0$ survives, and in this sector, $\rho_{Q_a, Q_b} \propto \prod_{j} (I-h_{b,j})  e^{-\beta_a \sum_j h_{a,j} } P_{Q_a}$. We will now provide two separate arguments that show that $\rho_{Q_a, Q_b}$ is a sym-LRE (i.e. not  sym-SRE) mixed state when $p_a > 0,\ p_b = 0$.

\paragraph*{\underline{First argument:}} We want to show that  $\rho_{Q_a, Q_b} \propto \prod_{j} (I-h_{b,j})  e^{-\beta_a \sum_j h_{a,j} } P_{Q_a}$ can not be written as $\sum_m p_m |\psi_m\rangle \langle \psi_m|$ where $|\psi_m\rangle$ are SRE states that can be prepared via a short-depth circuit consisting of symmetric, local gates. We utilize the result in Ref.\cite{levin2020constraints}, which shows that for an area-law entangled state in 1D (which we will take to be $|\psi_m\rangle$) which is symmetric under an Ising symmetry (which we will take here to be $U_a = \prod_j X_{a,j}$), both order and disorder parameters cannot vanish simultaneously.
Note that we are assuming that $|\psi_m\rangle$ has an area-law entanglement, as otherwise, it is certainly not SRE and there is nothing more to prove.

Therefore, following the results in Ref.\citep{levin2020constraints}, $|\psi_m\rangle$ must either 
(a) have a non-zero order parameter corresponding to the symmetry $U_a$ i.e. $\langle \psi_m |\tilde{Z}_{j} \tilde{Z}_{k} |\psi_m\rangle \neq 0$ where $|j-k| \gg 1$ and $\tilde{Z}$ is an operator that is odd under $U_a$ e.g. $\tilde{Z}_{i} = Z_{a,i}$, or, (b) it must have a non-zero `disorder parameter' corresponding to the symmetry $U_a$ i.e. $\langle \psi_m |O_L \left(\prod_{l=j}^k X_{a,l} \right) O_{R}|\psi_m\rangle \neq 0$ where $|j-k| \gg 1$, and $O_L, O_{R}$ are operators localized close to site $j$ and $k$ respectively that are \textit{either} both even or both odd under $U_a$.
In case (a), the system has a long-range GHZ type order since the state $|\psi_m\rangle$ is symmetric under $U_a$. In case (b), we now argue that the system has an SPT order.

For $\langle \psi_m |O_L \left(\prod_{l=j}^k X_{a,l} \right) O_{R}|\psi_m\rangle$ to be non-zero, the operator $O_L \otimes O_R$ must carry no charge under the symmetry $U_b$ as  $|\psi_m\rangle$ is an eigenstate of $U_b$.
Therefore, there are two disjoint possibilities for the operators $O_L$ and $O_R$: they are either both charged under the symmetry $U_b$ or neither of them are charged under $U_b$.
If neither of them are charged under  $U_b$, then $\langle \psi_m |O_L \left(\prod_{l=j}^k X_{a,l} \right) O_{R}|\psi_m \rangle $ must vanish. 
This is because $|\psi_m\rangle$ is an eigenstate of the string operator {$S_{b}(l,r) =  Z_{a,l} \left( \prod_{j=l}^{r}X_{b,j} \right) Z_{a,r+1}$}(this follows from the fact that $\rho_{Q_a, Q_b} \propto \prod_{j} (I-h_{b,j})$) which anticommutes with $O_L \left(\prod_{l=j}^k X_{a,l} \right) O_{R}$ for an appropriate choice of {$(l,r)$}, whenever neither of $O_L, O_R$ are charged under $U_b$.  
As a consequence, for $\langle \psi_m |O_L \left(\prod_{l=j}^k X_{a,l} \right) O_{R}|\psi_m\rangle $ to be non-zero, $O_L$ and $O_R$ must both be odd under $U_b$. If so, then the disorder order parameter precisely corresponds to one of the two SPT string order parameters, namely, $S_{a}(j,k) =  Z_{b,j-1} \left( \prod_{l=j}^{k}X_{a,l} \right) Z_{b,k+1}$ up to finite depth symmetric unitary transformation.  At the same time, the other SPT string order parameter $\langle \psi_m |S_{b}(j,k)|\psi_m \rangle$ is also non-zero (due to $\rho_{Q_a, Q_b} \propto \prod_{j} (I-h_{b,j})$), and therefore, we arrive at the conclusion that in case (b),  $|\psi_m\rangle$ must possess  non-trivial SPT order since string order parameters on both sublattices are non-zero.
Therefore, in either case (a) or (b), $|\psi_m\rangle$ cannot be prepared by a short-depth circuit composed of local gates that respect both  $U_a$ and $U_b$, starting with a symmetric product state. 

\vspace{0.2cm}
\paragraph*{\underline{Second argument:}} This argument is essentially the same as the one introduced in Ref.\citep{huang2015quantum} to show that the circuit depth of various states with a non-trivial string order parameter cannot be a system-size independent constant due to locality/Lieb-Robinson bound  \cite{lieb1972finite,hastings2010locality}. Again, recall that we want to show that  $\rho_{Q_a, Q_b} \propto \prod_{j} (I-h_{b,j})  e^{-\beta_a \sum_j h_{a,j} } P_{Q_a}$ can not be written as $\sum_m p_m |\psi_m\rangle \langle \psi_m|$ where $|\psi_m\rangle$ are SRE. Since  $\rho_{Q_a, Q_b}$ carries an exact symmetry charge of $U_a, U_b$, so do each of the pure states $|\psi_m\rangle$. As already discussed above, the expectation value of the string order parameter $S_{b}(j,k) =  \prod_{l = j}^k (-h_{b,l})  =  Z_{a,j} \left( \prod_{l=j}^{k}X_{b,l} \right) Z_{a,k+1}$ is unity with respect to $\rho_{Q_a, Q_b}$, which implies that its expectation value is also unity with respect to each of the states $|\psi_m\rangle$. Let us assume that $|\psi_m\rangle$ can be obtained from a symmetric product state (i.e. an eigenstate of Pauli $X$ on all sites) which we denote as $ |x_\mathbf{a,b}\rangle = \otimes_{j}|x_{a,j}, x_{b,j}\rangle$, i.e., $|\psi_m\rangle = V|x_\mathbf{a,b}\rangle$  (here $x_{a/b, j} = \pm 1$ are chosen so as to satisfy the symmetry $U_{a/b}|\psi_m\rangle = (-1)^{Q_{a/b}}|\psi_m\rangle$). Note that  $|x_\mathbf{a,b}\rangle$ not only satisfy the global symmetry $U_{a/b}$ but the `local' ones as well, i.e., $\prod_{j \in l} X_j |x_\mathbf{a,b}\rangle \propto  |x_\mathbf{a,b}\rangle$ for any string $l$. Since each end point of $S_b$ is charged under $U_a$ (i.e., $U_a Z_{a,j/k+1} U^\dagger_a = -Z_{a,j/k+1} $), the local symmetry of $|x_\mathbf{a,b}\rangle$ implies $\langle x_\mathbf{a,b}| S_{b}(j,k)|x_\mathbf{a,b}\rangle = 0$. Moreover, since $V$ is a finite-depth unitary,  the operator $V^\dagger S_{b}(j,k) V $ is still a string operator with each `end point operator' $V^\dagger Z_{a,j/k+1} V$ a  sum of local operators (due to the locality of $V$) that are charged under $U_a$ (due to $V$ being a symmetric unitary, i.e., $[V, U_a] = [V,U_b] = 0$). Due to these properties, the expectation value $\langle x_\mathbf{a,b}|V^\dagger S_{b}(j,k) V  |x_\mathbf{a,b}\rangle$ will be identically zero. However, $\langle x_\mathbf{a,b}|V^\dagger S_{b}(j,k) V  |x_\mathbf{a,b}\rangle$  is nothing but  $\langle \psi_m| S_{b}(j,k)|\psi_m\rangle$, which is unity, as discussed above. Therefore, we arrive at a contradiction. This implies that our assumption that $|\psi_m\rangle$ is a symmetric SRE state must be incorrect.

We now discuss the general case of both $p_a$ and $p_b$ being non-zero.
Based on our discussion above, it is instructive to evaluate the string order parameter with respect to each $\rho_{Q_a, Q_b}$, i.e., $\tr(\rho_{Q_a, Q_b}S_{a/b})/\tr(\rho_{Q_a, Q_b})$. One finds (see Appendix \ref{sec:appendix_1dcluster}) that both string order parameters can be mapped to two-point correlation functions of spins in the 1d classical Ising model at non-zero temperature and hence decay exponentially with the length of the strings. This result merely implies that the corresponding mixed state $\rho = \sum_{Q_a, Q_b} \rho_{Q_a, Q_b}$ doesn't satisfy the aforementioned sufficient condition for non-trivial sym-SRE, and does not guarantee that $\rho$ must be trivial. We now use the CDA in Eq.\eqref{Eq:metts} to argue that $\rho$ is indeed sym-SRE. In particular, we choose $\Gamma = \rho^{1/2}$ so that $\rho = \sum_m \Gamma |m\rangle \langle m| \Gamma^\dagger = \sum_m |\psi_m\rangle \langle \psi_m|$ with $|\psi_m\rangle \propto e^{-(\beta_a\sum_j h_{a,j} + \beta_b \sum_j h_{b,j})/2}|m\rangle$.
To ensure each $|\psi_m\rangle$ respects the global $\mathbb{Z}_2 \times \mathbb{Z}_2$ symmetry, we choose the set $\{ |m\rangle \}= \{|x_\mathbf{a}, x_\mathbf{b}\rangle_m\}$.
When $\beta_{a} = \beta_b = 0$, $|\psi_m\rangle = |x_\mathbf{a}, x_\mathbf{b}\rangle_m$ is a product state. To check whether $|\psi_m\rangle$ remains SRE for any non-infinite $\beta_a$ and $\beta_b$, let us consider the `partition function with respect to $|\psi_m\rangle$'
\be 
\mathcal{Z}_m (\beta_a, \beta_b) = \langle \psi_m|\psi_m\rangle \label{eq:partition_fn}
\ee 
as a function of $\beta$. As $\beta_a,\beta_b$ are increased from zero, if the state $|\psi_m\rangle$ becomes long-range entangled, one expects that it will lead to a non-analytic behavior of $\mathcal{Z}_m (\beta)$ as a function of $\beta_a, \beta_b$. The calculation for $\mathcal{Z}_m (\beta) =\langle x_\mathbf{a}, x_\mathbf{b}|\rho|x_\mathbf{a}, x_\mathbf{b} \rangle $ is quite similar to the one for $\tr(\rho_{Q_a, Q_b})$ detailed in Appendix \ref{sec:appendix_1dcluster}, and one finds that $\mathcal{Z}_m (\beta)$ is proportional to the product of two partition functions for the 1d classical Ising model at inverse temperatures $\beta_a, \beta_b$. Therefore, we expect that $|\psi_m  (\beta)\rangle$ remains an SRE state as long as both $\beta_a, \beta_b < \infty$, which confirms our expectation that $\rho$ is sym-SRE for non-infinite $\beta_a, \beta_b$  (i.e. $p_a, p_b > 0$).

One can also compute the string order parameters $S_{a} (S_b)$ for $|\psi_m\rangle$ and show its equivalence to $\langle z_j z_k\rangle_{\text{1D Ising}}$ at inverse temperature $\beta_a (\beta_b)$. Therefore, $|\psi_m\rangle$ does not develop string order as long as $\beta_{a/b}< \infty$. The triviality of $|\psi_m\rangle$  is also manifested by the \textit{non-zero} expectation value of disorder operator $U_{a/b}(k,j) = \prod_{l = j}^{k} X_{a/b,l}$. For example, consider the expectation value of the disorder operator on $a$ sublattice: $\langle U_{a}(k,j) \rangle_m = \langle \psi_m | U_{a}(k,j) |\psi_m \rangle/\langle \psi_m|\psi_m\rangle$. Using the fact that the only terms in $e^{-(\beta_a\sum_j h_{a,j} + \beta_b \sum_j h_{b,j})/2}$ that anticommutes with $U_a(k,j)$ are $h_{b,j-1}$ and $h_{b,k}$, we find that $\langle U_{a}(k,j) \rangle_m = (\prod_{l = j}^k x_l) \sech^{2}(\beta_a)$, which is non-vanishing except for $\beta_{a} = \infty$. This is of course expected based on the result of Ref.\cite{levin2020constraints}, since $|\psi_m\rangle$ does not have any GHZ type order. The result for $U_b(k,j)$ is similar.

It is also instructive to apply the aforementioned convex decomposition to the case $\beta_b = \infty, \beta_a \neq \infty$, i.e., the above discussed case of `average SPT order'. In this case we find that the corresponding state $|\psi_m\rangle$ develops GHZ type long-range entanglement. To see this, 
one can rewrite $|\psi_m\rangle$ as $|\psi_m\rangle \sim e^{-\beta_a \sum_k h_{a,j}/2}|\chi_m\rangle$, where $|\chi_m\rangle \sim \prod_j (I-h_{b,j})|m\rangle = |x_\mathbf{b}\rangle \otimes \prod_j (I-x_{b,j}Z_{a,j}Z_{a,j+1})|x_\mathbf{a}\rangle $  exhibits GHZ-type long-range entanglement characterized by $|\langle \chi_m|Z_{a,j} Z_{a,k} |\chi_m\rangle | = 1$.
Using the fact that the only terms in $e^{-\beta_a\sum_j h_{a,j}/2} $ that anticommute with $Z_{a,j} Z_{a,k}$ are $h_{a,j}$ and $h_{a,k}$, one finds that $|\langle \psi_m| Z_{a,j} Z_{a,k}|\psi_m \rangle | = \sech^{2}(\beta_a)|\langle \chi_m|Z_{a,j} Z_{a,k} |\chi_m\rangle |=\sech^{2}(\beta_a) $, which is non-vanishing except for $\beta_{a} = \infty$.

To summarize the results in this subsection, the decohered state $\rho$ as a function of $p_a$ and $p_b$ can be divided into four regimes (see Fig.\ref{Fig:cluster_main}):

\begin{enumerate}[(i)]
	\item $p_a = p_b =0$: $\tr(\rho_{Q_a, Q_b} S_a(j,k)) = 1 $ (in the $Q_a = 0$ sector) and $\tr(\rho_{Q_a, Q_b} S_b(j,k)) = 1$ (in the $Q_b = 0$ sector). This is just the pure state SPT.
	
	\item $p_a > 0$ and $p_b = 0 $: $\tr(\rho_{Q_a, Q_b} S_a (j,k))$ decays exponentially with $|j-k|$ and $\tr(\rho_{Q_a, Q_b} S_b (j,k)) = 1$ (in the $Q_b = 0$ sector). This regime is sym-LRE i.e. a non-trivial mixed state, in agreement with the non-trivial `average SPT' discussed in Ref.\cite{ma2023topological}.
	
	\item $p_a = 0$ and $p_b >  0$:  this is similar to the case (ii) with $a \leftrightarrow b$ and is again a sym-LRE state.
	
	\item $p_a, p_b > 0 $: both $\tr(\rho_{Q_a, Q_b} S_a (j,k))$  and $\tr(\rho_{Q_a, Q_b} S_b (j,k))$ decay exponentially with $|j-k|$. This is a sym-SRE state.
\end{enumerate}

Based on our discussion above, we also provide one possible `phase diagram' to express $\rho$ as a convex sum of symmetric states using CDA states $|\psi_m\rangle = \rho^{1/2} |x_\mathbf{a}, x_\mathbf{b}\rangle$, as summarized in the third column of Fig.\ref{Fig:cluster_main}(a).
Note that the boundary of the phase diagram using the employed CDA matches the boundary of regimes (i)-(iv), and therefore, the CDA is optimal in this sense. However, it's worth noting that the decomposition we chose is just one possible choice, and the label `GHZ' on the $x$ and $y$ axis in the third column of Fig.\ref{Fig:cluster_main}(a) is tied to this choice. One may also chose to expand $\rho$ as a convex sum of SPT states. Therefore, the result that is independent of any specific choice of CDA is that the regime (iv) is sym-SRE, while the regimes (i), (ii) and (iii) are sym-LRE.

\subsection{2d cluster state} \label{sec:2dcluster}

The 2d cluster state Hamiltonian $H_{\textrm{2d Cluster}}$ is:

\begin{equation}
	\label{Eq:2dcluster}
	\begin{aligned}
		H_{\textrm{2d Cluster}} & = -\sum_v X_v (\prod_{e \ni v} Z_e) - \sum_e X_e (\prod_{v \in e} Z_v) \\
		& = \sum_v h_{v} + \sum_e h_{e}.
	\end{aligned}
\end{equation}
Here the Hilbert space consists of qubits residing on both the vertices $v$ and the edges $e$ of a 2D square lattice, see Fig.\ref{Fig:cluster_main}(b).
The Hamiltonian has both a zero-form symmetry $Z^{(0)}_2$, and a one-form symmetry $Z^{(1)}_2 $ with the corresponding generators
\begin{equation}
	U^{(0)} = \prod_v X_v,\ U_{ p}^{(1)} = \prod_{e \in \partial p} X_e,
\end{equation}
where $p$ labels the plaquette on the lattice and $\partial p$ is the boundary of $p$. We assume periodic boundary conditions, so that $H$ has a unique, symmetric, gapped ground state.
Using Eqs.\eqref{Eq:ham_spt},\eqref{Eq:Oj_kraus}, if one subjects the ground state of  $H_{\textrm{2d Cluster}}$ to Kraus operators $O_{v/e} = Z_{v/e}$ with respective probabilities $p_{v/e}$, the resulting decohered density matrix is  $\rho =  \frac{1}{Z} e^{-\left(\beta_{v} \sum_v h_{v}^{(0)}+ \beta_{e}\sum_e h_{e}^{(1)}\right)}$
with $\tanh\beta_{e/v} = (1-2p_{e/v})$.

Let us decompose $\rho$ as a convex sum of symmetric states by writing $\rho = \sum_{Q^{(0)}, Q^{(1)}}\rho_{Q^{(0)}, Q^{(1)}}$, where each $\rho_{Q^{(0)}, Q^{(1)}}$ carries the exact symmetry: $U^{(0)} \rho_{Q^{(0)}, Q^{(1)}} = (-1)^{Q^{(0)}} \rho_{Q^{(0)}, Q^{(1)}}$, $U^{(1)}_{ p} \rho_{Q^{(0)}, Q^{(1)}} = (-1)^{Q^{(0)}_{p}} \rho_{Q^{(0)}, Q^{(1)}}$.
Here, the one-form symmetry charge is labeled by the set $Q^{(1)} = \{ Q^{(1)}_p\}$ with $Q^{(1)}_p =  0, 1$ defined on each plaquette $p$.  Crucially, the number of one-form symmetry sectors grows exponentially as a function of the system size, and this implies that the probability for a given sector $(Q^{(0)}, Q^{(1)})$, i.e., $\tr(\rho_{Q^{(0)}, Q^{(1)}})$, is exponentially small in general. 
It follows that even if there exists some $\rho_{Q^{(0)}, Q^{(1)}}$ that is not sym-SRE, the decohered state $\rho$ may still be well approximated by a sym-SRE mixed state as long as the total probability corresponding to the non-trivial sectors is exponentially small.
Therefore, the notion of $\rho$ being sym-SRE must take  into account the probability for each symmetry sector, and can only be made precise in a statistical sense (a similar situation arises for a certain non-optimal decomposition for decohered toric code \cite{chen2023separability}. We will return to this point in detail below. For now, let's focus on the physical observables in each symmetry sector.

The observables that characterize the 2d cluster ground state are the expectation value of the membrane operator $M_S = \prod_{v \in S} (-h_v^{(0)})$ with $S$ a surface (for simplicity, we will assume that the boundary $\partial S$ of this surface is contractible), and the string operator $S_C = \prod_{e \in C} (-h_e^{(1)})$ with $C$ a curve (the expectation value of either of these operators equals unity in the 2d cluster ground state). To detect whether $\rho_{Q^{(0)}, Q^{(1)}}$ is sym-SRE, i.e., it can be expanded as a convex sum of pure SRE states that each carries a definite symmetry charge $(Q^{(0)}, Q^{(1)})$, it is instructive to calculate the expectation value of these operators with respect to $\rho_{Q^{(0)}, Q^{(1)}}$, i.e., $\tr(\rho_{Q^{(0)}, Q^{(1)}} M_S)/\tr(\rho_{Q^{(0)}, Q^{(1)}})$ and $\tr(\rho_{Q^{(0)}, Q^{(1)}} S_C)/\tr(\rho_{Q^{(0)}, Q^{(1)}})$. To proceed, we first compute the denominator in these expressions, i.e., $\tr(\rho_{Q^{(0)}, Q^{(1)}})$. Similar to the 1d cluster state, this can be easily done by inserting the complete basis $\{ |x_\mathbf{e,v}\rangle \}$ and $\{ |z_\mathbf{e,v}\rangle \}$, where  $|{x}_\mathbf{e,v}\rangle = \otimes_{e,v} |{x}_{e},x_{v}\rangle$ and $|{z}_\mathbf{e,v}\rangle = \otimes_{e,v} |{z}_{e}, z_{v}\rangle$ denote the product state in Pauli-$X$ and $Z$ basis, respectively. Following a calculation quite similar to that in the 1D cluster state, one finds that $\tr(\rho_{Q^{(0)}, Q^{(1)}}) \propto \sum_{x_\mathbf{v} \in Q^{(0)}} \mathcal{Z}_{\text{2D gauge}, x_\mathbf{v}}
\sum_{x_\mathbf{e} \in Q^{(1)}} \mathcal{Z}_{\text{2D Ising}, x_\mathbf{e}} $.
Here, $ \mathcal{Z}_{\text{2D gauge}, x_\mathbf{v}} = \sum\limits_{z_\mathbf{e}} e^{\beta_v \sum_v x_v (\prod_{e \ni v} z_e)} $ is the partition function of the 2D Ising gauge theory with the sign of interaction on each vertex given by $x_\mathbf{v}$ while  $\mathcal{Z}_{\text{2D Ising}, x_\mathbf{e}} = \sum\limits_{z_\mathbf{v}} e^{\beta_e \sum_e x_e (\prod_{v \in e} z_v)}$ is the partition function of the 2D Ising model with the sign of Ising interaction given by $x_\mathbf{e}$. In the summation, the notation $x_\mathbf{v} \in Q^{(0)} $ denotes all possible $x_\mathbf{v}$ which satisfy $\prod_v x_\mathbf{v} = (-1)^{Q^{(0)}}$ while $x_\mathbf{e} \in Q^{(1)} $ denotes all possible $x_\mathbf{e}$ which satisfy $\prod_{e \in \partial p} x_\mathbf{e} = (-1)^{Q^{(1)}_p},\ \forall p$.
For a system with periodic boundary conditions, all possible $x_\mathbf{v} \in Q^{(0)} (x_\mathbf{e} \in Q^{(1)}) $ can be reached by the  transformation $x_v \rightarrow x_v \prod_{e \ni v} \sigma_e , \sigma_e = \pm 1$ ($x_{e} \rightarrow x_{e} \prod_{v \in e} s_v, s_v = \pm 1$).
One may verify that $\mathcal{Z}_{\text{2D gauge}, x_\mathbf{v}}$ ($\mathcal{Z}_{\text{2D Ising}, x_\mathbf{e}}$) is invariant under the aforementioned transformation by changing the dummy variables $z_e \rightarrow \sigma_e z_e$ ($z_v \rightarrow s_v z_v$).
It follows that $\mathcal{Z}_{\text{2D gauge}, x_\mathbf{v}}$ ($\mathcal{Z}_{\text{2D Ising}, x_\mathbf{e}}$)  is only a function of the charge $Q^{(0)} (Q^{(1)})$, and therefore we will label it as $\mathcal{Z}_{\text{2D gauge}, Q^{(0)}}$  ($\mathcal{Z}_{\text{2D Ising}, Q^{(1)}}$).
Therefore, $\tr(\rho_{Q^{(0)}, Q^{(1)}}) \propto \mathcal{Z}_{\text{2D gauge},  Q^{(0)}} \mathcal{Z}_{\text{2D Ising}, Q^{(1)}}$ (see footnote \footnote{ Here we ignore the non-contractible `charges' corresponding to $\prod_{\ell} x_e$ where $\ell$ is a non-contractible loop around the torus on which the system lives. This is because we will only be concerned with observables involving operators in the bulk of the system and such observables are insensitive to non-contractible charges.}).

One may similarly compute $\tr(\rho_{Q^{(0)}, Q^{(1)}} M_S)$ and $\tr( \rho_{Q^{(0)}, Q^{(1)}}S_C)$, the numerators in the expectation value for the membrane  and the  string operators.  Let us first consider the membrane order parameter in the sector $(Q_0, Q_1)$ which we denote as $\langle M_S \rangle_{Q_0,Q_1}$. One finds
\begin{equation}
	\begin{aligned}
		\label{Eq:2d_membrane_sector}
		& \langle M_S \rangle_{Q_0,Q_1} =\frac{\tr(\rho_{Q^{(0)}, Q^{(1)}} M_S)}{\tr(\rho_{Q^{(0)}, Q^{(1)}})}  \\
		& =   \frac{\sum_{z_\mathbf{e}} (\prod_{v \in S} x_v \prod_{e \in \partial S} z_e) e^{\beta_v \sum_v x_v (\prod_{e \ni v} z_e)}}{\mathcal{Z}_{\text{2D gauge}, Q^{(0)}}} \Bigg|_{x_\mathbf{v} \in Q^{(0)}} \\
		& = (\prod_{v \in S} x_v )\langle W_{\partial S} \rangle_{\text{2D gauge}, x_\mathbf{v}} \Big|_{x_\mathbf{v} \in Q^{(0)}} \\
		& \sim e^{-\kappa \textrm{Area}(S)} \,\,\,\textrm{for}\,\, \beta_v < \infty
	\end{aligned}
\end{equation}
where $\langle W_{\partial S} \rangle_{\text{2D gauge}, x_\mathbf{v}}$ is the expectation value of the Wilson loop operator along the curve $\partial S$ for the 2D Ising gauge theory with interaction $x_\mathbf{v}$ while Area($S$) is the area enclosed by the surface $S$. The area law follows because the 2d Ising gauge theory is confining at any non-zero temperature. We conclude that $\rho_{Q^{(0)}, Q^{(1)}}$ has no membrane order as long as $p_v >0$.

On the other hand, the string order parameter $\langle S_C \rangle_{Q_0,Q_1}$ is
\begin{equation}
	\label{Eq:2d_string_sector}
	\begin{aligned}
		& \langle S_C \rangle_{Q_0,Q_1} = \frac{\tr(\rho_{Q^{(0)}, Q^{(1)}} S_C)}{\tr(\rho_{Q^{(0)}, Q^{(1)}})} \\
		& =  \frac{\sum_{z_\mathbf{v}} (\prod_{e \in C} x_e)  z_{v_1} z_{v_2}  e^{\beta_e \sum_e x_e (\prod_{v \in e} z_v)}}{\mathcal{Z}_{\text{2D Ising}, Q^{(1)}}}\Bigg|_{x_\mathbf{e} \in Q^{(1)}}   \\
		& = (\prod_{e \in C} x_e) \langle z_{v_1} z_{v_2}  \rangle_{\text{2D Ising}, x_\mathbf{e}} \Big|_{x_\mathbf{e} \in Q^{(1)}},
	\end{aligned}
\end{equation}
where $v_1$ and $v_2$ label the end points of the curve $C$ and $\langle z_{v_1} z_{v_2}  \rangle_{\text{2D Ising},  x_\mathbf{e}}$ is the spin-spin correlation function of the 2D Ising model with the sign of the Ising interaction determined by $x_\mathbf{e}$. Clearly, $\langle S_C \rangle_{Q_0,Q_1}$ can show long-range order at low-temperature, and following the same argument as that for the 1d cluster state, long-range order for a given sector implies that the (unnormalized) density matrix $\rho_{Q^{(0)}, Q^{(1)}}$ is sym-LRE. For example, in the sector corresponding to all $x_e = 1$, the long range order sets in below 2d Ising critical temperature. However, since the ordering temperature clearly depends on the sector $Q^{(1)}$, to understand whether the full density matrix $\rho = \sum_{Q^{(0)}, Q^{(1)}}\rho_{Q^{(0)}, Q^{(1)}}$ is sym-LRE, one needs to statistically quantify the string order as a function of the error rate. To do so, we introduce the following `average string order parameter':

\begin{equation}
	\label{Eq:average_string_order}
	\begin{aligned}
		[\langle S_C \rangle^2 ]  =   \sum_{Q^{(0)}, Q^{(1)}} \tr(\rho_{Q^{(0)}, Q^{(1)}}) \left(\langle S_C \rangle_{Q_0,Q_1} \right)^2
	\end{aligned}
\end{equation} 

Eq.\eqref{Eq:average_string_order} is equivalent to the disorder averaged spin-spin correlation function of RBIM along the Nishimori line \cite{nishimori1981internal}. It follows that $[\langle S_C \rangle^2 ] $ decays exponentially as a function of $|C|$ when $p_e > p_c \approx 0.109$ \cite{honecker2001universality}. 

Based on above analysis, the decohered state $\rho$ as a function of $p_e$ and $p_v$ can be divided into four regimes using the qualitative behavior of membrane and average string orders [see Fig.\ref{Fig:cluster_main}(b)]: 

\begin{enumerate}[(i)]
	\item  $p_v = 0$ and $p_c > p_e \geq 0$: $\langle M_S \rangle_{Q_0,Q_1}  = 1$  (in the sector $Q^{(0)} = 0$) and $ [\langle S_C \rangle^2 ]$ is a non-zero constant as $|C| \rightarrow \infty$. In this regime  $\rho$ must be sym-LRE.
	
	\item $p_v = 0$ and $p_e > p_c$: $\langle M_S \rangle_{Q_0,Q_1}  = 1$  (in the sector $Q^{(0)} = 0$) and $ [\langle S_C \rangle^2 ]$ decays exponentially as a function of $|C|$. In this regime  $\rho$ must again be sym-LRE.
	
	\item $p_v > 0$ and $p_c > p_e \geq 0$:  $\langle M_S \rangle_{Q_0,Q_1} \sim e^{-\textrm{Area}(S)} $  and $ [\langle S_C \rangle^2 ]$ is a non-zero constant as $|C| \rightarrow \infty$. In this regime  $\rho$ must also be (statistically) sym-LRE.
	
	\item $p_v > 0 $ and $p_e > p_c$: $\langle M_S \rangle_{Q_0,Q_1} \sim e^{-\textrm{Area}(S)} $  and $ [\langle S_C \rangle^2 ] \sim e^{-|C|}$. This is suggestive that in this regime $\rho$ is (statistically) sym-SRE and we provide an argument in favor of this conclusion below using an explicit convex decomposition.
\end{enumerate}
We now use the CDA in Eq.\eqref{Eq:metts} with $\Gamma = \sqrt{\rho}$ to argue that the regime (iv) above, namely $p_v > 0 $ and $p_e > p_c$, is indeed sym-SRE. 
To ensure that each CDA state $|\psi_m\rangle$ satisfies the $Z^{(0)}_2 \times Z^{(1)}_2$ symmetry, we choose $\{ |m\rangle =  |x_{\mathbf{v}}, x_\mathbf{e}\rangle \}$.
Similar to the 1D case, we consider the singularity of `partition function' $\mathcal{Z}_m = \langle \psi_m |\psi_m\rangle$ as a diagnostic for transition from SRE to LRE as $\beta$ is increased from zero.
Since $\mathcal{Z}_m =\langle x_\mathbf{e}, x_\mathbf{v}|\rho|x_\mathbf{e}, x_\mathbf{v} \rangle $, a calculation similar to that for $\tr(\rho_{Q^{(0)}, Q^{(1)}})$ shows that  $\mathcal{Z}_m$ is proportional to $\mathcal{Z}_{\text{2D Ising gauge}, x_\mathbf{v}} \mathcal{Z}_{\text{2D Ising}, x_\mathbf{e}}$. One can also compute the expectation values of membrane and average string order operators with respect to $|\psi_m\rangle$ and obtain $\langle M_S \rangle_m$ is proportional to the expression in Eq.\eqref{Eq:2d_membrane_sector} while $[\langle S_C \rangle_m^2 ] $ is proportional to the expression in  Eq.\eqref{Eq:average_string_order}, and therefore both vanish when $p_v > 0 $ and $p_e > p_c$.

Alternatively, one may define an `average free energy' $[\log \mathcal{Z}] = \sum_m P_m \log(\mathcal{Z}_m) \propto \sum_m \mathcal{Z}_m \log(\mathcal{Z}_m)$ with respect to $|\psi_m\rangle $ to detect whether the ensemble $\{ \psi_m\rangle \}$  encounters a phase transition as a function of the error rate. When $\beta = 0$, $|\psi_m\rangle = |x_\mathbf{a}, x_\mathbf{b}\rangle_m$ is the trivial product state.  On the other hand, $|\psi_m\rangle$ becomes the 2D cluster state when $\beta \rightarrow \infty$. One  expects that the phase transition point can be located by the singular behavior of $[\log \mathcal{Z}]$. Since $[\log \mathcal{Z}]$ is proportional  to the disorder-averaged free energy of the 2d RBIM along the Nishimori line, it is singular at $p_e  \approx  0.109$. This leads to the same conclusion that $\{ |\psi_m\rangle \}$ remains SRE in the regime (iv) above.

Interestingly, if one adopts the aforementioned CDA in regimes (ii) and (iii), then $|\psi_m\rangle$ hosts intrinsic topological order and GHZ order, respectively. This can be argued by first considering the extreme case $(p_v, p_e) = (0,0.5)$ in regime  (ii)  and $(p_v, p_e) = (0.5,0)$ in regime (iii).
When $(p_v, p_e) = (0,0.5)$, $|\psi_m\rangle \propto \prod_v (I +h_{v}^{(0)})|m\rangle \propto (|x_\mathbf{v}\rangle \otimes \prod_v (I+x_v \prod_{e \ni v} Z_e)|x_\mathbf{e}\rangle)$, which is an eigenstate of toric code. On the other hand, when $(p_v, p_e) = (0.5,0)$, $|\psi_m\rangle \propto \prod_e (I +h_{e}^{(1)})|m\rangle \propto (|x_\mathbf{e}\rangle \otimes \prod_e (I+ x_e \prod_{v \in e} Z_v)|x_\mathbf{v}\rangle)$ is the 2D GHZ state. The argument based on the analyticity of the average free energy $[\log \mathcal{Z}] $ then indicates that regimes (ii)  and (iii) continue to host topological order and GHZ order, respectively. The phase diagram using the current decomposition is summarized in Fig.\ref{Fig:cluster_main}(b). 

Finally, we note that order parameters similar to $[\langle S_C \rangle^2 ]$ (Eq.\eqref{Eq:average_string_order}) and the connections between the decohered cluster states and RBIM have also appeared in the context of preparing long-range entangled states using measurement protocols in Refs.\citep{lee2022measurement,zhu2022nishimori}. In paricular, our  phase diagram (Fig.\ref{Fig:cluster_main}(b)) along the line $p_v = 0.5$ is similar to the finite-time measurement induced phase transtions in Ref.\citep{lee2022measurement,zhu2022nishimori}. However, one crucial  difference is that the mixed states in Refs. \citep{lee2022measurement,zhu2022nishimori} do not respect the $ Z_2^{(1)}$ symmetry and therefore the corresponding transitions can not be interpreted as separability transitions protected by $Z^{(0)}_2 \times Z^{(1)}_2$ symmetry between a sym-LRE phase and a sym-SRE phase.  Instead, the role of different sectors corresponding to the $ Z_2^{(1)}$ symmetry is played by the flux  $f_p = \prod_{e \in p} s_e$ through a plaquette $p$, where $s_e$ is the measurement outcome. One may then regard the transition in Ref.\citep{lee2022measurement,zhu2022nishimori} as a separability transition where in the non-trivial phase it is impossible to decompose the density matrix as a convex sum of SRE states which carry both definite $\mathcal{Z}_2^{(0)}$ charge and flux $f_p$. Similar statements hold true for the case of 3D cluster state, which we discuss next.

\subsection{3d cluster state} \label{sec:3dcluster}
The 3d cluster state Hamiltonian $H_{\textrm{3d Cluster}}$ is:

\begin{equation}
	\label{Eq:3dcluster} 
	\begin{aligned}
		H_{\textrm{3d Cluster}}  & = - \sum_e X_e \prod_{f \ni e} Z_f - \sum_f X_f \prod_{e \in f} Z_e \\
		& = \sum_e h_e + \sum_f h_f.
	\end{aligned}
\end{equation}
The Hilbert space consists of qubits residing at both the faces $f$ and the edges $e$ of a cubic lattice, see Fig.\ref{Fig:cluster_main}(c), or equivalently, at the edges of a cubic lattice, and the edges of its dual lattice (recall that each edge (plaquette) of the original lattice is in one-to-one correspondence with a plaquatte (edge) of the dual lattice). We assume periodic boundary conditions. This model has a $\mathbb{Z}^{(1)}_2 \times \mathbb{Z}^{(1')}_2$ symmetry whose generators are given by
\begin{equation}
	U^{(1')}_{c} = \prod_{f \in \partial c} X_f,\ U^{(1)}_{\tilde{c}} = \prod_{ e \in  \partial \tilde{c}} X_e.
\end{equation}
where $c (\tilde{c})$ specifies the cube in the lattice (dual lattice) and $\partial c ( \partial \tilde{c})$ denotes the faces on the boundary of $c (\tilde{c})$. Choosing Kraus operators $O_{e/f} = Z_{e/f}$ with respective probabilities $p_{e/f}$, using Eqs.\eqref{Eq:ham_spt},\eqref{Eq:Oj_kraus}, one obtains the decohered state $\rho =  \frac{1}{Z} e^{-\beta_e   \sum_e h_{e}^{(1)} - \beta_f \sum_f h_{f}^{(1')}}$ with $\tanh\beta_{e/f} = (1-2p_{e/f})$.

We now decompose $\rho$ as a convex sum of symmetric states by writing $\rho = \sum_{Q^{(1')}, Q^{(1)}}\rho_{Q^{(1')}, Q^{(1)}}$, where each $\rho_{Q^{(1')}, Q^{(1)}}$ carries exact symmetry: $U^{(1')}_c \rho_{Q^{(1')}, Q^{(1)}} = (-1)^{Q^{(1')}_c} \rho_{Q^{(1')}, Q^{(1)}}$, $U^{(1)}_{\tilde{c}} \rho_{Q^{(1')}, Q^{(1)}} = (-1)^{Q^{(1)}_{\tilde{c}}} \rho_{Q^{(1')}, Q^{(1)}}$.
Here, two one-form symmetry charges are labeled by $Q^{(1')} = \{ Q^{(1')}_c \}$ with $Q^{(1')}_c =  0, 1$ defined on each cube $c$ and 
$ Q^{(1)} = \{ Q^{(1)}_{\tilde{c}} \}$ with $Q^{(1)}_{\tilde{c}} = 0, 1$ defined on each cube $\tilde{c}$ in the dual lattice.
Now, let's focus on the physical observables that characterize each sector. These are the membrane operators $M_S = \prod_{f \in S} (-h_{f}^{(1')})$ with $S$ a contractible surface on the original lattice (by contractible surface we mean an open-membrane whose boundary $\partial S$ is non-zero and is a closed loop) and $M_{\tilde{S}} = \prod_{e \in \tilde{S} } (-h_{e}^{(1')})$ with $\tilde{S}$ a non-contractible surface on the dual lattice. Thus, we want to compute $\tr(\rho_{Q^{(1')}, Q^{(1)}} M_S)/\tr(\rho_{Q^{(1')}, Q^{(1)}})$ and $\tr(\rho_{Q^{(1')}, Q^{(1)}} M_{\tilde{S}} )/\tr(\rho_{Q^{(1')}, Q^{(1)}})$. 

Similar to the  cases in previous sections, we first compute the denominator $\tr(\rho_{Q^{(1')}, Q^{(1)}})$ in these expressions by inserting the complete complete basis $\{ |x_{\mathbf{f,e}} \rangle \}$ and $\{ |z_\mathbf{f,e} \rangle \}$, and obtain $\tr(\rho_{Q^{(1')}, Q^{(1)}}) \sim \sum_{x_\mathbf{f} \in Q^{(1')}}\mathcal{Z}_{\text{3D gauge}, x_{\mathbf{f}}} \sum_{x_\mathbf{e} \in Q^{(1)} }\mathcal{Z}_{\text{3D gauge}, x_\mathbf{e}}$.
Here, $\mathcal{Z}_{\text{3D gauge}, x_\mathbf{f}} = \sum_{z_\mathbf{e}}  e^{\beta_f \sum_f x_f (\prod_{e \in f} z_e)}$ is the partition function of the 3D Ising guage theory with the sign of the interaction on each face labeled by $x_{\mathbf{f}}$, and $x_\mathbf{f} \in Q^{(1')}$ denotes all possible $x_\mathbf{f}$ satisfying $\prod_{f \in \partial c} x_\mathbf{f} = (-1)^{Q^{(1')}_c}$.
For a system with periodic boundary condition, all possible $x_f \in Q^{(1')}$ can be reached by the  transformation $x_f \rightarrow x_f \prod_{e \ni f} \sigma_e , \sigma_e = \pm 1$. Further, one may verify that $\mathcal{Z}_{\text{3D gauge}, x_\mathbf{f}}$ is invariant under the aforementioned transformation by changing the dummy variables $z_e \rightarrow \sigma_e z_e$. It follows that $\mathcal{Z}_{\text{3D gauge}, x_\mathbf{f}} = \mathcal{Z}_{\text{3D gauge}, Q^{(1')}}$ is only a function of charge $Q^{(1')}$.  Analogous statements hold true for $\mathcal{Z}_{\text{3D gauge}, x_\mathbf{e}}$. Therefore, we write
\begin{equation}
	\tr(\rho_{Q^{(1')}, Q^{(1)}}) \propto \mathcal{Z}_{\text{3D gauge}, Q^{(1')}} \mathcal{Z}_{\text{3D gauge}, Q^{(1)}}.
\end{equation}
One may similarly compute $\tr(\rho_{Q^{(1')}, Q^{(1)}} M_S)$, and obtain the following expressions:
\begin{equation}
	\begin{aligned}
		\label{Eq:3d_membrane_sector}
		& \langle M_S \rangle_{Q^{(1')}, Q^{(1)}}  = \frac{\tr(\rho_{Q^{(1')}, Q^{(1)}} M_S)}{\tr(\rho_{Q^{(0)}, Q^{(1)}})}  \\
		& =   \frac{\sum_{z_\mathbf{e}} (\prod_{f \in S} x_f \prod_{e \in \partial S} z_e) e^{\beta_f \sum_f x_f (\prod_{e \in f} z_e)}}{\mathcal{Z}_{\text{3D gauge}, Q^{(1')}}} \Bigg|_{x_\mathbf{f} \in Q^{(1')}} \\
		& = (\prod_{f \in S} x_f )\langle W_{\partial S} \rangle_{\text{3D gauge}, x_\mathbf{f}} \Big|_{x_\mathbf{f} \in Q^{(1')}},
	\end{aligned}
\end{equation}
where $\langle W_{\partial S} \rangle_{\text{3D gauge}, x_\mathbf{f}}$ is the expectation value of the Wilson loop operator ($= \prod z_e$ along a closed curve) along the boundary of $S$ for the 3d classical Ising gauge theory whose Hamiltonian is defined by the term that multiplies $\beta_f$ in the exponential in the second line of Eq.\eqref{Eq:3d_membrane_sector}. Since the plaquette interaction term in this Ising gauge theory depends on $x_\mathbf{f} \in Q^{(1')}$, similar to the discussion for 2d cluster state, we introduce an average membrane order parameter
\begin{equation}
	\label{Eq:3D_average_membrane}
	[\langle M_S\rangle^2 ]   =  \sum_{Q^{(1')}, Q^{(1)}} \tr(\rho_{Q^{(1')}, Q^{(1)}}) \left(\langle M_S \rangle_{Q_1',Q_1} \right)^2
\end{equation}
Eq.\eqref{Eq:3D_average_membrane} precisely corresponds to the disorder averaged Wilson loop of the 3D random plaquette gauge model (RPGM) along the Nishimori line \cite{wang2003confinement}.
It follows that  $[\langle M_S\rangle^2 ]  \sim e^{-\kappa |\partial S|}$ (`perimeter-law') when $p_f < p_c \approx 0.029$ while $[\langle M_S\rangle^2 ]  \sim e^{-\kappa | S|}$ (`area-law') when $p_f > p_c $.
One can also define the average membrane order parameter $[\langle M_{\tilde{S}}\rangle^2 ]$ for $M_{\tilde{S}}$, and the results are analogous with the same critical error rate $p_c$.

Therefore, using the qualitative behaviors of $[\langle M_S\rangle^2 ]  $ and $[\langle M_{\tilde{S}}\rangle^2 ]  $, one can divide the decohered state $\rho$ as a function of $p_f$ and $p_e$ into four regimes, see Fig.\ref{Fig:cluster_main}(c):
\begin{enumerate}[(i)]
	\item $p_f, p_e < p_c$: both $[\langle M_S\rangle^2 ]  $ and $[\langle M_{\tilde{S}}\rangle^2 ] $ satisfy perimeter law.
	
	\item $p_f < p_c, p_e > p_c$: $[\langle M_S\rangle^2 ]  $ satisfies perimeter-law while $[\langle M_{\tilde{S}}\rangle^2 ] $  satisfies area-law.
	\item $p_f > p_c, p_e < p_c$: $[\langle M_S\rangle^2 ]  $ satisfies area-law while $[\langle M_{\tilde{S}}\rangle^2 ] $ satisfies perimeter-law.
	\item  $p_f, p_e > p_c$: Both $[\langle M_S\rangle^2 ]  $ and $[\langle M_{\tilde{S}}\rangle^2 ] $ satisfy area-law.
\end{enumerate}

Using an argument similar to  Ref.\citep{lu2023mixed}, and also similar to those already used in previous subsections for 1d and 2d cluster states, one can show that in regimes (i)-(iii), $\rho$ cannot be a convex sum of symmetric pure states where membrane operators only exhibit an area-law. This suggests that these three regimes are sym-LRE.
In regime (iv), $\rho$ does not develop any average membrane orders, which strongly suggests that it is a sym-SRE state. We now use a CDA to support this expectation.

We again choose a CDA (Eq.\eqref{Eq:metts})  with $\Gamma = \sqrt{\rho}$. To ensure that each $|\psi_m\rangle$ that enters the CDA satisfies $\mathbb{Z}^{(1)}_2 \times \mathbb{Z}^{(1')}_2$ symmetry, we choose the basis $\{ |m\rangle =  |x_{\mathbf{e}}, x_\mathbf{f}\rangle \}$. Similar to the previous cases, we consider the `partition function' $\mathcal{Z}_m = \langle \psi_m |\psi_m\rangle$ whose singularties are expected to indicate the presence of a phase transition. The evaluation of $\mathcal{Z}_m =\langle x_\mathbf{f}, x_\mathbf{e}|\rho|x_\mathbf{f}, x_\mathbf{e} \rangle $ is quite  similar to that for $\tr(\rho_{Q^{(1')}, Q^{(1)}})$ and one finds that $\mathcal{Z}_m \sim \mathcal{Z}_{\text{3D gauge}, x_{\mathbf{f}}} \mathcal{Z}_{\text{3D gauge}, x_\mathbf{e}}$. One may also compute the expectation values of two membrane operators and find $\langle \psi_m |M_{S}| \psi_m \rangle = (\prod_{f \in S} x_f) \langle W_{\partial S}\rangle_{\text{3D gauge}, x_\mathbf{f}}$ and $\langle \psi_m |M_{\tilde{S}}| \psi_m \rangle = (\prod_{e \in \tilde{S}} x_e) \langle W_{\partial \tilde{S}}\rangle_{\text{3D gauge}, x_\mathbf{e}}$. Using these one may then define an average membrane order parameters $[\langle M_S \rangle^2] = \sum_m P_m \langle \psi_m |M_{S}| \psi_m\rangle ^2 $ and  $[\langle M_{\tilde{S}} \rangle^2] = \sum_m P_m \langle \psi_m |M_{\tilde{S}}| \psi_m\rangle ^2 $. Using same arguments as those following Eq.\eqref{Eq:3D_average_membrane}, one concludes that both of these order parameters vanish in regime (iv).

One may also conclude that the aforementioned decomposition in regimes (ii) and (iii) correspond to topologically ordered phases.
This can be argued by first considering the extreme case $(p_f, p_e) = (0,0.5)$ in (ii) and $(p_f, p_e) = (0.5,0)$ in (iii).
When $(p_f, p_e) = (0,0.5)$, $|\psi_m\rangle \sim \prod_f (I +h_{f}^{(0)})|m\rangle \sim (|x_\mathbf{f}\rangle \otimes \prod_f (I+x_f \prod_{e \in f} Z_e)|x_\mathbf{e}\rangle)$, which is an eigenstate of the 3D toric code. The argument based on the singularity of average free energy $[\log \mathcal{Z}]$ then indicates that in regime (ii) CDA states are topologically ordered. Similar arguments hold for regime (iii). The phase diagram using such a convex decomposition is summarized in the third column of Fig.\ref{Fig:cluster_main}(c).

It is interesting to compare our results with Ref.\cite{yoshida2016topological} where the Gibbs state of 3d cluster Hamiltonian was studied. The main difference between the decohered state we study, which is also takes the Gibbs form, with the state studied in Ref.\cite{yoshida2011} is that in Ref.\cite{yoshida2011}, the Gibbs state is projected to a \textit{single} charge sector of both 1-form symmetries (and therefore possesses an exact symmetry, see comment $\#3$ in Sec.\ref{sec:separability}), which results in a phase transition as a function of temperature that is in the 3d Ising universality. In contrast, the decoherence we are considering leads only to an average (instead of exact) symmetry, and therefore, we obtain an \textit{ensemble} of density matrices $\rho_{Q^{(1')}, Q^{(1)}}$ labeled by the symmetry charges $Q^{(1')}, Q^{(1)}$. As discussed above, this implies that the universality class of the transition is related to the 3d random plaquette gauge model (and not 3d Ising transition).

\subsection{1d and 2d topological  phases protected by a $Z_2^{(0)}$ symmetry} \label{sec:zeroformSPT}
Aside from the cluster states in several dimensions, Eq.\eqref{Eq:ham_spt} and Eq.\eqref{Eq:Oj_kraus} also holds for various stabilizer models realizing 1d and 2d  SPT phases protected by a $Z_2^{(0)}$ symmetry, which we now discuss briefly. An example in 1d is the non-trivial phase of the Kitaev chain \cite{kitaev2001unpaired}:
\begin{equation}
	H = - i \sum_j \gamma_{2j-1} \gamma_{2j} 
\end{equation} 
where $\gamma_j$ denotes the majorana operator satisfying $\{ \gamma_j, \gamma_k\} = 2\delta_{ij}$.
It is straightforward to see that the Hamiltonian satisfies Eq.\eqref{Eq:ham_spt} and one can choose $O_j$ as  $\gamma_{2j-1}$ or $\gamma_{2j}$ such that Eq.\eqref{Eq:Oj_kraus} is satisfied.
Therefore, under the composition of the channel $\mathcal{E}_j[\rho] = (1-p)\rho + p  \gamma_{2j-1} \rho \gamma_{2j-1}$, the pure state density matrix becomes the finite temperature Gibbs state with $\tanh \beta = 1-2p$.
A 2d example is the Levin-Gu state  \citep{levin2012braiding}, where the  Hamiltonian is defined on the triangular lattice and can be written as
\begin{equation}
	H = - \sum_p B_p,\ B_p = -X_p \prod_{\langle pqq' \rangle} i^{\frac{1-Z_q Z_{q'}}{2}},
\end{equation}
where the product runs over the six triangles $\langle p q q' \rangle$ containing the site $p$. The ground state has non-trivial SPT order for the $Z_2^{(0)}$ symmetry generated by $U = \prod_p X_p$.
One can verify $[B_p, B_{p'}] = 0$ and $B^2_p = 1$ by straightforward algebra, and thus Eq.\eqref{Eq:ham_spt} is satisfied.
Besides, one can choose $O_j = Z_j$ such that Eq.\eqref{Eq:Oj_kraus} is satisfied.
Therefore, under the composition of the channel $\mathcal{E}_j[\rho] = (1-p)\rho + p  Z_j \rho Z_j$, the pure state density matrix becomes the finite temperature Gibbs states with $\tanh \beta = 1-2p$.
Using the CDA in Eq.\eqref{Eq:metts}, one may then argue that both the decohered Kitaev chain and Levin-Gu state are sym-SRE for any non-zero $p$ (we assume periodic boundary conditions so that there are no boundary modes).

\section{Separability transitions for 2d chiral topological states}
\label{sec:fermions}
\subsection{Setup and motivation}
In this subsection, we consider subjecting chiral fermions in 2d to local decoherence. The starting pure state we consider is the ground state of a $p_x+ip_y$ superconductor ($p+ip$ SC in short), although we expect that the results will qualitatively carry over to other non-interacting chiral states. 

Our motivation is as follows: it is generally believed that the 2d $p+ip$ SC cannot be prepared from a product state using a constant-depth unitary circuit (as suggested by the fact the thermal Hall conductance of a $p+ip$ SC is non-zero while that for a trivial, gapped paramagnet is zero). Indeed, one may think of a $p+ip$ SC as an SPT phase protected by the conservation of fermion parity \cite{wen2012symmetry}. Therefore, it is natural to ask what happens if one applies a quantum channel to this system where Kraus operators  anticommute with the fermion parity.  This is conceptually similar to our discussion in Sec.\ref{sec:trivial} where we subjected a non-trivial SPT ground state to Kraus operators odd under the symmetry responsible for the existence of (pure) SPT ground state. An example of such a Kraus operator is the fermion creation/annihilation operator, and we will study this case in detail. Alternatively, one may consider subjecting $p+ip$ ground state to decoherence with Kraus operators \textit{bilinear} in fermion creation/annihilation operators. In this latter case, the fermion parity remains an exact symmetry. Based on our discussion in Sec.\ref{sec:trivial}, one may expect a qualitative difference in these two cases, namely, Kraus operators linear Vs bilinear  in fermion creation/annihilation operators. Let us briefly outline such a qualitative difference as suggested by field-theoretic considerations whose details are presented in Sec.\ref{sec:bilinearkraus}.

Let us first consider Kraus operators linear in fermion operators. This is equivalent to bringing in ancillae fermions and entangling them with the fermions of the $p+ip$ SC by a finite-depth unitary. Since this is a finite depth unitary operation on the enlarged Hilbert space (= ancillae + original $p+ip$ SC), the expectation value of any observable, including non-local ones that detect chiral topological order \cite{kvorning2020nonlocal,girvin1987off}, cannot become zero. At the same time, intuitively, the resulting mixed state for the electrons belonging to the original $p+ip$ SC must somehow ``lose its chirality'' at infinitesimal coupling to the ancillae.  This is indicated by treating the density matrix as a pure state in the doubled Hilbert space using C-J isomorphism, which we discuss below in detail, where we also clarify subtleties pertinent to the mapping of Kraus operators linear in fermion operators. Under the C-J map, the effect of the channel becomes a coupling bilinear in fermion operators between two chiral Ising CFTs with opposite chirality, and which, therefore, gaps out the counter-propagating chiral CFTs. The gapping out of the edge states in the double state is also manifested in the entanglement spectrum of the double state, which we also study. In particular, we show that infinitesimal decoherence leads to a gap in the entanglement spectrum.

Although working with the double state using C-J map is insightful, it does not directly tell us the nature of the decohered mixed state. One of our central aims is to understand the difference between the original pure (non-decohered) state and the decohered state not in terms of the  double state obtained via the C-J map, or in terms of non-linear functions of density matrix, but directly in terms of the separability properties of the mixed state. Our main result is that the resulting mixed state can be expressed as a convex sum of non-chiral states, and in this sense, is non-chiral (i.e. it can be prepared using an ensemble of finite-depth unitaries that commute with fermion parity).

Let us next consider Kraus operators bilinear in the fermion operators. We study this problem only using the double-state formalism (i.e. the aforementioned C-J map), and obtain an effective action consisting of two counter-propagating free, chiral Majorana CFTs coupled via a four-fermion interaction. Such a Hamiltonian has already been studied in the past (see e.g. Refs.\cite{grover2014emergent, rahmani2015emergent}), and we  simply borrow the previous results to conclude that unlike the case for Kraus operators linear in Majorana operators, this system is \textit{stable} against infinitesimal decoherence. Furthermore, the field-theory corresponding to the double state indicates that this system undergoes a spontaneous symmetry breaking where the gapless modes corresponding to the CFT are gapped out. The university class for this transition lies in the (supersymmetric) $c=7/10$ tricritical Ising model. We discuss this below in detail in Sec.\ref{sec:bilinearkraus}.  We note that recently, Ref.\cite{su2023conformal} studied chiral topological phases subjected to decoherence using a generalization of strange correlator \cite{you2014wave} to mixed states \cite{lee2022symmetry,zhang2022strange}. Although Ref.\cite{su2023conformal} did not study the problem of our interest (namely, $p+ip$ SC subjected to Kraus operators bilinear in Majorana fermions), the overall structure of the field theories obtained in Ref.\cite{su2023conformal} using strange correlator bears resemblance to the one we motivate using entanglement spectrum in Sec.\ref{sec:bilinearkraus}.

\subsection{Separability of $p+ip$ SC subjected to fermionic Kraus operators} \label{sec:chiral_free}
Our starting point is the ground state of the $p+ip$ superconductor  \cite{read2000paired} described by the following Hamiltonian on a square lattice
\begin{widetext}
	\begin{equation}
		\label{Eq:ppip_ham}
		H = \sum_{x,y}
		-t
		(\mathbf{c}^\dagger_{x+1,y}\mathbf{c}_{x,y} + \mathbf{c}^\dagger_{x,y+1} \mathbf{c}_{x,y}+ h.c.) 
		+
		\Delta (\mathbf{c}^\dagger_{x+1,y}\mathbf{c}^\dagger_{x,y} +i \mathbf{c}^\dagger_{x,y+1}\mathbf{c}^\dagger_{x,y} + h.c. ) 
		-(\mu - 4t)\mathbf{c}^\dagger_{x,y}\mathbf{c}	_{x,y}.
	\end{equation}
\end{widetext}
When $t = \Delta = 1/2$ and the chemical potential $\mu = 1$, the system is in the topologically non-trivial phase. This can be diagnosed, for example, by studying the entanglement spectrum which will exhibit chiral propagating modes \cite{li2008entanglement,Kitaev06_1}, or, by studying the modular commutator \cite{kim2022chiral, kim2022modular, zou2022modular, fan2022entanglement} which is proportional to the chiral central charge of the edge modes that appear if the system had boundaries. Relatedly, in the topological phase, the ground state cannot be written as a Slater determinant of exponentially localized Wannier single-particle states \cite{thouless1984wannier,read2000paired,schindler2020pairing}. In our discussion, we assume periodic boundary conditions, so that there are no physical edge modes.

We are interested in subjecting the ground state of Eq.\eqref{Eq:ppip_ham} to the composition of the following single-majorana channel on all sites:
\begin{equation}
	\label{Eq:single_maj_channel}
	\mathcal{E}_j[\rho] = (1-p)\rho+p\gamma_j \rho \gamma_j.
\end{equation}
Is the chiral nature of the ground state $\rho_0$ stable under the channel?  More precisely, can we express the decohered density matrix as a convex sum of pure states, where each of these pure states now does not exhibit chiral states in its entanglement spectrum, and relatedly, has a vanishing modular commutator in the thermodynamic limit?

Under the aforementioned channel (Eq.\eqref{Eq:single_maj_channel}), the density matrix will continue to remain Gaussian, and is fully determined by the covariance matrix $M$ defined as $M_{j k} = -i \tr(\rho (\gamma_j \gamma_k - \delta_{jk}))$. As shown in Appendix \ref{sec:covariance}, under the channel in Eq.\ref{Eq:single_maj_channel}, $M$ evolves as $\mathcal{E}(M) = (1-2p)^2 M  $. We write the decohered density matrix $\rho$ as $\rho(p) = e^{-H_\rho(p)}$, where $H_\rho(p)$ can be determined explicitly in terms of $\mathcal{E}(M) =  (1-2p)^2 M $ as detailed in the Appendix \ref{sec:covariance}.

To write the decohered mixed state $\rho$ as a convex sum of pure states, we consider the decomposition in Eq.\eqref{Eq:metts}, and write 
\bea 
\label{eq:CDA_pwave}
\rho(p) & = & \sum_m e^{-H_\rho(p)/2} |m\rangle \langle m| e^{-H_\rho(p)/2} \nonumber \\
& = & \sum_m |\psi_m\rangle \langle \psi_m|
\eea 
where $|m\rangle$ are product states in the occupation number basis: $|m\rangle = |m_1, ..., m_N\rangle,\ m_j = 0,1$ and $|\psi_m\rangle = \sqrt{\rho}|m\rangle = e^{-H_\rho(p)/2} |m\rangle$. To build intuition for the states $|\psi_m\rangle$, let's consider the particular state $|\psi_0\rangle = \sqrt{\rho}|0\rangle$ where $|0\rangle$ is a state with no fermions. One can analytically show at any non-zero decoherence, the real-space wavefunction for this state is a Slater determinant of localized Wannier orbitals, unlike the (undecohered) ground state of $p+ip$ SC \cite{thouless1984wannier,read2000paired,schindler2020pairing}. 
The argument is as follows. One may write $|\psi_0\rangle \propto e^{- \beta \sum_{\mathbf{k}} \alpha^{\dagger}_{\mathbf{k}} \alpha_k}|0\rangle$ where $\tanh(\beta) = (1-2p)^2$ and $\alpha^{\dagger}_{\mathbf{k}} = u_{\mathbf{k}} c^{\dagger}_{\mathbf{k}} + v^{*}_{\mathbf{k}} c_{-{\mathbf{k}}}$ are the same (complex) fermionic operators that diagonalize the original $p+ip$ BCS Hamiltonian (see Appendix \ref{sec:covariance}), with $|u_{\mathbf{k}}|^2 + |v_{\mathbf{k}}|^2 = 1$ due to unitarity. Since $c_{\mathbf{k}}|0\rangle = 0$, this implies that 
\be 
|\psi_0\rangle  \propto \prod_{\mathbf{k}} \left[1 + \left(e^{-\beta}- 1\right) \left(|v_{\mathbf{k}}|^2 + u_{\mathbf{k}} v_{\mathbf{k}} c^{\dagger}_{\mathbf{k}} c^{\dagger}_{-{\mathbf{k}}}\right) \right] |0\rangle
\ee
This expression may then be exponentiated to obtain the standard BCS-like form for $|\psi_0\rangle \propto e^{\sum_{\mathbf{k}} h({\mathbf{k}})c^{\dagger}_{\mathbf{k}} c^{\dagger}_{-{\mathbf{k}}}}|0\rangle$, where

\be 
h({\mathbf{k}}) = \frac{u_{\mathbf{k}} v_{\mathbf{k}} \left(e^{-\beta}- 1\right)}{|u_{\mathbf{k}}|^2 + |v_{\mathbf{k}}|^2 e^{-\beta}}
\ee 
As $p\rightarrow 0$, $\beta \rightarrow \infty$ (recall $\tanh(\beta) = (1-2p)^2$), and one recovers the $p+ip$ ground state where $h({\mathbf{k}}) \sim v_{\mathbf{k}}/u_{\mathbf{k}}$ diverges as $1/(k_x + i k_y)$ and results in a power-law decay of Wannier orbitals \cite{read2000paired}. In contrast, at any non-infinite $\beta$ (i.e. non-zero decoherence rate $p$), $h({\mathbf{k}})$ is non-infinite for any ${\mathbf{k}}$ (since $|u_{\mathbf{k}}|^2 + |v_{\mathbf{k}}|^2 = 1$), and therefore, the Wannier orbitals corresponding the state $|\psi_0\rangle$ are exponentially localized. As an aside, this same argument also applies to the decohered 1d Kitaev chain (Sec.\ref{sec:zeroformSPT}), and more generally, to other decohered non-interacting fermionic topological superconductors.

The above argument only applies to the translationally invariant state $|\psi_0\rangle$ that enters the convex decomposition in Eq.\eqref{eq:CDA_pwave}. To make progress for general $|\psi_m\rangle$, we found it more helpful to consider diagnostics which directly access the topological character (or lack thereof) of a wavefunction, and which are also more amenable to finite-size scaling. In particular, we employ the `modular commutator' introduced in Refs. \citep{kim2022chiral, kim2022modular, zou2022modular, fan2022entanglement}. Modular commutator is a multipartite entanglement measure that quantifies the chiral central charge for a \textit{pure state}, and can be completely determined by the many-body wavefuntion \citep{kim2022chiral, kim2022modular, zou2022modular, fan2022entanglement}.
Specifically, it is defined as $J_{ABC} := i\tr(\rho_{ABC} [\ln \rho_{AC}, \ln \rho_{BC}])$ with $\rho_{X}$ the reduced density matrix in region $X$ obtained from a pure state $|\psi\rangle$ (i.e. $\rho_{X} = \tr_{\overline{X}} |\psi\rangle \langle \psi|$). 

\begin{figure}
	\centering
	\includegraphics[width=\linewidth]{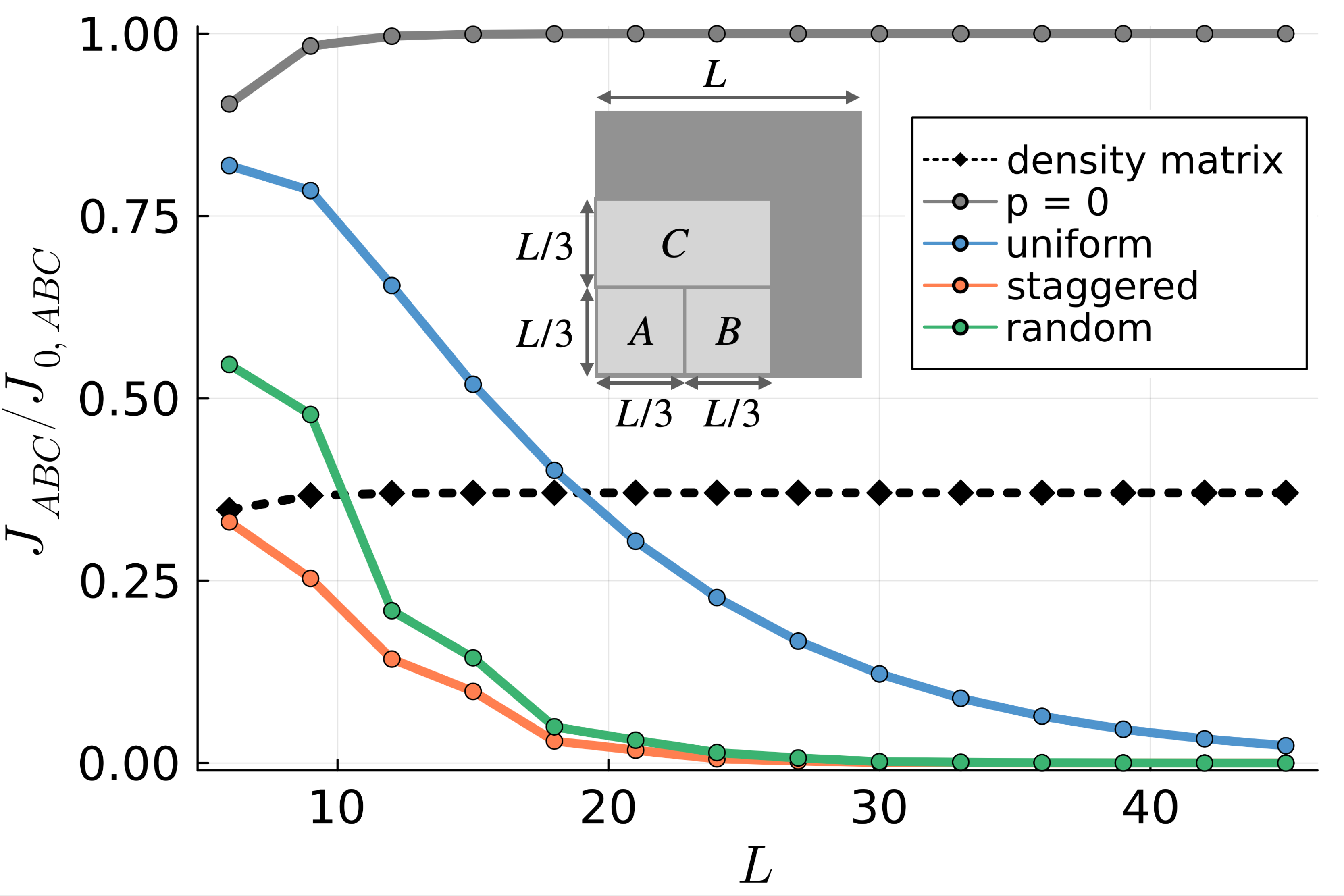}
	\caption{Modular commutator $J_{ABC}/J_{0,ABC} $ on a $L\times L$ torus as a function of $L$ corresponding to several different pure states $|\psi_m\rangle$ that enter the convex decomposition of the $p+ip$ SC subjected to decoherence with Kraus operators linear in Majorana fermions (Eq.\eqref{eq:CDA_pwave}), as well as the modular commutator of the decohered mixed state itself. We choose error rate $p = 0.04$, and the following initial states $|m\rangle$ in Eq.\eqref{eq:CDA_pwave}: $|m\rangle = |0, ..., 0\rangle$ (uniform), $|0,1,0,1,..,0,1\rangle$ (staggered), and $|m\rangle = $ a random bit string in the occupational number basis.
		The inset shows the geometry of regions $A,B,C$ used to define the modular commutator. 
		We use anti-periodic boundary conditions along both directions so that the ground state is unique.} 
	\label{Fig:JABC_p04_W36}
\end{figure}

In the absence of decoherence, the modular commutator of $|\psi_m\rangle$ for this setup is $J_{0,ABC} = \pi c/3 = \pi/6$, as the chiral central charge $c = 1/2$ for the $p+ip$ superconductor.
Fig.\ref{Fig:JABC_p04_W36} shows the modular commutator $J_{ABC}/J_{0,ABC} $ on  a $L\times L$ torus as a function of $L$. 
We choose the error rate $p = 0.04$ and several different initial states, including $|m\rangle = |0, ..., 0\rangle$ (uniform), $|0,1,0,1,..,0,1\rangle$ (staggered), and also a random bit string in the occupational number basis. We find that in all cases, $J_{ABC}$ vanishes in the thermodynamic limit. We also studied other values of $p$, and our results are again consistent with the claim that  at any non-zero $p$, the modular commutator for the states $|\psi_m\rangle$ vanishes in the thermodynamic limit. This provides  numerical evidence that at any non-zero error rate, the decohered mixed state can be expressed as a convex sum of states that do not have any chiral topological order, and hence must be representable as Slater determinants of single-particle localized Wannier states \cite{thouless1984wannier} (note that all states $|\psi_m\rangle$ are area-law entangled).

It is important to note that in contrast to the pure states $|\psi_m\rangle$, the modular commutator for the decohered mixed state $\rho$ \textit{does not} show any abrupt behavior change at $p = 0$ (dashed plot in Fig.\ref{Fig:JABC_p04_W36}). This is consistent with the fact that the arguments relating modular commutator to the chiral central charge rely on the overall state being pure \cite{kim2022chiral, kim2022modular, zou2022modular, fan2022entanglement}, and therefore, we don't expect that modular commutator for the mixed-state $\rho$ captures the separability transition at $p = 0$. This again highlights the utility of the convex decomposition of $\rho$ into pure states.

\begin{figure}
	\centering
	\includegraphics[width=\linewidth]{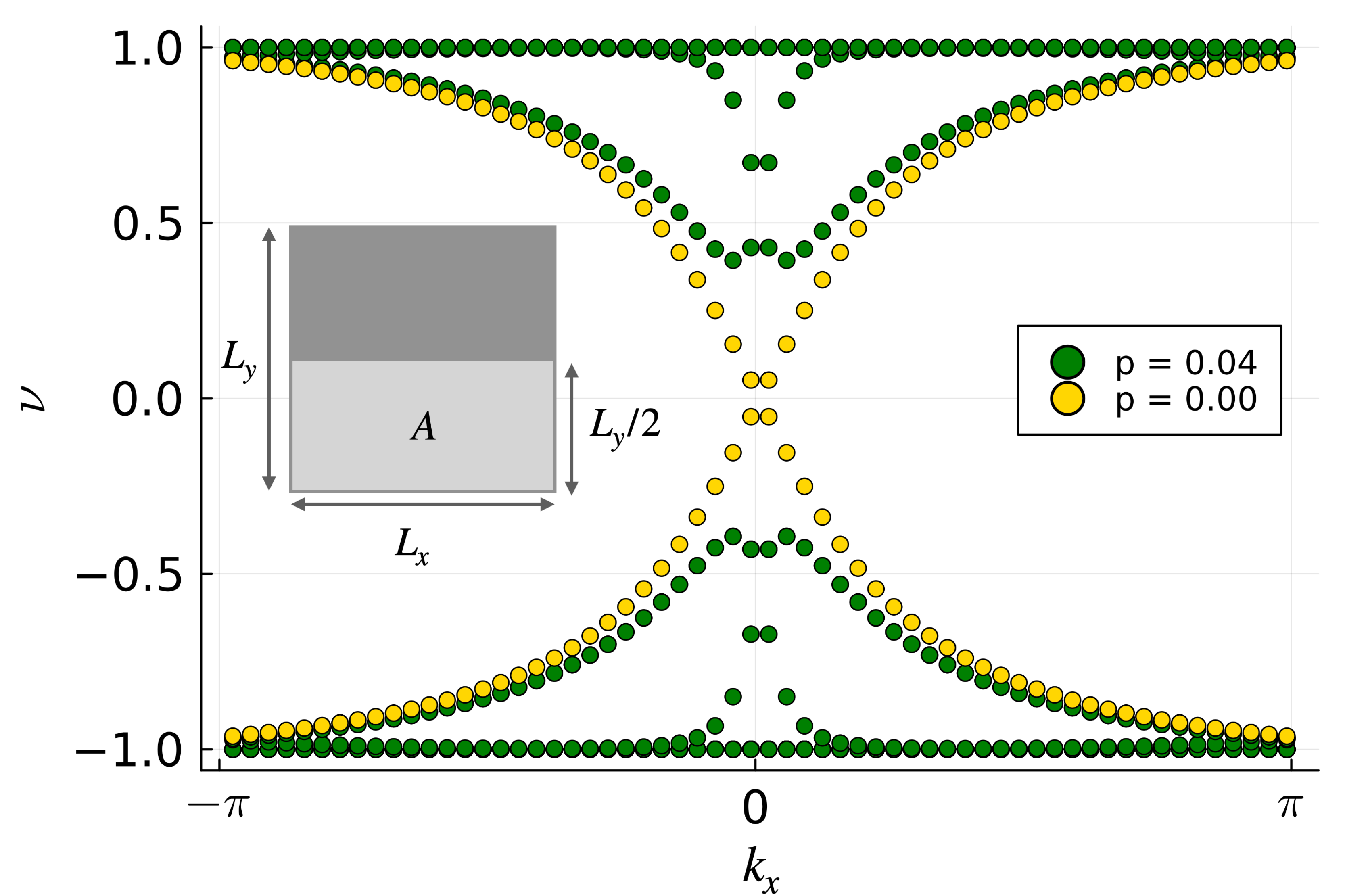}
	\caption{The spectrum of $i M_{A}$ (=  restriction of the covariance matrix to region $A$ in the inset) for a state $|\psi_m\rangle$ obtained from $|m\rangle = |0, ..., 0\rangle$ (see Eq.\eqref{eq:CDA_pwave}) as a function of the momentum $k_y$ for error rates $p = 0$ (i.e. non-decohered) and $p = 0.04$ (i.e. decohered).
		Here, we put the system on a $L_x \times L_y $ torus with $L_x = 60$ and $L_y = 30$.} 
	\label{Fig:mets_ES_L60W48}
\end{figure}

In addition, we also numerically compute the entanglement spectrum of $|\psi_m\rangle$ with $|m\rangle$ the uniform product state (so that momentum along the entanglement bipartition is a good quantum number).  For a chiral topological state, one expects that the edge spectrum of a physical edge will be imprinted on the entanglement spectrum of a subregion \cite{li2008entanglement}.  Since  $|\psi_m\rangle$ is Gaussian, the entanglement spectrum is encoded in the spectrum of the matrix $i M_{A}$, where $M_{A}$ is the restriction of the covariance matrix $M$ to the region $A$ in the inset of Fig.\ref{Fig:mets_ES_L60W48}.
Fig.\ref{Fig:mets_ES_L60W48} shows the  spectrum of $i M_{ABC}$ (denoted as $\nu$) as a function of the momentum $k_y$ with error rate $p = 0$ and $p = 0.04$. The geometry is again chosen as a torus, with length $L_x = 60$, and height $L_y = 30$.
In the absence of error ($p = 0$), all states $|\psi_m\rangle$ are projected to the $p+ip$ ground state, and thus the spectrum shows chirality, resembling the edge states of the $p+ip$ SC (note that we have two entanglement boundaries resulting in counter-propagating chiral states in the entanglement spectrum).
After the decoherence is introduced, one  finds that the chiral mode in the entanglement spectrum is gapped out, see  Fig.\ref{Fig:mets_ES_L60W48}. We also confirmed that the gap between the two `bands' of the entanglement spectrum increases with the system size (not shown). Overall, both the modular commutator and the entanglement spectrum provide numerical evidence that the decohered density matrix can be written as a convex sum of free-fermion, pure states that have no chiral topological order.

\subsection{Double-state formalism for fermions} \label{sec:CJ_fermions_maintext}

Previous subsection focused on the single-Majorana channel that breaks the fermion parity symmetry of the initial density matrix from exact ($U \rho = \rho U = \rho$) down to average ($U^\dagger \rho U = \rho$). As briefly mentioned above, if one instead uses a channel where Kraus operators are bilinear in Majorana operators (so that the fermion parity remains an exact symmetry), one might expect a more interesting behavior, in particular the possibility of a phase transition between different non-trivial mixed states. One way to make progress on this case is to study appropriate non-linear functions of the density matrix \cite{lee2022symmetry,lee2023quantum, fan2023diagnostics,bao2023mixed,zou2023channeling,su2023conformal}. Relatedly, one may employ the double state obtained using C-J map, which has been used in \citep{bao2023mixed, lee2023quantum, bao2023mixed, lee2023quantum, xue2024tensor, ma2024symmetry, guo2024locally, li2024replica} to study decoherence in bosonic problems. Specifically, given a density matrix $\rho_\mathcal{H}$ acting on the Hilbert space $\mathcal{H}$, one can define a state vector $|\rho\rangle_{\mathcal{H} \otimes \bar{\mathcal{H}}}$ in the doubled Hilbert space $\mathcal{H} \otimes \bar{\mathcal{H}}$ (with $\bar{\mathcal{H}}$ having the same dimension as $\mathcal{H}$) using the C-J map \citep{jamiolkowski1972linear, choi1975completely} (see footnote
\footnote{We note that the C-J isomorphism discussed here is a bit different from the original C-J isomorphism between channels $\mathcal{E}[\cdot]$ and operators $\mathcal{E}\otimes I[|\Phi\rangle \langle \Phi |]$ introduced in Ref.\citep{jamiolkowski1972linear, choi1975completely}, and is along the lines of super-operator formalism in Refs.\cite{schmutz1978real,prosen2008third}.  We use C-J isomorphism as a mnemonic  to transform bra(ket) to ket(bra) spaces using maximally-entangled states. See App.\ref{sec:CJ_fermions} for more discussion.
}):
\begin{equation}
	\label{Eq:double_state}
	|\rho\rangle_{\mathcal{H} \otimes \bar{\mathcal{H}}} = \rho_{\mathcal{H}} \otimes I_{\bar{\mathcal{H}}} |\Phi\rangle_{\mathcal{H} \otimes \bar{\mathcal{H}}}.
\end{equation}
Here $I_{\bar{\mathcal{H}}}$ denotes the identity in $\bar{\mathcal{H}}$ and $|\Phi\rangle_{\mathcal{H} \otimes \bar{\mathcal{H}}} $ is the product of (unnormalized) maximally entangled pairs connecting $\mathcal{H}$ and $\bar{\mathcal{H}}$, i.e., $	|\Phi\rangle_{\mathcal{H} \otimes \bar{\mathcal{H}}}  = \otimes_j |\phi\rangle_{j, \mathcal{H} \otimes \bar{\mathcal{H}}} $ with $|\phi\rangle_ {j, \mathcal{H} \otimes \bar{\mathcal{H}}}= \otimes_j ( \sum_{p = 1}^{d} |p_\mathcal{H},  p_{\bar{\mathcal{H}}}\rangle_j)$ and
$d$ the Hilbert space dimension on a single site.
Henceforth,  for notational simplicity we omit the subscript labeling the Hilbert space if there is no confusion.
For bosons, it is straightforward to see that under Eq.\eqref{Eq:double_state}, the density matrix $\rho = \sum_{p,q} \rho^p_q |p\rangle \langle q|$  is mapped to $|\rho\rangle =  \sum_{p,q} \rho^p_q |p,q\rangle$.
On the other hand, the channel $\mathcal{E}[\cdot] = \sum_\alpha K_\alpha (\cdot) K_\alpha^\dagger$ is mapped to the operator
\begin{equation}
	\label{Eq:DBS_kraus_bosons}
	\mathcal{N}_\mathcal{E} = \sum_{\alpha} K_\alpha \otimes \bar{K}_\alpha.
\end{equation}
This can be derived by expressing $| \mathcal{E} [\rho] \rangle$ as an operator acting on $|\rho\rangle$, i.e., $| \mathcal{E} [\rho] \rangle = \mathcal{N}_{\mathcal{E}} |\rho\rangle$. See App.\ref{sec:CJ_fermions} for details.
However, a similar correspondence for fermions is a bit subtle. For example, naively applying Eq.\eqref{Eq:DBS_kraus_bosons} to the single-majorana channel in Eq.\eqref{Eq:single_maj_channel} gives
\begin{equation}
	\label{Eq:fermions_wrong}
	\begin{aligned}
		\mathcal{E}_j|\rho \rangle & \stackrel{?}{=} [ (1-p)I_j \otimes I_j +p\gamma_j \otimes \bar{\gamma}_j) ] |\rho\rangle \\
		& = [(1-p)I +p\gamma_j \eta_j) ] |\rho\rangle \\
		& \sim e^{-i\mu (i \gamma_j \eta_j)}|\rho\rangle,\ \mu = \tan(p/(1-p)),
	\end{aligned}
\end{equation}
where we denote $\eta = \bar{\gamma}$ as the Majorana operators in the Hilbert space $\bar{\mathcal{H}}$.
Eq.\eqref{Eq:fermions_wrong} suggests that the channel generates a \textit{real} time evolution for the double state, which contradicts our intuition that the channel instead gives rise to an imaginary time evolution.
Another hint that Eq.\eqref{Eq:fermions_wrong} is incorrect comes from setting $p = 1/2$, where the relation $\mathcal{E}_j[\mathcal{E}_j[\rho]] = \mathcal{E}_j[\rho]$ holds. 
However, Eq.\eqref{Eq:fermions_wrong}  gives $\mathcal{E}_j \mathcal{E}_j |\rho\rangle = \gamma_j \eta_j |\rho\rangle /2$, which is not equal to $\mathcal{E}_j |\rho\rangle$.
Therefore, to find the correct correspondence between $ \mathcal{E}[\cdot]$ and $\mathcal{N}_{\mathcal{E}}$ for fermions, one should begin with the more fundamental definition of the double state, i.e, $|\rho\rangle = \rho \otimes I |\Phi\rangle$.
{
Due to the linearlity of Eq.\eqref{Eq:double_state}, one can consider each $K_\alpha (\cdot) K_\alpha^\dagger$ individually.
%
Now, using $|K_\alpha \rho K^\dagger_\alpha \rangle = (K_\alpha \rho K^\dagger_\alpha) \otimes I |\Phi\rangle = K_\alpha (\rho K_\alpha^\dagger \otimes I |\Phi\rangle) = K_\alpha |\rho K^\dagger_\alpha\rangle$, one finds $K_\alpha$ is unchanged under Eq.\eqref{Eq:double_state}.
On the other hand, since one can always write $K^\dagger_\alpha$ as a function of $\mathbf{c}$ and $\mathbf{c}^\dagger$, it suffices to consider how to express $|\rho \mathbf{c}\rangle$ and  $|\rho \mathbf{c}^\dagger\rangle$ as an operator applying to $|\rho\rangle$.
In Appendix \ref{sec:CJ_fermions},  we find that
\begin{equation}
	\label{Eq:rho_c_rho_cdagger}
\begin{aligned}
|\rho \mathbf{c}\rangle =  \mathbf{d}^\dagger|\rho\rangle,\
|\rho \mathbf{c}^\dagger\rangle =  -\mathbf{d}|\rho \rangle.
\end{aligned}
\end{equation}
One can then use Eq.\eqref{Eq:rho_c_rho_cdagger} to derive $\mathcal{N}_{\mathcal{E}}$ given $\mathcal{E}[\cdot]$.
For example, for the Kraus operator given by $K = \gamma_1 \equiv (\mathbf{c}+\mathbf{c}^\dagger)$, one finds
\begin{equation}
\begin{aligned}
|(\mathbf{c}+\mathbf{c}^\dagger) \rho (\mathbf{c}+\mathbf{c}^\dagger) \rangle & = (\mathbf{c}+\mathbf{c}^\dagger) | \rho (\mathbf{c}+\mathbf{c}^\dagger) \rangle \\
&  = (\mathbf{c}+\mathbf{c}^\dagger) (\mathbf{d}^\dagger- \mathbf{d})| \rho  \rangle,
\end{aligned}
\end{equation}
This implies the C-J transformed operator $\mathcal{N}_{\mathcal{E}} = (\mathbf{c}+\mathbf{c}^\dagger) (\mathbf{d}^\dagger- \mathbf{d}) = - i \gamma_1 \eta_1$, where $\eta_1 = (\mathbf{d}-\mathbf{d}^\dagger)/i$.
}

\subsection{Phase transition induced by an interacting channel in a $p+ip$ SC} \label{sec:bilinearkraus}
Being equipped with the correspondence between $\mathcal{E}[\cdot]$ and $\mathcal{N}_\mathcal{E}$, we now return to our discussion of decoherence induced transitions in chiral topological states of fermions. We first revisit the problem discussed in \ref{sec:chiral_free}, and then consider a more interesting problem where the Kraus operators are bilinear in fermions so that the decohered density matrix is not Gaussian.

There are different ways to employ the double state to probe the effect of decoherence. For example, one may consider non-linear functions such as the normalization of the double state \cite{lee2022symmetry, zou2023channeling, lee2023quantum, bao2023mixed}.  Here we will motivate the entanglement spectrum of a state obtained from the double state $|\rho\rangle$ (after space-time rotation) as a probe of the decoherence-induced phase transitions.

To begin with, consider the normalization of the double state
\begin{equation}
	\langle \rho|\rho \rangle = \langle \rho_0 |\mathcal{E}^\dagger \mathcal{E} |\rho_0\rangle.
\end{equation}
If the bulk action describing $|\rho_0\rangle = |\Psi_0, \Psi^*_0 \rangle$ is rotationally invariant, one can map $\langle \rho|\rho \rangle$ to the path integral of the $(1+1)$-D boundary fields following Ref.\citep{bao2023mixed}:
\begin{equation}
	\label{eqn:rho_path_integral}
	\begin{aligned}
		\langle \rho|\rho \rangle = \int & \mathcal{D}(\psi_L, \psi^*_L, \psi_R, \psi^*_R) \\
		& e^{-S_{0,L}(\psi_L, \psi^*_L)-S_{0,R}(\psi_R, \psi^*_R)-S_\text{int}(\psi_L, \psi^*_L, \psi_R, \psi^*_R)}.
	\end{aligned}
\end{equation}
Here, $\psi_L$ and $\psi^*_L$ denote the low-energy field variables in {$\mathcal{H}$ and $\bar{\mathcal{H}}$}, respectively. $S_{0,L}$ is the partition function on the left side of the spatial interface $x=0^-$ (the meaning of $(\psi_R, \psi^*_R)$ and $S_{0,R}$ are similar with left $\leftrightarrow$ right).
$S_\text{int}$ describes the effect of the channel $\mathcal{E}^\dagger \mathcal{E}$ and has two contributions:
\begin{equation}
	S_\text{int} = S_1 + S_\mathcal{E}.
\end{equation}
Here, $S_1$ denotes the action that exists even in the absences of decoherence. In particular, $S_1$ strongly couples the fields $\psi_L (\psi^*_L)$ and $\psi_R (\psi^*_R)$ such that $\psi_L = \psi_R (\psi^*_L = \psi^*_R)$ in the absence of decoherence. 
On the other hand, $S_\mathcal{E}$ describes the action that merely comes from the decoherence and vanishes when the error rate $p = 0$.
{
We note that a similar field theory has also been discussed in Ref. \cite{ashida2023system} in a different context when evaluating the system-environment entanglement in the (1+1)D system.
}
In general, the exact form of $S_\mathcal{E}$ involves four fields $(\psi_L, \psi^*_L, \psi_R, \psi^*_R)$ and may be schematically captured by the following Hamiltonian: 

\begin{equation}
	H = (H_{0,L} + H_{\mathcal{E},L})+ (H_{0,R} +  H_{\mathcal{E},R}) + H_1.
\end{equation}
where $H_1$ strongly couples the $L$ and $R$ fields. One may then consider the reduced density matrix for $L$ fields that is obtained after tracing out the $R$ fields. One expects \cite{peschel2011relation,qi2012general} that the corresponding entanglement Hamiltonian (= logarithm of the reduced density matrix) will essentially correspond to $H_{0,L} + H_{\mathcal{E},L}$. Working with entanglement Hamiltonian has the advantage that the number of fields one needs to keep track of are now halved. Similar simplification occurs if one considers the fidelity $\tr(\rho_d\, \rho_0)$ between the decohered density matrix  $\rho_d$ and the non-decohered density matrix $\rho_0$, see Ref.\cite{su2023conformal}. Since we are now working only with the $L$ fields, in the following we will omit the subscript $L$ for notational simplicity.

\begin{figure}
	\centering
	\includegraphics[width=\linewidth]{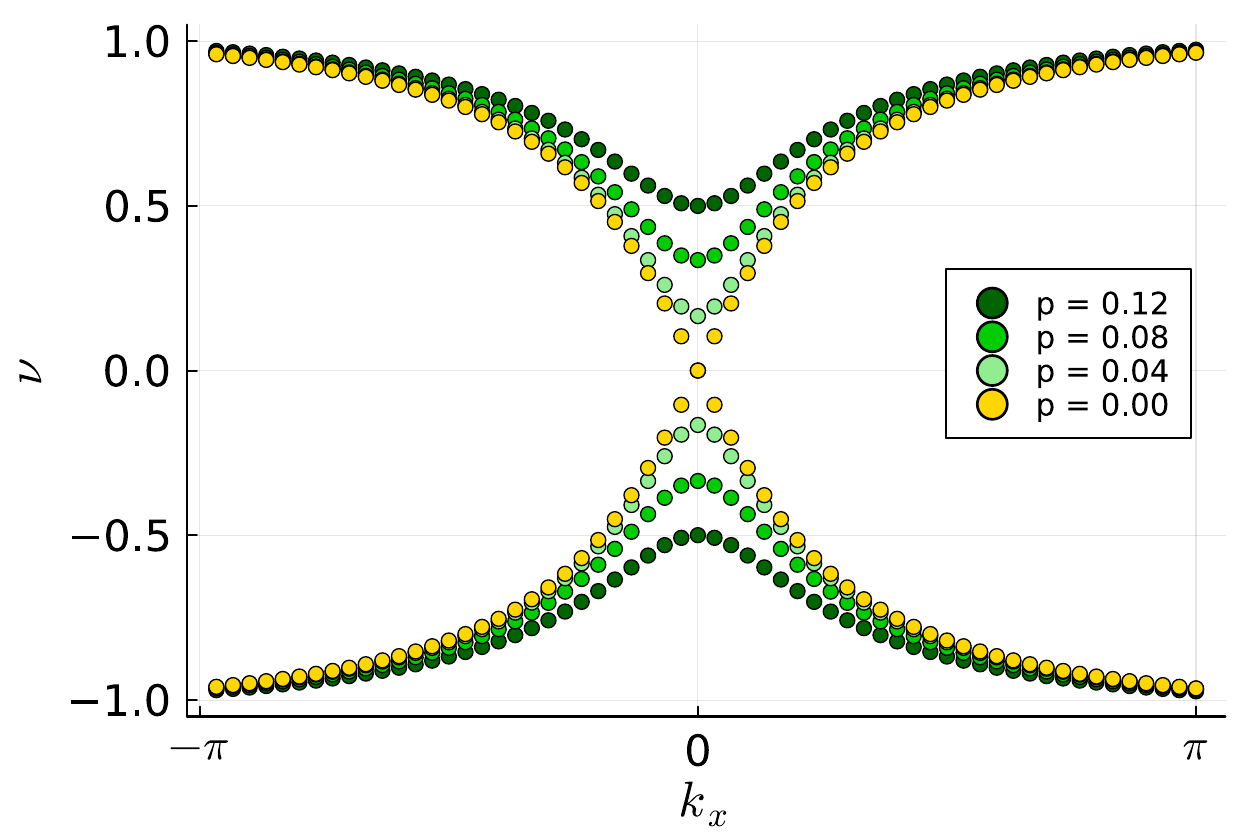}
	\caption{The spectrum of $i M_L$ for the double state $|\rho\rangle$, where $M_L$ is the restriction of the covariance matrix $M$ to the region $L$, as a function of the momentum $k_y$ for different error rates $p$.
		Here, we put the system on a cylinder with circumference $L_x = 60$, and height $L_y = 16$.} 
	\label{Fig:db_ES_L60W16}
\end{figure}

As an example, let us first revisit the case of $p+ip$ superconductor perturbed by a channel that is linear in Majorana fermions (Sec.\ref{sec:chiral_free}). Recall that here the Kraus map corresponds to the composition of the following map on all sites: $\mathcal{E}_\mathbf{x}[\rho] = (1-p)\rho + p \gamma_\mathbf{x} \rho \gamma_\mathbf{x}$. 
Based on our discussion above on the C-J map for fermions, this translates to a term of the form $H_{\mathcal{E}} = i g \int dy\, \gamma \,\eta$ where $p \sim g$, and $\gamma, \eta$ respectively denote the fields corresponding to $\mathcal{H}$ and $\bar{\mathcal{H}}$ of the $L$ fields. 
In the absence of any decoherence, the spatial boundary of the $p+ip$ superconductor has a simple description in terms of a chiral Majorana fermion. The entanglement Hamiltonian in the doubled Hilbert space then corresponds to stacking the boundary of $p+ip$ and $p-ip $ superconductors, and is given by $H_{0} = i \int dy (\gamma \partial_y \gamma - \eta \partial_y \eta)$. Therefore, one expects that the entanglement Hamiltonian for the $L$ fields in the presence of decoherence takes the form:

\be 
H_E =  i \int dy (\gamma \partial_y \gamma - \eta \partial_y \eta) +  i g \int dy \gamma \eta
\ee 
The counter-propagating edge modes are gapped out for any non-zero $g\, (\propto p)$, in line with our earlier discussion where we provided evidence that at any non-zero $p$, the density matrix can be written as a convex sum of pure states that are SRE. The gapping out of the edge modes can also be seen by numerically evaluating the entanglement spectrum of the double state obtained via C-J map. Fig.\ref{Fig:db_ES_L60W16} shows the  spectrum of $i M_L$ (denoted as $\nu$) as a function of the momentum $k_y$ with different error rate $p$. Here, we put the system on the cylinder with circumference $L_x = 60$, and height $L_y = 16$. In the absence of error ($p = 0$), there are two counter-propagating modes, resembling the edge states of the initial double state $|\rho_0\rangle$. After the decoherence is introduced, one can clearly see from Fig.\ref{Fig:db_ES_L60W16} that these counter-propagating modes are gapped out for arbitrary small error rate. Note that we \textit{did not} perform any space-time rotation to obtain Fig.\ref{Fig:db_ES_L60W16}. This is suggestive that the entanglement Hamiltonian of the double state $|\rho\rangle$ may already have the same qualitative behavior as the one obtained after space-time rotation. We leave further investigation of this point to the future.

Let us return to the problem of our main interest in this subsection, namely that of Kraus operators that \textit{commute} with the fermion parity operator. The simplest possibility is the composition of the following Kraus map on all nearest-neighbor bonds $\langle \mathbf{x}, \mathbf{y}\rangle$ of the square lattice:

\begin{equation}
	\mathcal{E}_{\langle \mathbf{x}, \mathbf{y}\rangle}[\rho] = (1-p)\rho + p \gamma_\mathbf{x} \gamma_\mathbf{y} \rho \gamma_\mathbf{y} \gamma_\mathbf{x} 
\end{equation}
The interaction term $H_{\mathcal{E}}$ induced by such a Kraus map
in the double state should respect the following $\mathbb{Z}_2 \times \mathbb{Z}_2$ symmetries: $\gamma \rightarrow -\gamma$ and/or $\eta \rightarrow -\eta$. Since Majorana fermions square to identity, the simplest term that is bilinear in both $\gamma$ and $\eta$ and respects all the symmetries involves derivatives:

\begin{equation}
	H_{\mathcal{E}} =  g \int dy (\gamma \partial_y \gamma) (\eta \partial_y \eta).
\end{equation}
where $g \propto p$. Therefore, the full entanglement Hamiltonian for the $L$ fields in the presence of decoherence is given by:

\be 
H_E =  i \int dy (\gamma \partial_y \gamma - \eta \partial_y \eta) + g \int dy (\gamma \partial_y \gamma) (\eta \partial_y \eta).
\ee 
This field theory has been studied earlier in Refs.\cite{grover2014emergent,rahmani2015emergent}. At a particular $g = g_c$, the system undergoes a phase transition in the tricitical Ising university class with central charge $c = 7/10$. For $g < g_c$, the interaction term is irrelevant, while above $g_c$, the system spontaneously breaks the $\mathbb{Z}_2\times \mathbb{Z}_2$ symmetry down to the diagonal $\mathbb{Z}_2$ symmetry. Physically, this means that the exact fermion-parity symmetry (i.e. $U \rho = \rho$ where $U$ is the generator of the fermion parity), has been spontaneously broken down to an average symmetry (i.e. $U\rho U^{\dagger} = \rho$). We note that a class of 2d chiral topological phases subjected to decoherence with fermion-bilinear Kraus operators was also studied in Ref.\cite{su2023conformal}. One notable difference between Ref.\cite{su2023conformal} and our problem is that in the examples considered in Ref.\cite{su2023conformal}, the decoherence always reduces the effective central charge of the action corresponding to the double state. In contrast, in our problem, the effective central charge increases from $c = 1/2$ to $c = 7/10$.

It is interesting to contemplate the implications of the above phase transition in terms of the separability properties of the original mixed state $\rho$ (instead of the double state $|\rho\rangle$). We conjecture that for $p \ll  p_c$, there exists no decomposition of the density matrix as a convex sum of area-law entangled pure states without any chirality, while $p > p_c$ the density matrix is expressible as a convex sum of area-law entangled pure states with GHZ-like entanglement (due to spontaneous breaking of fermion parity). Similar to the case of intrinsic topological orders subjected to local decoherence \cite{fan2022entanglement,bao2023mixed,lee2023quantum,chen2023separability}, we anticipate that the universality class as well as the location of the critical point obtained from the double-state formalism will differ from that of  the `intrinsic' mixed-state transition for the density matrix, e.g., when viewed from the perspective of separability. We do not know the universality for the latter transition and we leave it as an open question.

\section{Separability transition in Gibbs states of NLTS Hamiltonian}
\label{sec:ldpc}
In this section we will consider an exotic separability transition in a Gibbs state relevant to certain quantum codes. Although this transition does not require any symmetry, which has been a main ingredient in the rest of this work, the argument below to deduce the existence of a separability transition  is broadly similar in spirit to that in Secs.\ref{sec:SSB} and \ref{sec:trivial}. 

Recently, there has been discovery of `good LDPC codes' where the code distance as well as the number of logical qubits scale with the total number of qubits \cite{panteleev2022asymptotically,leverrier2022quantum,dinur2023good}. Moreover, Ref.\cite{anshu2022nlts} showed that the construction of a good LDPC code in Ref.\cite{leverrier2022quantum} satisfies Freedman-Hastings’ No Low-Energy Trivial States (NLTS) conjecture \cite{freedman2013quantum} which, when satisfied by a Hamiltonian, means that any state $|\psi\rangle$ with energy density less than a non-zero value $\e_c$ can not be prepared by a constant depth unitary circuit (the energy density $e$ of a state $|\psi\rangle$ is defined as $e = \lim_{N \to \infty} \left(\langle \psi| H |\psi \rangle - E_0\right)/N$ where $E_0$ is the groundspace energy of $H$). Here we ask: does the Gibbs state of a NLTS Hamiltonian show a separability transition at a non-zero temperature? That is, does there exist a $T_c > 0$ so that for $ T < T_c$, the Gibbs state can not be written as a convex sum of SRE pure states?

Firstly, we note Ref.\cite{anshu2022nlts} already proved that any mixed state whose energy density is less than an $e_c > 0$ cannot be purified to a pure SRE state by a short-depth channel, i.e., it cannot be prepared by first enlarging the Hilbert space to include ancillae, which are initially all in a product state, followed by a finite-depth unitary that entangles the `system' qubits (which are also initially in a product state) with the ancillae qubits, and eventually integrating out the ancillae.  However, as already discussed in Sec.\ref{sec:separability}, inability to purify to an SRE state via a short-depth channel does not imply that a mixed state is SRE using our definition (= expressibility of a mixed state as a convex sum of SRE pure states). Just to briefly re-iterate the example discussed in Sec.\ref{sec:separability} that illustrates these two different notions of mixed-state entanglement (see comment \#4 in Sec.\ref{sec:separability}), any Gibbs state that exhibits spontaneous symmetry breaking (which therefore has long-range correlations for the operator corresponding to the order parameter), can not be purified to an SRE pure state via  a short-depth channel. Therefore such a mixed state will be SRE using our definition and LRE using the definition in Refs.\cite{anshu2022nlts,anshu2020circuit}. Here we provide a simple argument that the Gibbs state of an NLTS staisfying Hamiltonian shows a separability transition at a non-zero temperature.

Let us assume that the Gibbs state of an NLTS satisfying Hamiltonian $H$ can be expressed as a convex sum of SRE pure states for \textit{any} temperature $ T > 0$, i.e., $\rho(T) = e^{-H/T}/Z = \sum_i p_i |\psi_i\rangle \langle \psi_i|$ where each $|\psi_i \rangle$ can be prepared via a unitary whose depth is independent of the number of qubits $N$. For simplicity of notation, we set the groundspace energy $E_0$ to zero (this can always be achieved by adding a constant $N c$ to the Hamiltonian, where $c$ is a constant). We will show that this assumption leads to a contradiction. Since all pure states $|\psi_i\rangle $ are SRE, by NLTS condition, they must all satisfy $\langle \psi_i | H |\psi_i \rangle/N > e_c$ as $N \to \infty$. Therefore $\tr( \rho(T) H)/N = \sum_i p_i \langle \psi_i |H|\psi_i\rangle/N > \sum_i p_i e_c = e_c$. This implies that if the Gibbs state can be expressed as a convex sum of SRE pure states, then its energy density is non-zero. However, non-zero energy density necessarily implies non-zero temperature. This is equivalent to showing that as $ T \to 0$,  $\tr( \rho(T) H)/N \to 0$. This is indeed the case because as $T \to 0$, $\tr( \rho(T) H)/N \approx E_1 e^{-E_1/T}/N$ which indeed vanishes as $T \to 0$ ($E_1$ denotes the energy of the first-excited state, which is a constant independent of $N$ since the LDPC code Hamiltonian under discussion is a sum of commuting projectors). Therefore, if we assume that the Gibbs state is separable for all non-zero temperatures, we arrive at a contradiction. Hence the Gibbs state must be long-range entangled upto a non-zero temperature $T$. It seems reasonable to assume that at sufficiently high temperature, the Gibbs state is SRE. Therefore, one expects a separability transition at some temperature $T_c$ that satisfies $0 < T_c < \infty$.

It is important to note that the above-argued separability transition does not necessarily imply that the Gibbs state has a \textit{thermodynamic} phase transition, i.e., it need not be accompanied by a singularity of the partition function.

\section{Some connections between separability and other measures of mixed-state complexity} \label{sec:connections}
In this section, we comment on some connections among the separability criteria, purifications, double states, and strange correlators.

\subsection*{Connections among separability, purification, and double state}
In Sec.\ref{sec:CJ_fermions_maintext}, we used the double-state formalism to probe decoherence-induced transitions. However, the connection between $\rho$ being sym-SRE and the double state $|\rho\rangle$ being trivial remains unclear. In this subsection, we will attempt to bridge the gap between them using purification of the mixed state.

We first recall the idea of purification: given a mixed state $\rho_\mathcal{H}$ in the Hilbert space $\mathcal{H}$, there exists a purification in the double Hilbert space $\mathcal{H} \otimes \bar{\mathcal{H}}$ with $\bar{\mathcal{H}}$ having the same dimension as $\mathcal{H}$:
\begin{equation}
	\label{Eq:purification}
	|\rho^{1/2} \rangle = {\rho}^{1/2}_\mathcal{H} \otimes I_{\bar{\mathcal{H}}}|\Phi\rangle_{\mathcal{H} \otimes \bar{\mathcal{H}}}.
\end{equation}
where $|\Phi\rangle_{\mathcal{H} \otimes \bar{\mathcal{H}}}$ is a maximally entangled state between $\mathcal{H}$ and $\bar{\mathcal{H}}$. It is straightforward to see that $\tr_{\bar{\mathcal{H}}}(|\rho^{1/2}\rangle \langle \rho^{1/2}|)  = \rho_\mathcal{H}$.
Besides, we note that Eq.\eqref{Eq:purification} is somtimes called `standard purification' \cite{terhal2002entanglement}, and all possible purifications are equivalent up to an isometry applied merely in $\bar{\mathcal{H}}$. If one uses Eq.\eqref{eq:SREmixeddef2} as a definition of an SRE mixed state $\rho$, then $|\rho^{1/2} \rangle$ being an SRE implies that $\rho$ is SRE.
However, it is not obvious to us how to show that this implies that $\rho$ can be written a convex sum of SRE states (Eq.\eqref{eq:SREmixeddef1}). Instead, we are only able to show that if $|\rho^{1/2}\rangle$ is SRE, one can write the mixed state $\rho \otimes I/\text{dim}(\bar{\mathcal{H}})$ (which lives in the Hilbert space system$\otimes$ancillae) as a convex sum of SRE states. To see this, we first note that a complete basis for the Hilbert space $\mathcal{H} \otimes \bar{\mathcal{H}}$  can be obtained from a single maximally entangled state  by applying local unitaries \textit{merely} in $\bar{\mathcal{H}}$. Specifically, denoting the complete basis of Bell pairs for spin-1/2 system as $\{ |\phi_{m,n} \rangle,\ m, n = 0, 1 \}$, all of them are related to $|\phi \rangle = (|00\rangle + |11\rangle)/\sqrt{2}$ through $|\phi_{m,n}\rangle = (Z_{\bar{\mathcal{H}}})^{m} 
(X_{\bar{\mathcal{H}}})^{n}|\phi\rangle$. It then follows that a complete basis for $\mathcal{H} \otimes \bar{\mathcal{H}}$ can be written as
\begin{equation}
	|\Phi_{m,n}\rangle = \prod_j (Z_{j, \bar{\mathcal{H}}})^{m_j} (X_{j, \bar{\mathcal{H}}})^{n_j}|\Phi\rangle
\end{equation}
with $m = (m_1, m_2, ...)$ and $n = (n_1, n_2, ...)$. Since $|\Phi_{m,n}\rangle$ are obtained by applying local unitary in $\mathcal{H}$ to a maximally entangled state, they are all also maximally entangled. Now, we use the same idea as we used to define CDA states (Eq.\eqref{Eq:metts}) by writing $\rho \otimes I/\text{dim}(H)$ as $\frac{1}{\text{dim}(\bar{\mathcal{H}})}  \sum_{m,n} \left(\rho^{1/2} \otimes I\right) 	|\Phi_{m,n}\rangle \langle \Phi_{m,n}| \left(\rho^{1/2} \otimes I\right) $:
\begin{widetext}
	\begin{equation} \label{eq:convexsum_purification}
		\begin{aligned}
			\rho \otimes \frac{I}{\text{dim}(\bar{\mathcal{H}})} 
			& = \frac{1}{\text{dim}(\bar{\mathcal{H}})} 
			\sum_{m,n} \rho^{1/2}\otimes I 
			\Big[\prod_{j}(Z_{j, \bar{\mathcal{H}}})^{m_j} (X_{j, \bar{\mathcal{H}}})^{n_j} \Big]
			| \Phi\rangle \langle \Phi|
			\Big[ \prod_{k} (Z_{k, \bar{\mathcal{H}}})^{m_k} 
			(X_{k, \bar{\mathcal{H}}})^{n_k} \Big]
			(\rho^{1/2} \otimes I) \\
			& = \frac{1}{\text{dim}(\bar{\mathcal{H}})} \sum_{m,n}
			\Big[\prod_{j} (Z_{j, \bar{\mathcal{H}}})^{m_j} (X_{j, \bar{\mathcal{H}}})^{n_j} \Big]
			( \rho^{1/2} \otimes I) | \Phi\rangle \langle \Phi| (\rho^{1/2}  \otimes I )
			\Big[\prod_{k}(Z_{k, \bar{\mathcal{H}}})^{m_k} (X_{k, \bar{\mathcal{H}}})^{n_k}\Big] \\
			& = 
			\frac{1}{\text{dim}(\bar{\mathcal{H}})}
			\sum_{m,n}
			\Big[\prod_{j} (Z_{j, \bar{\mathcal{H}}})^{m_j} (X_{j, \bar{\mathcal{H}}})^{n_j} \Big]
			| \rho^{1/2} \rangle \langle \rho^{1/2}|
			\Big[\prod_{k}(Z_{k, \bar{\mathcal{H}}})^{m_k} (X_{k, \bar{\mathcal{H}}})^{n_k}\Big] \\
			& = \frac{1}{\text{dim}(\bar{\mathcal{H}})}
			\sum_{m,n}|{\rho}^{1/2}_{m,n}\rangle \langle {\rho}^{1/2}_{m,n}| ,\ \ |{\rho}^{1/2}_{m,n}\rangle = (\prod_j (Z_{j, \bar{\mathcal{H}}})^{m_j} (X_{j, \bar{\mathcal{H}}})^{n_j}|{\rho}^{1/2}\rangle.
		\end{aligned}
	\end{equation}
\end{widetext}
In the second line, we use the property that $\prod_j (Z_{j, \bar{\mathcal{H}}})^{m_j} (X_{j, \bar{\mathcal{H}}})^{n_j}$ and $\rho^{1/2}$ commute, as they act on different Hilbert space.
Since $|{\rho}^{1/2}_{m,n}\rangle$ is related to $|{\rho}^{1/2}\rangle$ by a  unitary acting solely in $\bar{\mathcal{H}}$ , if $|{\rho}^{1/2}\rangle$ is SRE, then so is $|{\rho}^{1/2}_{m,n}\rangle$. Therefore, if there exists an SRE purification for $\rho$ (Eq.\eqref{Eq:purification}), then $\rho \otimes I/\text{dim}(\bar{\mathcal{H}})$ can be written as a convex sum of SRE pure states (Eq.\eqref{eq:convexsum_purification}). 

However, we emphasize that the converse is not true: if $|{\rho}^{1/2} \rangle$ is not an SRE, it does not rule out the possibility that the mixed state $\rho$ is still an SRE. This can be most easily seen by considering the following counterexample that also appeared in Sec.\ref{sec:separability}. Let $\rho$ be the convex sum of two product state $|0\rangle^{N}$ and $|1\rangle^{N}$, i.e.,
\begin{equation}
	\rho = \frac{1}{2} [ ( |0\rangle\langle 0|)^{\otimes N} +(|1\rangle\langle 1|)^{\otimes N}],
\end{equation}
It follows that the purified state is the GHZ state:
\begin{equation}
	|\rho^{1/2}\rangle = \frac{1}{\sqrt{2}}[|0 0\rangle^{\otimes N} + |1 1\rangle^{\otimes N} ],
\end{equation}
which is clearly long-range entangled. This implies that $|\rho^{1/2}\rangle$ being trivial is a sufficient but not necessary condition for $\rho \otimes I/\text{dim}(\bar{\mathcal{H}})$ being trivial.

The advantage of studying $\rho$ using its purification is obvious: instead of finding the decomposition in Eq.\eqref{Eq:metts}, one only needs to deal with a single pure state $|\rho^{1/2}\rangle$.
However, it is in general difficult to compute $|\rho^{1/2}\rangle$, as taking a square root of $\rho$ is non-trivial if $H_\rho = - \log(\rho)$ doesn't admit a simple compact form.
An alternative is to consider the double state $|\rho\rangle = \rho \otimes I |\Phi\rangle$ in Eq.\eqref{Eq:double_state} (note that if the original density is pure, i.e., $\rho^2 = \rho$, then the double state $|\rho\rangle$ is equivalent to the purified state $|\rho^{1/2}\rangle$). Heuristically, since the coefficient in front of $H_\rho$ for $|\rho\rangle$ is higher than the coefficient for $|\rho^{1/2}\rangle$, we expect that if $|\rho\rangle$ is SRE, then $|\rho^{1/2}\rangle$ is SRE as well, but we do not know how to prove this. 
This is consistent with the result in Ref.\cite{bao2023mixed}, where the critical error rate for $|\rho\rangle$ being trivial is higher than the error rate that the topological entanglement negativity drops to zero, and also consistent with the results in Ref.\cite{chen2023separability} on the error threshold for separability for topologically ordered mixed states.

Finally, as a small application of the state $|\rho^{1/2}\rangle$, let us briefly consider the 2d toric code under the phase-flip channel. We will now argue that under such a decoherence process, $|\rho^{1/2}\rangle$ undergoes a transition from being double topologically ordered to single topologically ordered precisely when the decoherence rate corresponds to the multicritical Nishimori point in RBIM, which further coincides with the separability transition for $\rho$ \cite{chen2023separability}, as well as the decoding transition and other metrics of mixed state entanglement \cite{dennis2002,bao2023mixed,fan2023diagnostics,lee2023quantum}.
The calculation is almost identical to that for a closely related pure state studied in Ref.\cite{chen2023separability}, and therefore we will only sketch the calculation, and refer the interested readers to Ref.\cite{chen2023separability} for details. 
The density matrix of the 2d toric code (on a torus of size $L \times L$) under the phase-flip channel can be written as \cite{lee2023quantum, chen2023separability}: $\rho = \sum_{Q^{(1)}} \mathcal{Z}_{\text{2d Ising} ,  Q^{(1)}}(\beta) |\Omega_{ Q^{(1)} } \rangle \langle \Omega_{ Q^{(1)}}|$, where $\tanh{\beta} = 1-2p$, $|\Omega_{ Q^{(1)} } \rangle \propto \prod_{v} (I + \prod_{e \ni v} Z_e) |x_\mathbf{e}\rangle$ and the set $Q^{(1)} = \{Q^{(1)}_p \}$ defines the flux configuration through all  plaquettes $p$ of the square lattice via the equation $Q^{(1)}_p  = \prod_{e \in \partial p } x_e$ ($ \prod_{e \in \partial p }$ means that the product is taken around the edges belonging to the plaquette $p$). Further, we choose $L_x = \prod_{e \in \ell, e \parallel \hat{x}} x_e = L_y = \prod_{e \in \ell, e \parallel \hat{y}} x_e = 1$ with $\ell$ a non-contractible loop along $\hat{x}/\hat{y}$ direction.
Therefore, one finds $|\rho^{1/2}\rangle \propto \sum_{Q^{(1)}} \sqrt{\mathcal{Z}_{\text{2d Ising} ,  Q^{(1)}}} |\Omega_{ Q^{(1)}} \bar{\Omega}_{ Q^{(1)}} \rangle$. The Higgs transition for this pure state can be detected by the t'Hooft loop $T_{\tilde{l}} = \prod_{e \in \tilde{l}} Z_e \bar{Z}_e$ where $\tilde{l}$ denotes a homologically non-contractible loop on the dual lattice.
Specifically, a straightforward calculation shows that $\langle \rho^{1/2} | T_{\tilde{l}} |\rho^{1/2} \rangle /\langle \rho^{1/2}  |\rho^{1/2} \rangle  = \sum_{Q^{(1)}} \mathcal{Z}_{Q^{(1)}} e^{-\Delta F( Q^{(1)}_{\tilde{l} } )/2 } / \sum_{Q^{(1)}} \mathcal{Z}_{Q^{(1)}} = \langle e^{-\Delta F_{\tilde{l}}/2} \rangle $, where $\Delta F( Q^{(1)}_{\tilde{l} })$ is the free energy cost of inserting a domain wall of size $\tilde{l} \sim L$.
It follows that $\langle T_{\tilde{l}}\rangle = 0 \, \, (\text{a non-zero constant})$ in the thermodynamic limit when $p< p_c = p_{\text{2d RBIM}} \, \, (p > p_c)$, indicating the transition of $|\rho^{1/2}\rangle$ from being double topologically ordered to single topologically ordered.

\subsection*{Connections between convex decomposition and strange correlator} 
In Sec.\ref{sec:trivial}, we studied separability transitions for cluster state SPTs in various dimensions using CDA (Eq.\eqref{Eq:metts}) with the initial basis $\{ |m\rangle \}$ as product states satisfying the corresponding  symmetry of the cluster state SPT (which was Pauli-$X$ basis in all the cases we considered). Fortuitously, as we discussed, the threshold for the CDA states being sym-SRE exactly corresponded to the error rate beyond which $\rho$ must be sym-LRE using general arguments, indicating that our choice of CDA is optimal.

Intriguingly, the symmetric product state basis to generate CDA has an apparently close connection with the strange correlator \cite{you2014wave}, which was originally devised as a diagnosis of the SPT pure states and has recently been used to probe the non-trivial SPT mixed states \cite{lee2022symmetry,zhang2022strange}.
To see the connection between them, we briefly review the original strange correlator for SPT pure states and two types of strange correlator introduced in Ref.\cite{lee2022symmetry}.
Choosing $|m\rangle$ as the disordered product state respecting the symmetry group $G$, the strange correlator for a pure state $|\psi\rangle$ is defined as  \cite{you2014wave}
\begin{equation}
	\label{Eq:pure_stange_correlator}
	C_m(j-k) = \frac{\langle m| O_j O_k |\psi \rangle }{\langle m| \psi\rangle},
\end{equation}
where $O$ is some operator that transforms non-trivially under $G$.
The basic idea of strange correlator is that the temporal edge of SPT pure state (when the many-body wavefunction is expressed as an imaginary time path integral) mimics its spatial edge.
Since at least 2d SPT possess nontrivial spatial edge states (in 3d, there also exists a possibility of boundary topological order), one may also use the temporal correlation defined in Eq.\eqref{Eq:pure_stange_correlator} to probe non-trivial SPT.
To generalize the strange correlator from pure states to mixed states, two types of strange correlator were introduced in Ref.\cite{lee2022symmetry}. The type-I strange correlator is defined as
\begin{equation}
	\label{Eq:type_1}
	C^I_m(j-k) = \frac{\langle m|\rho O_j O_k |m\rangle }{\langle m| \rho |m\rangle}.
\end{equation}
In the pure state limit $\rho = |\psi\rangle \langle \psi|$, the type-I strange correlator reduces to Eq.\eqref{Eq:pure_stange_correlator}.
Therefore, in the case of subjecting local decoherence to an SPT pure state, $C^I_m$ can be intuitively regarded as asking whether the local decoherence destroys the temporal edge states.
However, it has been shown in Ref.\cite{lee2022symmetry} that the type-I strange correlator is unable to detect the average SPT order mentioned in Ref.\cite{ma2022average}.
Instead, it was argued that the non-triviality of such an SPT order should be detected by the type-II strange correlator, defined as
\begin{equation}
	\label{Eq:type_2}
	C^{II}_m(j-k) = \frac{\langle m|  O^\dagger_k O^\dagger_j \rho O_j O_k |m\rangle }{\langle m| \rho |m\rangle}.
\end{equation}
In the pure state limit, it reduces to ${|\langle m| O_j O_k |\psi \rangle|^2 }/{\langle m| \psi\rangle}$.
Roughly speaking, the type-II strange correlator is devised to capture the case that $\rho$ can be written as an incoherent sum of pure state $|\psi_p\rangle$, where $\langle m|O_j O_k|\psi_p\rangle$ is non-trivial but can be either positive or negative depending on $|\psi_p\rangle$.

On the other side, the necessary condition for the mixed state to be non-trivial using separability criteria is the non-triviality of CDA states $|\psi_m\rangle$, which may be probed by several physical observables $S$ as discussed in Sec.\ref{sec:trivial}:
\begin{equation}
	\label{Eq:metts_obs}
	\frac{\langle \psi_m |S|\psi_m\rangle}{\langle \psi_m| \psi_m\rangle} 
	=
	\frac{\langle m|\rho^{1/2} S \rho^{1/2} |m\rangle}{\langle m| \rho|m\rangle}.
\end{equation}
Comparing Eq.\eqref{Eq:type_1}, Eq.\eqref{Eq:type_2} and Eq.\eqref{Eq:metts_obs}, one finds that  the denominator is always the fidelity between a symmetric product state and the mixed state of interest:
\begin{equation}
	\begin{aligned}
		\mathcal{Z}_m & = \tr(\rho |m\rangle\langle m|) \\
		& = \langle m| \rho|m\rangle= \langle \psi_m| \psi_m\rangle.
	\end{aligned}
\end{equation}
For the numerator, Eq.\eqref{Eq:metts_obs} involves inserting an operator between $\langle m| \rho^{1/2}$ and $\rho^{1/2}|m\rangle$, while the strange correlator involves inserting an operator between $\langle m| \rho$ and $|m\rangle$.

\section{Summary and discussion} \label{sec:summary}
In this work we explored the interplay of complexity and symmetry for many-body mixed states. Specifically, we asked whether a given mixed state can be expressed as a convex sum of symmetric short-ranged entangled pure states, which we took as a definition of an SRE mixed state subject to a given symmetry  (a `sym-SRE' mixed state, Sec.\ref{sec:separability}). Our primary aim was to identify `many-body separability transitions' as a function of an appropriate tuning parameter (e.g decoherence rate or temperature) across which the nature of the mixed state changes qualitatively - on one side of transition the mixed state is sym-SRE, and on the other side it is sym-LRE (= not sym-SRE). Analogous phase diagrams  for intrinsic topological orders subject to local decoherence  \cite{lee2023quantum, fan2023diagnostics,bao2023mixed,lu2023mixed,su2023conformal} were recently studied in Ref.\cite{chen2023separability}.
Our general approach was to first seek constraints that imply that a mixed state is necessarily long-range entangled, and absent such constraints, we developed tools to find the regime where a mixed state can be shown to be sym-SRE. One of the tools that allowed us to make progress was that local decoherence converts ground states of several SPTs, e.g. cluster states in various dimensions, to a Gibbs state.

In the context of SPTs subjected to local decoherence, we focussed on cluster states in various dimensions and obtained their `separability phase diagram' as shown in Fig.\ref{Fig:cluster_main}. As evident from the figure, the phase diagram gets progressively richer as one moves up in spatial dimensionality. The paths solely along the $x$ and $y$ axes in these phase diagrams correspond to the special case  of `average SPT' mixed states where one of the symmetries is exact while the other is average \cite{ma2022average,ma2023topological,de2022symmetry,lee2022symmetry,zhang2022strange}. 
It is crucial to note that although the decohered mixed state takes a Gibbs form, the corresponding partition function is \textit{not} singular at any non-zero temperature for any of these cluster states. 
{
This is because any local channel can be Stinespring dilated as a local unitary circuit in the enlarged system, and thus any physical observables $\text{tr}(\rho O)$ with $O$ acting on a large but finite region must be a smooth function of the error rate \cite{fan2023diagnostics,lee2023quantum,ashida2023system}. 
%
} 
Therefore, the different phases in Fig.\ref{Fig:cluster_main} arise \textit{only} because we are requiring that the density matrix be expressible as a convex sum of pure, symmetric states. 
As a consequence, these transitions are conceptually distinct from thermal phase transitions, and are more akin to `complexity phase transitions' for the mixed state, when a symmetry is enforced. We briefly discussed relation with other approaches to classifying mixed-state SPTs \cite{ma2022average,ma2023topological,de2022symmetry,lee2022symmetry,zhang2022strange} in Sec.\ref{sec:connections}.

It is also interesting to contrast the symmetry-enforced separability transitions in decohered 2d and 3d cluster states with decoherence induced separability transitions in 2d and 3d toric codes, studied in Ref.\cite{chen2023separability}. In both cases, one finds the appearance of same statistical mechanics models (e.g. RBIM in 2d). This similarity can be traced to the fact that the ground state of toric codes can be obtained from the ground state of the cluster states by performing appropriate projective measurements \cite{raussendorf2005long,aguado2008creation,verresen2021efficiently,tantivasadakarn2021long}, along with the equivalence between local and thermal decoherence for cluster states (this statement holds true also for the fractonic X-cube model \cite{vijay2016fracton} and its parent cluster state \cite{verresen2021efficiently}).

We also studied non-stabilizer topological states subjected to local decoherence. In particular, for free fermion chiral states corresponding to a $p+ip$ superconductor, we argued that if the quantum channel responsible for decoherence breaks the fermion parity, the resulting Gibbs state can be expressed as a convex sum of non-chiral states, and is therefore SRE at any non-zero decoherence rate (Sec.\ref{sec:fermions}). We also studied a case where the channel respects the fermion parity and identified a mixed state phase transition as a function of the decoherence rate using the double-state formalism. This transition can be thought of as corresponding to spontaneous breaking of the fermion parity, and as far as we know, does not have a pure-state counterpart. Intuitively, in a pure-state context, breaking fermion parity spontaneously essentially requires assigning a non-zero expectation value to fermionic operators, which is unphysical. In contrast, in the context of a mixed state, breaking fermion parity spontaneously means that the environment can exchange fermions with the system `spontaneously', which is not unphysical (in the double-state formalism, this corresponds to assigning non-zero expectation value to the bosonic order-parameter $\eta\, \gamma$ where $\eta$ and $\gamma$  respectively denote the fields corresponding to bra and ket Hilbert spaces).

We also analyzed  separability transitions in the Gibbs state of the quantum Ising model and argued that the Gibbs state is SRE at any non-zero temperature, and sym-SRE only for $T > T_c$, where $T_c$ is the critical temperature corresponding to the spontaneous symmetry breaking (Sec.\ref{sec:SSB}). We expect similar results to hold for other models whose Gibbs state shows a spontaneous breaking of zero-form discrete symmetry.

Finally, in Sec.\ref{sec:ldpc}, we  provided a short argument that the Gibbs states of  Hamiltonians that satisfy NLTS (no low-energy state) condition \cite{freedman2013quantum} must exhibit a separability transition at a non-zero temperature.

In the rest of this section, we discuss various aspects of our results and motivate questions for further exploration.

\textbf{1. SPT and chiral states:} The technique we used to study phase diagrams of various cluster states relied on the fact the quantum channel resulted in Gibbs states (Eqs.\eqref{Eq:ham_spt},\eqref{Eq:Oj_kraus}). It is not obvious how to generalize it to other SPT states. On that note, the following $\mathbb{Z}_N$ generalization may be helpful to study $\mathbb{Z}_N$ cluster states and topological orders produced by partial measurement of such states. Let us consider a commuting projector Hamiltonian of the form $H = \sum P_i$, where $P_i$ are projector ($P^2_i = 1$) written as $P_i = \frac{1}{N} \sum_{n=0}^{N-1} h^n_i$, with $h^N_i = 1$. Let us now introduce the following set of Kraus operators on each site $i$, $K_1(i) = \sqrt{1-p} \mathds{1}, K_2(i) = \sqrt{\frac{p}{2}} K(i), K_3(i) = \sqrt{\frac{p}{2}} K^{\dagger}(i)$, where $K^{\dagger}(i) K(i) = K(i) K^{\dagger}(i) = \mathds{1}$, and $K(i)$ are clock operators that satisfy $K(i) h_i K^{\dagger} (i)= e^{2 \pi i/N} h_i,  K^{\dagger} (i) h_i K(i) = e^{-2 \pi i/N} h_i$, one may verify that the application of this channel on all sites again results in a Gibbs state for $H$.

It might also be interesting to study `intrinsically mixed' SPT states introduced in Refs.\cite{ma2022average,ma2023topological} from the point of view of separability. These are SPT states that can exist only in the presence of decoherence. Conversely, it will be interesting to understand our results on non-trivial mixed SPTs protected by higher-form symmetries, such as 2d and 3d cluster state, using the techniques in Refs.\cite{ma2022average,ma2023topological} which primarily focussed on zero-form symmetry SPTs.

In the context of chiral states, we studied a phase transition driven by a channel where the Kraus operators were Majorana bilinears (Sec.\ref{sec:CJ_fermions_maintext}). We analyzed this problem only using the double-state formalism. As suggested by the problem of decoherence in toric code, the double state is likely to overestimate the threshold for the actual transition, and it will be interesting to find a description of the aforementioned transition in $p+ip$ SC directly in terms of the separability properties of the mixed state.

One important subtlety we would like to point out is that we assumed periodic boundary condition in our discussion of the SPT and chiral states. If instead one considers open boundaries such that the boundaries do not break the symmetry responsible for non-trivial SPT/chiral topological order, then the pure (non-decohered) state is always LRE, e.g., due to propagating edge modes or topolgical order at the boundary. In the presence of decoherence, our naive expectation is that the resulting mixed state is not sym-SRE, even if the decoherence breaks the symmetry from exact down to average. It will be interesting to study this aspect in the future.

\textbf{2. Symmetry broken states:} The first example we discussed, primarily to illustrate the distinction between SRE and sym-SRE states, was the Gibbs state of the transverse-field Ising model in any dimension (Sec.\ref{sec:SSB}). We discussed an explicit  decomposition of  this state at a non-zero temperature as a convex sum of pure states which we argued are SRE at any non-zero temperature. This conclusion is consistent with numerical results on Renyi negativity \cite{wu2020entanglement}, and mean-field arguments \cite{lu2020structure,wald2020entanglement}. On the other hand, if one imposes the Ising symmetry on the pure states into which the Gibbs state is being decomposed, we adopted an argument from Ref.\cite{lu2023mixed} to show that these pure states must be long-range entangled for $T < T_c$. This implies that the Gibbs state is sym-LRE for $ T \leq T_c$. In contrast, for $T > T_c$ we provided an explicit sym-SRE decomposition of the Gibbs state. The basic idea of the argument is to write $e^{-\beta H}$ as $\sum_{\phi} e^{-\beta H/2}|\phi \rangle \langle \phi| e^{-\beta H/2}$ where $\{\phi\}$ are chosen as  complete set of states in $z (x)$ basis if one wants to expand the Gibbs state as a sum of (sym-)SRE pure states.

There are several open questions along this direction. Firstly, the argument we provided for the aforementioned pure states being (sym-)SRE is not mathematically rigorous. To explicitly show that a state is SRE, one needs to construct a finite-depth circuit that prepares it starting from a product state. We only provided arguments in the continuum limit that the pure states under consideration have short-range correlations.  It will be worthwhile to study the entanglement structure of the pure states we claimed to be SRE using numerical methods (e.g. quantum Monte Carlo), or using a detailed field-theoretic analysis. Secondly, as we discussed, the transverse field Ising model in $d \geq 2$ must exhibit a separability transition from a sym-LRE to sym-SRE as a function of temperature. It will be interesting to study the symmetry-resolved negativity \cite{cornfeld2018imbalance} to quantify the nature of long-range entanglement across this transition. Finally, our arguments only apply to Gibbs states that break a discrete symmetry spontaneously, and it will be interesting to consider generalization to systems with spontaneously broken continuous symmetries that host Goldstone modes at a non-zero temperature.

\textbf{3. Experimental and numerical implications:} 
It is interesting to contemplate experimental implications of a symmetry-enforced separability transition. One perspective is that symmetry resolved versions of mixed-state entanglement measures such as entanglement negativity or entanglement of formation, that are specifically designed to quantify the lack of separability, would likely experience a singularity across such a phase transition. For example, for the Gibbs state $\rho$ of the transverse-field Ising model (Sec.\ref{sec:SSB}), one can in principle prepare the states $\rho_{\pm} = P_{\pm}\rho$ where $P_{\pm}$ are the projectors onto the even and odd sectors of the Ising symmetry. This can be done, e.g., by entangling an ancilla qubit with the system qubits sequentially using CNOT gates, and by measuring the ancilla qubit at the end. As discussed in Sec.\ref{sec:SSB}, the resulting mixed state (i.e. $\rho_{+}$ or $\rho_{-}$, depending on the outcome of the measurement on the ancilla) will show long-range mixed-state entanglement for $T < T_c$, in contrast to the original density matrix $\rho$, which will be short-ranged entangled for any $T > 0$. The long-range entanglement of $\rho_{\pm}$ can in principle be quantified experimentally using the Renyi negativity \cite{elben2020mixed}.

One may also imagine a very patient, gedanken experimentalist who has access to local unitary gates with a finite fidelity, so that they have the ability to  prepare only an ensemble of SRE pure states (i.e. pure states preparable with a constant depth unitary). If so, then a separability transition from an SRE to an LRE mixed state is equivalent to the transition from success to failure in preparing the ensemble corresponding to the mixed state. One may similarly characterize a transition from a sym-SRE to a sym-LRE by putting symmetry constraints on the local gates that form the circuit.

Perhaps a more practical implication of our results is that it may allow efficient classical simulation of a class of mixed states. For example, in the context of Gibbs state of the quantum Ising model, we argued that it admits a convex decomposition in terms of SRE pure states at any non-zero temperature if one does not impose any symmetry constraint on the pure states. 
Since SRE states are typically easier to study, {such a representation facilitates the task of simulating the corresponding mixed state.}
%
In contrast, if one tries to prepare the Gibbs state of the quantum Ising model starting with a \textit{product state} (assisted with ancillae), then long-range correlations below $T_c$ imply that one necessarily requires a deep quantum circuit \cite{brandao2019finite}.
{
We note that the decompositions we study generically involve an exponentially large number of pure states, which may lead to another difficulty in preparation. We can imagine at least two distinct ways to address this. Firstly, if a mixed state is SRE, and does not contain any classical long-range correlations, then it is reasonable to expect that it can be purified into an SRE pure state (using ancilla degrees of freedom). This equivalence between SRE mixed state and SRE purification is  discussed for Gibbs states in Ref.\cite{brandao2019finite}, and suggested more generally in Ref.\cite{hastings2011topological} although we are not aware of an explicit proof/construction in the non-Gibbs case. If one can indeed find an SRE purification, then an SRE mixed state can be prepared by a finite depth unitary circuit acting on system and ancilla degrees of freedom. An alternative route that is more generally available is to sample from the ensemble of SRE states that enter a given decomposition (assuming that the density matrix is SRE) using Monte Carlo algorithms, instead of preparing each and every SRE state that enters the SRE decomposition. While the sampling task may still be generally hard even in classical mechanics \cite{barahona1982on, lucas2014ising}, the decomposition we have provided clearly simplifies the ``quantum hardness'' of simulating a mixed state, analogous to the METTS algorithm \cite{white2009minimally}.
}

On a different note, one way to prepare an ensemble of pure states that may show a mixed-state separability phase transition is via a judicious combination of unitaries and measurements \cite{raussendorf2005long,aguado2008creation,piroli2021quantum,verresen2021efficiently,tantivasadakarn2021long,lee2022measurement,zhu2022nishimori,lu2023mixed}. For example, Refs.\cite{lee2022measurement,zhu2022nishimori}  provide a construction of mixed states that are closely related to the mixed states discussed in Sec.\ref{sec:separability}, and which have also been implemented in a recent experiment (Ref.\cite{chen2023realizing}). It will be interesting to design experiments that probe the phase diagram in Fig.\ref{Fig:cluster_main} using similar ideas, although we suspect it may be comparatively more challenging as it requires measuring non-local observables supplemented with appropriate decoding scheme \cite{lee2022measurement}.

\begin{acknowledgments}
	The authors thank Dan Arovas,  Tim Hsieh, John McGreevy, Bowen Shi for helpful discussions, and Tsung-Cheng Lu,  Shengqi Sang, William Witczak-Krempa for helpful comments on the draft. TG is supported by the National Science Foundation under Grant No. DMR-1752417. This research was supported in part by the National Science Foundation under Grant No. NSF PHY-1748958.
\end{acknowledgments}


\begin{thebibliography}{108}%
	\makeatletter
	\providecommand \@ifxundefined [1]{%
		\@ifx{#1\undefined}
	}%
	\providecommand \@ifnum [1]{%
		\ifnum #1\expandafter \@firstoftwo
		\else \expandafter \@secondoftwo
		\fi
	}%
	\providecommand \@ifx [1]{%
		\ifx #1\expandafter \@firstoftwo
		\else \expandafter \@secondoftwo
		\fi
	}%
	\providecommand \natexlab [1]{#1}%
	\providecommand \enquote  [1]{``#1''}%
	\providecommand \bibnamefont  [1]{#1}%
	\providecommand \bibfnamefont [1]{#1}%
	\providecommand \citenamefont [1]{#1}%
	\providecommand \href@noop [0]{\@secondoftwo}%
	\providecommand \href [0]{\begingroup \@sanitize@url \@href}%
	\providecommand \@href[1]{\@@startlink{#1}\@@href}%
	\providecommand \@@href[1]{\endgroup#1\@@endlink}%
	\providecommand \@sanitize@url [0]{\catcode `\\12\catcode `\$12\catcode
		`\&12\catcode `\#12\catcode `\^12\catcode `\_12\catcode `\%12\relax}%
	\providecommand \@@startlink[1]{}%
	\providecommand \@@endlink[0]{}%
	\providecommand \url  [0]{\begingroup\@sanitize@url \@url }%
	\providecommand \@url [1]{\endgroup\@href {#1}{\urlprefix }}%
	\providecommand \urlprefix  [0]{URL }%
	\providecommand \Eprint [0]{\href }%
	\providecommand \doibase [0]{https://doi.org/}%
	\providecommand \selectlanguage [0]{\@gobble}%
	\providecommand \bibinfo  [0]{\@secondoftwo}%
	\providecommand \bibfield  [0]{\@secondoftwo}%
	\providecommand \translation [1]{[#1]}%
	\providecommand \BibitemOpen [0]{}%
	\providecommand \bibitemStop [0]{}%
	\providecommand \bibitemNoStop [0]{.\EOS\space}%
	\providecommand \EOS [0]{\spacefactor3000\relax}%
	\providecommand \BibitemShut  [1]{\csname bibitem#1\endcsname}%
	\let\auto@bib@innerbib\@empty
	\bibitem [{\citenamefont {Werner}(1989)}]{werner1989}%
	\BibitemOpen
	\bibfield  {author} {\bibinfo {author} {\bibfnamefont {R.~F.}\ \bibnamefont
			{Werner}},\ }\bibfield  {title} {\bibinfo {title} {Quantum states with
			einstein-podolsky-rosen correlations admitting a hidden-variable model},\
	}\href {https://doi.org/10.1103/PhysRevA.40.4277} {\bibfield  {journal}
		{\bibinfo  {journal} {Phys. Rev. A}\ }\textbf {\bibinfo {volume} {40}},\
		\bibinfo {pages} {4277} (\bibinfo {year} {1989})}\BibitemShut {NoStop}%
	\bibitem [{\citenamefont {Hastings}(2011)}]{hastings2011topological}%
	\BibitemOpen
	\bibfield  {author} {\bibinfo {author} {\bibfnamefont {M.~B.}\ \bibnamefont
			{Hastings}},\ }\bibfield  {title} {\bibinfo {title} {Topological order at
			nonzero temperature},\ }\href
	{https://doi.org/10.1103%2Fphysrevlett.107.210501} {\bibfield  {journal}
		{\bibinfo  {journal} {Physical review letters}\ }\textbf {\bibinfo {volume}
			{107}},\ \bibinfo {pages} {210501} (\bibinfo {year} {2011})}\BibitemShut
	{NoStop}%
	\bibitem [{\citenamefont {Verstraete}\ \emph {et~al.}(2005)\citenamefont
		{Verstraete}, \citenamefont {Cirac}, \citenamefont {Latorre}, \citenamefont
		{Rico},\ and\ \citenamefont {Wolf}}]{verstraete2005renormalization}%
	\BibitemOpen
	\bibfield  {author} {\bibinfo {author} {\bibfnamefont {F.}~\bibnamefont
			{Verstraete}}, \bibinfo {author} {\bibfnamefont {J.~I.}\ \bibnamefont
			{Cirac}}, \bibinfo {author} {\bibfnamefont {J.~I.}\ \bibnamefont {Latorre}},
		\bibinfo {author} {\bibfnamefont {E.}~\bibnamefont {Rico}},\ and\ \bibinfo
		{author} {\bibfnamefont {M.~M.}\ \bibnamefont {Wolf}},\ }\bibfield  {title}
	{\bibinfo {title} {Renormalization-group transformations on quantum states},\
	}\href {https://doi.org/10.1103/PhysRevLett.94.140601} {\bibfield  {journal}
		{\bibinfo  {journal} {Phys. Rev. Lett.}\ }\textbf {\bibinfo {volume} {94}},\
		\bibinfo {pages} {140601} (\bibinfo {year} {2005})}\BibitemShut {NoStop}%
	\bibitem [{\citenamefont {Chen}\ \emph {et~al.}(2010)\citenamefont {Chen},
		\citenamefont {Gu},\ and\ \citenamefont {Wen}}]{chen2010local}%
	\BibitemOpen
	\bibfield  {author} {\bibinfo {author} {\bibfnamefont {X.}~\bibnamefont
			{Chen}}, \bibinfo {author} {\bibfnamefont {Z.-C.}\ \bibnamefont {Gu}},\ and\
		\bibinfo {author} {\bibfnamefont {X.-G.}\ \bibnamefont {Wen}},\ }\bibfield
	{title} {\bibinfo {title} {Local unitary transformation, long-range quantum
			entanglement, wave function renormalization, and topological order},\ }\href
	{https://doi.org/10.1103/PhysRevB.82.155138} {\bibfield  {journal} {\bibinfo
			{journal} {Phys. Rev. B}\ }\textbf {\bibinfo {volume} {82}},\ \bibinfo
		{pages} {155138} (\bibinfo {year} {2010})}\BibitemShut {NoStop}%
	\bibitem [{\citenamefont {Hastings}\ and\ \citenamefont
		{Wen}(2005)}]{hastings2005quasiadiabatic}%
	\BibitemOpen
	\bibfield  {author} {\bibinfo {author} {\bibfnamefont {M.~B.}\ \bibnamefont
			{Hastings}}\ and\ \bibinfo {author} {\bibfnamefont {X.-G.}\ \bibnamefont
			{Wen}},\ }\bibfield  {title} {\bibinfo {title} {Quasiadiabatic continuation
			of quantum states: The stability of topological ground-state degeneracy and
			emergent gauge invariance},\ }\href@noop {} {\bibfield  {journal} {\bibinfo
			{journal} {Physical review b}\ }\textbf {\bibinfo {volume} {72}},\ \bibinfo
		{pages} {045141} (\bibinfo {year} {2005})}\BibitemShut {NoStop}%
	\bibitem [{\citenamefont {Bravyi}\ \emph {et~al.}(2006)\citenamefont {Bravyi},
		\citenamefont {Hastings},\ and\ \citenamefont {Verstraete}}]{bravyi2006lieb}%
	\BibitemOpen
	\bibfield  {author} {\bibinfo {author} {\bibfnamefont {S.}~\bibnamefont
			{Bravyi}}, \bibinfo {author} {\bibfnamefont {M.~B.}\ \bibnamefont
			{Hastings}},\ and\ \bibinfo {author} {\bibfnamefont {F.}~\bibnamefont
			{Verstraete}},\ }\bibfield  {title} {\bibinfo {title} {Lieb-robinson bounds
			and the generation of correlations and topological quantum order},\ }\href
	{https://doi.org/10.1103/PhysRevLett.97.050401} {\bibfield  {journal}
		{\bibinfo  {journal} {Phys. Rev. Lett.}\ }\textbf {\bibinfo {volume} {97}},\
		\bibinfo {pages} {050401} (\bibinfo {year} {2006})}\BibitemShut {NoStop}%
	\bibitem [{\citenamefont {Chen}\ \emph {et~al.}(2011)\citenamefont {Chen},
		\citenamefont {Gu},\ and\ \citenamefont {Wen}}]{chen2011classification}%
	\BibitemOpen
	\bibfield  {author} {\bibinfo {author} {\bibfnamefont {X.}~\bibnamefont
			{Chen}}, \bibinfo {author} {\bibfnamefont {Z.-C.}\ \bibnamefont {Gu}},\ and\
		\bibinfo {author} {\bibfnamefont {X.-G.}\ \bibnamefont {Wen}},\ }\bibfield
	{title} {\bibinfo {title} {Classification of gapped symmetric phases in
			one-dimensional spin systems},\ }\href
	{https://doi.org/10.1103/PhysRevB.83.035107} {\bibfield  {journal} {\bibinfo
			{journal} {Phys. Rev. B}\ }\textbf {\bibinfo {volume} {83}},\ \bibinfo
		{pages} {035107} (\bibinfo {year} {2011})}\BibitemShut {NoStop}%
	\bibitem [{\citenamefont {Zeng}\ and\ \citenamefont
		{Wen}(2015)}]{zeng2015gapped}%
	\BibitemOpen
	\bibfield  {author} {\bibinfo {author} {\bibfnamefont {B.}~\bibnamefont
			{Zeng}}\ and\ \bibinfo {author} {\bibfnamefont {X.-G.}\ \bibnamefont {Wen}},\
	}\bibfield  {title} {\bibinfo {title} {Gapped quantum liquids and topological
			order, stochastic local transformations and emergence of unitarity},\ }\href
	{https://doi.org/10.1103/PhysRevB.91.125121} {\bibfield  {journal} {\bibinfo
			{journal} {Phys. Rev. B}\ }\textbf {\bibinfo {volume} {91}},\ \bibinfo
		{pages} {125121} (\bibinfo {year} {2015})}\BibitemShut {NoStop}%
	\bibitem [{\citenamefont {Gu}\ and\ \citenamefont {Wen}(2009)}]{gu2009tensor}%
	\BibitemOpen
	\bibfield  {author} {\bibinfo {author} {\bibfnamefont {Z.-C.}\ \bibnamefont
			{Gu}}\ and\ \bibinfo {author} {\bibfnamefont {X.-G.}\ \bibnamefont {Wen}},\
	}\bibfield  {title} {\bibinfo {title} {Tensor-entanglement-filtering
			renormalization approach and symmetry-protected topological order},\ }\href
	{https://doi.org/10.1103/PhysRevB.80.155131} {\bibfield  {journal} {\bibinfo
			{journal} {Phys. Rev. B}\ }\textbf {\bibinfo {volume} {80}},\ \bibinfo
		{pages} {155131} (\bibinfo {year} {2009})}\BibitemShut {NoStop}%
	\bibitem [{\citenamefont {Pollmann}\ \emph {et~al.}(2012)\citenamefont
		{Pollmann}, \citenamefont {Berg}, \citenamefont {Turner},\ and\ \citenamefont
		{Oshikawa}}]{pollmann2012symmetry}%
	\BibitemOpen
	\bibfield  {author} {\bibinfo {author} {\bibfnamefont {F.}~\bibnamefont
			{Pollmann}}, \bibinfo {author} {\bibfnamefont {E.}~\bibnamefont {Berg}},
		\bibinfo {author} {\bibfnamefont {A.~M.}\ \bibnamefont {Turner}},\ and\
		\bibinfo {author} {\bibfnamefont {M.}~\bibnamefont {Oshikawa}},\ }\bibfield
	{title} {\bibinfo {title} {Symmetry protection of topological phases in
			one-dimensional quantum spin systems},\ }\href
	{https://doi.org/10.1103/PhysRevB.85.075125} {\bibfield  {journal} {\bibinfo
			{journal} {Phys. Rev. B}\ }\textbf {\bibinfo {volume} {85}},\ \bibinfo
		{pages} {075125} (\bibinfo {year} {2012})}\BibitemShut {NoStop}%
	\bibitem [{\citenamefont {Chen}\ \emph {et~al.}(2013)\citenamefont {Chen},
		\citenamefont {Gu}, \citenamefont {Liu},\ and\ \citenamefont
		{Wen}}]{chen2013symmetry}%
	\BibitemOpen
	\bibfield  {author} {\bibinfo {author} {\bibfnamefont {X.}~\bibnamefont
			{Chen}}, \bibinfo {author} {\bibfnamefont {Z.-C.}\ \bibnamefont {Gu}},
		\bibinfo {author} {\bibfnamefont {Z.-X.}\ \bibnamefont {Liu}},\ and\ \bibinfo
		{author} {\bibfnamefont {X.-G.}\ \bibnamefont {Wen}},\ }\bibfield  {title}
	{\bibinfo {title} {Symmetry protected topological orders and the group
			cohomology of their symmetry group},\ }\href
	{https://doi.org/10.1103/PhysRevB.87.155114} {\bibfield  {journal} {\bibinfo
			{journal} {Phys. Rev. B}\ }\textbf {\bibinfo {volume} {87}},\ \bibinfo
		{pages} {155114} (\bibinfo {year} {2013})}\BibitemShut {NoStop}%
	\bibitem [{\citenamefont {Schuch}\ \emph {et~al.}(2011)\citenamefont {Schuch},
		\citenamefont {P\'erez-Garc\'{\i}a},\ and\ \citenamefont
		{Cirac}}]{schuch2011classifying}%
	\BibitemOpen
	\bibfield  {author} {\bibinfo {author} {\bibfnamefont {N.}~\bibnamefont
			{Schuch}}, \bibinfo {author} {\bibfnamefont {D.}~\bibnamefont
			{P\'erez-Garc\'{\i}a}},\ and\ \bibinfo {author} {\bibfnamefont
			{I.}~\bibnamefont {Cirac}},\ }\bibfield  {title} {\bibinfo {title}
		{Classifying quantum phases using matrix product states and projected
			entangled pair states},\ }\href {https://doi.org/10.1103/PhysRevB.84.165139}
	{\bibfield  {journal} {\bibinfo  {journal} {Phys. Rev. B}\ }\textbf {\bibinfo
			{volume} {84}},\ \bibinfo {pages} {165139} (\bibinfo {year}
		{2011})}\BibitemShut {NoStop}%
	\bibitem [{\citenamefont {Bruzewicz}\ \emph {et~al.}(2019)\citenamefont
		{Bruzewicz}, \citenamefont {Chiaverini}, \citenamefont {McConnell},\ and\
		\citenamefont {Sage}}]{bruzewicz2019trapped}%
	\BibitemOpen
	\bibfield  {author} {\bibinfo {author} {\bibfnamefont {C.~D.}\ \bibnamefont
			{Bruzewicz}}, \bibinfo {author} {\bibfnamefont {J.}~\bibnamefont
			{Chiaverini}}, \bibinfo {author} {\bibfnamefont {R.}~\bibnamefont
			{McConnell}},\ and\ \bibinfo {author} {\bibfnamefont {J.~M.}\ \bibnamefont
			{Sage}},\ }\bibfield  {title} {\bibinfo {title} {Trapped-ion quantum
			computing: Progress and challenges},\ }\href
	{https://pubs.aip.org/aip/apr/article-abstract/6/2/021314/570103/Trapped-ion-quantum-computing-Progress-and?redirectedFrom=fulltext}
	{\bibfield  {journal} {\bibinfo  {journal} {Applied Physics Reviews}\
		}\textbf {\bibinfo {volume} {6}} (\bibinfo {year} {2019})}\BibitemShut
	{NoStop}%
	\bibitem [{\citenamefont {Henriet}\ \emph {et~al.}(2020)\citenamefont
		{Henriet}, \citenamefont {Beguin}, \citenamefont {Signoles}, \citenamefont
		{Lahaye}, \citenamefont {Browaeys}, \citenamefont {Reymond},\ and\
		\citenamefont {Jurczak}}]{henriet2020quantum}%
	\BibitemOpen
	\bibfield  {author} {\bibinfo {author} {\bibfnamefont {L.}~\bibnamefont
			{Henriet}}, \bibinfo {author} {\bibfnamefont {L.}~\bibnamefont {Beguin}},
		\bibinfo {author} {\bibfnamefont {A.}~\bibnamefont {Signoles}}, \bibinfo
		{author} {\bibfnamefont {T.}~\bibnamefont {Lahaye}}, \bibinfo {author}
		{\bibfnamefont {A.}~\bibnamefont {Browaeys}}, \bibinfo {author}
		{\bibfnamefont {G.-O.}\ \bibnamefont {Reymond}},\ and\ \bibinfo {author}
		{\bibfnamefont {C.}~\bibnamefont {Jurczak}},\ }\bibfield  {title} {\bibinfo
		{title} {Quantum computing with neutral atoms},\ }\href
	{https://quantum-journal.org/papers/q-2020-09-21-327/} {\bibfield  {journal}
		{\bibinfo  {journal} {Quantum}\ }\textbf {\bibinfo {volume} {4}},\ \bibinfo
		{pages} {327} (\bibinfo {year} {2020})}\BibitemShut {NoStop}%
	\bibitem [{\citenamefont {Kjaergaard}\ \emph {et~al.}(2020)\citenamefont
		{Kjaergaard}, \citenamefont {Schwartz}, \citenamefont {Braum{\"u}ller},
		\citenamefont {Krantz}, \citenamefont {Wang}, \citenamefont {Gustavsson},\
		and\ \citenamefont {Oliver}}]{kjaergaard2020superconducting}%
	\BibitemOpen
	\bibfield  {author} {\bibinfo {author} {\bibfnamefont {M.}~\bibnamefont
			{Kjaergaard}}, \bibinfo {author} {\bibfnamefont {M.~E.}\ \bibnamefont
			{Schwartz}}, \bibinfo {author} {\bibfnamefont {J.}~\bibnamefont
			{Braum{\"u}ller}}, \bibinfo {author} {\bibfnamefont {P.}~\bibnamefont
			{Krantz}}, \bibinfo {author} {\bibfnamefont {J.~I.-J.}\ \bibnamefont {Wang}},
		\bibinfo {author} {\bibfnamefont {S.}~\bibnamefont {Gustavsson}},\ and\
		\bibinfo {author} {\bibfnamefont {W.~D.}\ \bibnamefont {Oliver}},\ }\bibfield
	{title} {\bibinfo {title} {Superconducting qubits: Current state of play},\
	}\href
	{https://www.annualreviews.org/doi/abs/10.1146/annurev-conmatphys-031119-050605}
	{\bibfield  {journal} {\bibinfo  {journal} {Annual Review of Condensed Matter
				Physics}\ }\textbf {\bibinfo {volume} {11}},\ \bibinfo {pages} {369}
		(\bibinfo {year} {2020})}\BibitemShut {NoStop}%
	\bibitem [{\citenamefont {Preskill}(2018)}]{preskill2018quantum}%
	\BibitemOpen
	\bibfield  {author} {\bibinfo {author} {\bibfnamefont {J.}~\bibnamefont
			{Preskill}},\ }\bibfield  {title} {\bibinfo {title} {Quantum computing in the
			nisq era and beyond},\ }\href
	{https://quantum-journal.org/papers/q-2018-08-06-79/} {\bibfield  {journal}
		{\bibinfo  {journal} {Quantum}\ }\textbf {\bibinfo {volume} {2}},\ \bibinfo
		{pages} {79} (\bibinfo {year} {2018})}\BibitemShut {NoStop}%
	\bibitem [{\citenamefont {Chen}\ and\ \citenamefont
		{Grover}(2023)}]{chen2023separability}%
	\BibitemOpen
	\bibfield  {author} {\bibinfo {author} {\bibfnamefont {Y.-H.}\ \bibnamefont
			{Chen}}\ and\ \bibinfo {author} {\bibfnamefont {T.}~\bibnamefont {Grover}},\
	}\bibfield  {title} {\bibinfo {title} {Separability transitions in
			topological states induced by local decoherence},\ }\href
	{https://arxiv.org/abs/2309.11879} {\bibfield  {journal} {\bibinfo  {journal}
			{arXiv preprint arXiv:2309.11879}\ } (\bibinfo {year} {2023})}\BibitemShut
	{NoStop}%
	\bibitem [{\citenamefont {Lee}\ \emph {et~al.}(2023)\citenamefont {Lee},
		\citenamefont {Jian},\ and\ \citenamefont {Xu}}]{lee2023quantum}%
	\BibitemOpen
	\bibfield  {author} {\bibinfo {author} {\bibfnamefont {J.~Y.}\ \bibnamefont
			{Lee}}, \bibinfo {author} {\bibfnamefont {C.-M.}\ \bibnamefont {Jian}},\ and\
		\bibinfo {author} {\bibfnamefont {C.}~\bibnamefont {Xu}},\ }\bibfield
	{title} {\bibinfo {title} {Quantum criticality under decoherence or weak
			measurement},\ }\href {https://doi.org/10.1103/PRXQuantum.4.030317}
	{\bibfield  {journal} {\bibinfo  {journal} {PRX Quantum}\ }\textbf {\bibinfo
			{volume} {4}},\ \bibinfo {pages} {030317} (\bibinfo {year}
		{2023})}\BibitemShut {NoStop}%
	\bibitem [{\citenamefont {Fan}\ \emph {et~al.}(2023)\citenamefont {Fan},
		\citenamefont {Bao}, \citenamefont {Altman},\ and\ \citenamefont
		{Vishwanath}}]{fan2023diagnostics}%
	\BibitemOpen
	\bibfield  {author} {\bibinfo {author} {\bibfnamefont {R.}~\bibnamefont
			{Fan}}, \bibinfo {author} {\bibfnamefont {Y.}~\bibnamefont {Bao}}, \bibinfo
		{author} {\bibfnamefont {E.}~\bibnamefont {Altman}},\ and\ \bibinfo {author}
		{\bibfnamefont {A.}~\bibnamefont {Vishwanath}},\ }\bibfield  {title}
	{\bibinfo {title} {Diagnostics of mixed-state topological order and breakdown
			of quantum memory},\ }\href {https://arxiv.org/abs/2301.05689} {\bibfield
		{journal} {\bibinfo  {journal} {arXiv preprint arXiv:2301.05689}\ } (\bibinfo
		{year} {2023})}\BibitemShut {NoStop}%
	\bibitem [{\citenamefont {Bao}\ \emph {et~al.}(2023)\citenamefont {Bao},
		\citenamefont {Fan}, \citenamefont {Vishwanath},\ and\ \citenamefont
		{Altman}}]{bao2023mixed}%
	\BibitemOpen
	\bibfield  {author} {\bibinfo {author} {\bibfnamefont {Y.}~\bibnamefont
			{Bao}}, \bibinfo {author} {\bibfnamefont {R.}~\bibnamefont {Fan}}, \bibinfo
		{author} {\bibfnamefont {A.}~\bibnamefont {Vishwanath}},\ and\ \bibinfo
		{author} {\bibfnamefont {E.}~\bibnamefont {Altman}},\ }\bibfield  {title}
	{\bibinfo {title} {Mixed-state topological order and the errorfield double
			formulation of decoherence-induced transitions},\ }\href
	{https://arxiv.org/abs/2301.05687} {\bibfield  {journal} {\bibinfo  {journal}
			{arXiv preprint arXiv:2301.05687}\ } (\bibinfo {year} {2023})}\BibitemShut
	{NoStop}%
	\bibitem [{\citenamefont {Lu}\ \emph {et~al.}(2023)\citenamefont {Lu},
		\citenamefont {Zhang}, \citenamefont {Vijay},\ and\ \citenamefont
		{Hsieh}}]{lu2023mixed}%
	\BibitemOpen
	\bibfield  {author} {\bibinfo {author} {\bibfnamefont {T.-C.}\ \bibnamefont
			{Lu}}, \bibinfo {author} {\bibfnamefont {Z.}~\bibnamefont {Zhang}}, \bibinfo
		{author} {\bibfnamefont {S.}~\bibnamefont {Vijay}},\ and\ \bibinfo {author}
		{\bibfnamefont {T.~H.}\ \bibnamefont {Hsieh}},\ }\bibfield  {title} {\bibinfo
		{title} {Mixed-state long-range order and criticality from measurement and
			feedback},\ }\href {https://doi.org/10.1103/PRXQuantum.4.030318} {\bibfield
		{journal} {\bibinfo  {journal} {PRX Quantum}\ }\textbf {\bibinfo {volume}
			{4}},\ \bibinfo {pages} {030318} (\bibinfo {year} {2023})}\BibitemShut
	{NoStop}%
	\bibitem [{\citenamefont {Su}\ \emph {et~al.}(2023)\citenamefont {Su},
		\citenamefont {Myerson-Jain},\ and\ \citenamefont {Xu}}]{su2023conformal}%
	\BibitemOpen
	\bibfield  {author} {\bibinfo {author} {\bibfnamefont {K.}~\bibnamefont
			{Su}}, \bibinfo {author} {\bibfnamefont {N.}~\bibnamefont {Myerson-Jain}},\
		and\ \bibinfo {author} {\bibfnamefont {C.}~\bibnamefont {Xu}},\ }\bibfield
	{title} {\bibinfo {title} {Conformal field theories generated by chern
			insulators under quantum decoherence},\ }\href
	{https://arxiv.org/abs/2305.13410} {\bibfield  {journal} {\bibinfo  {journal}
			{arXiv preprint arXiv:2305.13410}\ } (\bibinfo {year} {2023})}\BibitemShut
	{NoStop}%
	\bibitem [{\citenamefont {Wang}\ \emph {et~al.}(2023)\citenamefont {Wang},
		\citenamefont {Wu},\ and\ \citenamefont {Wang}}]{wang2023intrinsic}%
	\BibitemOpen
	\bibfield  {author} {\bibinfo {author} {\bibfnamefont {Z.}~\bibnamefont
			{Wang}}, \bibinfo {author} {\bibfnamefont {Z.}~\bibnamefont {Wu}},\ and\
		\bibinfo {author} {\bibfnamefont {Z.}~\bibnamefont {Wang}},\ }\bibfield
	{title} {\bibinfo {title} {Intrinsic mixed-state topological order without
			quantum memory},\ }\href@noop {} {\bibfield  {journal} {\bibinfo  {journal}
			{arXiv preprint arXiv:2307.13758}\ } (\bibinfo {year} {2023})}\BibitemShut
	{NoStop}%
	\bibitem [{\citenamefont {Sang}\ \emph {et~al.}(2023)\citenamefont {Sang},
		\citenamefont {Zou},\ and\ \citenamefont {Hsieh}}]{sang2023mixed}%
	\BibitemOpen
	\bibfield  {author} {\bibinfo {author} {\bibfnamefont {S.}~\bibnamefont
			{Sang}}, \bibinfo {author} {\bibfnamefont {Y.}~\bibnamefont {Zou}},\ and\
		\bibinfo {author} {\bibfnamefont {T.~H.}\ \bibnamefont {Hsieh}},\ }\bibfield
	{title} {\bibinfo {title} {Mixed-state quantum phases: Renormalization and
			quantum error correction},\ }\href@noop {} {\bibfield  {journal} {\bibinfo
			{journal} {arXiv preprint arXiv:2310.08639}\ } (\bibinfo {year}
		{2023})}\BibitemShut {NoStop}%
	\bibitem [{\citenamefont {Lu}\ and\ \citenamefont
		{Grover}(2020)}]{lu2020structure}%
	\BibitemOpen
	\bibfield  {author} {\bibinfo {author} {\bibfnamefont {T.-C.}\ \bibnamefont
			{Lu}}\ and\ \bibinfo {author} {\bibfnamefont {T.}~\bibnamefont {Grover}},\
	}\bibfield  {title} {\bibinfo {title} {Structure of quantum entanglement at a
			finite temperature critical point},\ }\href
	{https://doi.org/10.1103/PhysRevResearch.2.043345} {\bibfield  {journal}
		{\bibinfo  {journal} {Phys. Rev. Res.}\ }\textbf {\bibinfo {volume} {2}},\
		\bibinfo {pages} {043345} (\bibinfo {year} {2020})}\BibitemShut {NoStop}%
	\bibitem [{\citenamefont {Wu}\ \emph {et~al.}(2020)\citenamefont {Wu},
		\citenamefont {Lu}, \citenamefont {Chung}, \citenamefont {Kao},\ and\
		\citenamefont {Grover}}]{wu2020entanglement}%
	\BibitemOpen
	\bibfield  {author} {\bibinfo {author} {\bibfnamefont {K.-H.}\ \bibnamefont
			{Wu}}, \bibinfo {author} {\bibfnamefont {T.-C.}\ \bibnamefont {Lu}}, \bibinfo
		{author} {\bibfnamefont {C.-M.}\ \bibnamefont {Chung}}, \bibinfo {author}
		{\bibfnamefont {Y.-J.}\ \bibnamefont {Kao}},\ and\ \bibinfo {author}
		{\bibfnamefont {T.}~\bibnamefont {Grover}},\ }\bibfield  {title} {\bibinfo
		{title} {Entanglement renyi negativity across a finite temperature
			transition: A monte carlo study},\ }\href
	{https://doi.org/10.1103/PhysRevLett.125.140603} {\bibfield  {journal}
		{\bibinfo  {journal} {Phys. Rev. Lett.}\ }\textbf {\bibinfo {volume} {125}},\
		\bibinfo {pages} {140603} (\bibinfo {year} {2020})}\BibitemShut {NoStop}%
	\bibitem [{\citenamefont {Wald}\ \emph {et~al.}(2020)\citenamefont {Wald},
		\citenamefont {Arias},\ and\ \citenamefont {Alba}}]{wald2020entanglement}%
	\BibitemOpen
	\bibfield  {author} {\bibinfo {author} {\bibfnamefont {S.}~\bibnamefont
			{Wald}}, \bibinfo {author} {\bibfnamefont {R.}~\bibnamefont {Arias}},\ and\
		\bibinfo {author} {\bibfnamefont {V.}~\bibnamefont {Alba}},\ }\bibfield
	{title} {\bibinfo {title} {Entanglement and classical fluctuations at
			finite-temperature critical points},\ }\href
	{https://iopscience.iop.org/article/10.1088/1742-5468/ab6b19/meta} {\bibfield
		{journal} {\bibinfo  {journal} {Journal of Statistical Mechanics: Theory and
				Experiment}\ }\textbf {\bibinfo {volume} {2020}},\ \bibinfo {pages} {033105}
		(\bibinfo {year} {2020})}\BibitemShut {NoStop}%
	\bibitem [{\citenamefont {Ma}\ and\ \citenamefont
		{Wang}(2023)}]{ma2022average}%
	\BibitemOpen
	\bibfield  {author} {\bibinfo {author} {\bibfnamefont {R.}~\bibnamefont
			{Ma}}\ and\ \bibinfo {author} {\bibfnamefont {C.}~\bibnamefont {Wang}},\
	}\bibfield  {title} {\bibinfo {title} {Average symmetry-protected topological
			phases},\ }\href {https://doi.org/10.1103/PhysRevX.13.031016} {\bibfield
		{journal} {\bibinfo  {journal} {Phys. Rev. X}\ }\textbf {\bibinfo {volume}
			{13}},\ \bibinfo {pages} {031016} (\bibinfo {year} {2023})}\BibitemShut
	{NoStop}%
	\bibitem [{\citenamefont {Ma}\ \emph {et~al.}(2023)\citenamefont {Ma},
		\citenamefont {Zhang}, \citenamefont {Bi}, \citenamefont {Cheng},\ and\
		\citenamefont {Wang}}]{ma2023topological}%
	\BibitemOpen
	\bibfield  {author} {\bibinfo {author} {\bibfnamefont {R.}~\bibnamefont
			{Ma}}, \bibinfo {author} {\bibfnamefont {J.-H.}\ \bibnamefont {Zhang}},
		\bibinfo {author} {\bibfnamefont {Z.}~\bibnamefont {Bi}}, \bibinfo {author}
		{\bibfnamefont {M.}~\bibnamefont {Cheng}},\ and\ \bibinfo {author}
		{\bibfnamefont {C.}~\bibnamefont {Wang}},\ }\bibfield  {title} {\bibinfo
		{title} {Topological phases with average symmetries: the decohered, the
			disordered, and the intrinsic},\ }\href {https://arxiv.org/abs/2305.16399}
	{\bibfield  {journal} {\bibinfo  {journal} {arXiv preprint arXiv:2305.16399}\
		} (\bibinfo {year} {2023})}\BibitemShut {NoStop}%
	\bibitem [{\citenamefont {de~Groot}\ \emph {et~al.}(2022)\citenamefont
		{de~Groot}, \citenamefont {Turz~illo},\ and\ \citenamefont
		{Schuch}}]{de2022symmetry}%
	\BibitemOpen
	\bibfield  {author} {\bibinfo {author} {\bibfnamefont {C.}~\bibnamefont
			{de~Groot}}, \bibinfo {author} {\bibfnamefont {A.}~\bibnamefont
			{Turz~illo}},\ and\ \bibinfo {author} {\bibfnamefont {N.}~\bibnamefont
			{Schuch}},\ }\bibfield  {title} {\bibinfo {title} {Symmetry {P}rotected
			{T}opological {O}rder in {O}pen {Q}uantum {S}ystems},\ }\href
	{https://doi.org/10.22331/q-2022-11-10-856} {\bibfield  {journal} {\bibinfo
			{journal} {{Quantum}}\ }\textbf {\bibinfo {volume} {6}},\ \bibinfo {pages}
		{856} (\bibinfo {year} {2022})}\BibitemShut {NoStop}%
	\bibitem [{\citenamefont {Lee}\ \emph {et~al.}(2022{\natexlab{a}})\citenamefont
		{Lee}, \citenamefont {You},\ and\ \citenamefont {Xu}}]{lee2022symmetry}%
	\BibitemOpen
	\bibfield  {author} {\bibinfo {author} {\bibfnamefont {J.~Y.}\ \bibnamefont
			{Lee}}, \bibinfo {author} {\bibfnamefont {Y.-Z.}\ \bibnamefont {You}},\ and\
		\bibinfo {author} {\bibfnamefont {C.}~\bibnamefont {Xu}},\ }\bibfield
	{title} {\bibinfo {title} {Symmetry protected topological phases under
			decoherence},\ }\href {https://arxiv.org/abs/2210.16323} {\bibfield
		{journal} {\bibinfo  {journal} {arXiv preprint arXiv:2210.16323}\ } (\bibinfo
		{year} {2022}{\natexlab{a}})}\BibitemShut {NoStop}%
	\bibitem [{\citenamefont {Zhang}\ \emph {et~al.}(2022)\citenamefont {Zhang},
		\citenamefont {Qi},\ and\ \citenamefont {Bi}}]{zhang2022strange}%
	\BibitemOpen
	\bibfield  {author} {\bibinfo {author} {\bibfnamefont {J.-H.}\ \bibnamefont
			{Zhang}}, \bibinfo {author} {\bibfnamefont {Y.}~\bibnamefont {Qi}},\ and\
		\bibinfo {author} {\bibfnamefont {Z.}~\bibnamefont {Bi}},\ }\bibfield
	{title} {\bibinfo {title} {Strange correlation function for average
			symmetry-protected topological phases},\ }\href
	{https://arxiv.org/abs/2210.17485} {\bibfield  {journal} {\bibinfo  {journal}
			{arXiv preprint arXiv:2210.17485}\ } (\bibinfo {year} {2022})}\BibitemShut
	{NoStop}%
	\bibitem [{\citenamefont {den Nijs}\ and\ \citenamefont
		{Rommelse}(1989)}]{den1989preroughening}%
	\BibitemOpen
	\bibfield  {author} {\bibinfo {author} {\bibfnamefont {M.}~\bibnamefont {den
				Nijs}}\ and\ \bibinfo {author} {\bibfnamefont {K.}~\bibnamefont {Rommelse}},\
	}\bibfield  {title} {\bibinfo {title} {Preroughening transitions in crystal
			surfaces and valence-bond phases in quantum spin chains},\ }\href
	{https://doi.org/10.1103/PhysRevB.40.4709} {\bibfield  {journal} {\bibinfo
			{journal} {Phys. Rev. B}\ }\textbf {\bibinfo {volume} {40}},\ \bibinfo
		{pages} {4709} (\bibinfo {year} {1989})}\BibitemShut {NoStop}%
	\bibitem [{\citenamefont {You}\ \emph {et~al.}(2014)\citenamefont {You},
		\citenamefont {Bi}, \citenamefont {Rasmussen}, \citenamefont {Slagle},\ and\
		\citenamefont {Xu}}]{you2014wave}%
	\BibitemOpen
	\bibfield  {author} {\bibinfo {author} {\bibfnamefont {Y.-Z.}\ \bibnamefont
			{You}}, \bibinfo {author} {\bibfnamefont {Z.}~\bibnamefont {Bi}}, \bibinfo
		{author} {\bibfnamefont {A.}~\bibnamefont {Rasmussen}}, \bibinfo {author}
		{\bibfnamefont {K.}~\bibnamefont {Slagle}},\ and\ \bibinfo {author}
		{\bibfnamefont {C.}~\bibnamefont {Xu}},\ }\bibfield  {title} {\bibinfo
		{title} {Wave function and strange correlator of short-range entangled
			states},\ }\href {https://doi.org/10.1103/PhysRevLett.112.247202} {\bibfield
		{journal} {\bibinfo  {journal} {Phys. Rev. Lett.}\ }\textbf {\bibinfo
			{volume} {112}},\ \bibinfo {pages} {247202} (\bibinfo {year}
		{2014})}\BibitemShut {NoStop}%
	\bibitem [{\citenamefont {Nishimori}(1981)}]{nishimori1981internal}%
	\BibitemOpen
	\bibfield  {author} {\bibinfo {author} {\bibfnamefont {H.}~\bibnamefont
			{Nishimori}},\ }\bibfield  {title} {\bibinfo {title} {Internal energy,
			specific heat and correlation function of the bond-random ising model},\
	}\href {https://academic.oup.com/ptp/article/66/4/1169/1860861} {\bibfield
		{journal} {\bibinfo  {journal} {Progress of Theoretical Physics}\ }\textbf
		{\bibinfo {volume} {66}},\ \bibinfo {pages} {1169} (\bibinfo {year}
		{1981})}\BibitemShut {NoStop}%
	\bibitem [{\citenamefont {Lee}\ \emph {et~al.}(2022{\natexlab{b}})\citenamefont
		{Lee}, \citenamefont {Ji}, \citenamefont {Bi},\ and\ \citenamefont
		{Fisher}}]{lee2022measurement}%
	\BibitemOpen
	\bibfield  {author} {\bibinfo {author} {\bibfnamefont {J.~Y.}\ \bibnamefont
			{Lee}}, \bibinfo {author} {\bibfnamefont {W.}~\bibnamefont {Ji}}, \bibinfo
		{author} {\bibfnamefont {Z.}~\bibnamefont {Bi}},\ and\ \bibinfo {author}
		{\bibfnamefont {M.}~\bibnamefont {Fisher}},\ }\bibfield  {title} {\bibinfo
		{title} {Measurement-prepared quantum criticality: from ising model to gauge
			theory, and beyond},\ }\href {https://arxiv.org/abs/2208.11699} {\bibfield
		{journal} {\bibinfo  {journal} {arXiv preprint arXiv:2208.11699}\ } (\bibinfo
		{year} {2022}{\natexlab{b}})}\BibitemShut {NoStop}%
	\bibitem [{\citenamefont {Zhu}\ \emph {et~al.}(2023)\citenamefont {Zhu},
		\citenamefont {Tantivasadakarn}, \citenamefont {Vishwanath}, \citenamefont
		{Trebst},\ and\ \citenamefont {Verresen}}]{zhu2022nishimori}%
	\BibitemOpen
	\bibfield  {author} {\bibinfo {author} {\bibfnamefont {G.-Y.}\ \bibnamefont
			{Zhu}}, \bibinfo {author} {\bibfnamefont {N.}~\bibnamefont
			{Tantivasadakarn}}, \bibinfo {author} {\bibfnamefont {A.}~\bibnamefont
			{Vishwanath}}, \bibinfo {author} {\bibfnamefont {S.}~\bibnamefont {Trebst}},\
		and\ \bibinfo {author} {\bibfnamefont {R.}~\bibnamefont {Verresen}},\
	}\bibfield  {title} {\bibinfo {title} {Nishimori's cat: Stable long-range
			entanglement from finite-depth unitaries and weak measurements},\ }\href
	{https://doi.org/10.1103/PhysRevLett.131.200201} {\bibfield  {journal}
		{\bibinfo  {journal} {Phys. Rev. Lett.}\ }\textbf {\bibinfo {volume} {131}},\
		\bibinfo {pages} {200201} (\bibinfo {year} {2023})}\BibitemShut {NoStop}%
	\bibitem [{\citenamefont {Lieb}\ and\ \citenamefont
		{Robinson}(1972)}]{lieb1972finite}%
	\BibitemOpen
	\bibfield  {author} {\bibinfo {author} {\bibfnamefont {E.~H.}\ \bibnamefont
			{Lieb}}\ and\ \bibinfo {author} {\bibfnamefont {D.~W.}\ \bibnamefont
			{Robinson}},\ }\bibfield  {title} {\bibinfo {title} {The finite group
			velocity of quantum spin systems},\ }\href
	{https://link.springer.com/article/10.1007/BF01645779} {\bibfield  {journal}
		{\bibinfo  {journal} {Communications in mathematical physics}\ }\textbf
		{\bibinfo {volume} {28}},\ \bibinfo {pages} {251} (\bibinfo {year}
		{1972})}\BibitemShut {NoStop}%
	\bibitem [{\citenamefont {Hastings}(2010)}]{hastings2010locality}%
	\BibitemOpen
	\bibfield  {author} {\bibinfo {author} {\bibfnamefont {M.~B.}\ \bibnamefont
			{Hastings}},\ }\bibfield  {title} {\bibinfo {title} {Locality in quantum
			systems},\ }\href
	{https://academic.oup.com/book/32653/chapter-abstract/270590458?redirectedFrom=fulltext}
	{\bibfield  {journal} {\bibinfo  {journal} {Quantum Theory from Small to
				Large Scales}\ }\textbf {\bibinfo {volume} {95}},\ \bibinfo {pages} {171}
		(\bibinfo {year} {2010})}\BibitemShut {NoStop}%
	\bibitem [{\citenamefont {Huang}\ and\ \citenamefont
		{Chen}(2015)}]{huang2015quantum}%
	\BibitemOpen
	\bibfield  {author} {\bibinfo {author} {\bibfnamefont {Y.}~\bibnamefont
			{Huang}}\ and\ \bibinfo {author} {\bibfnamefont {X.}~\bibnamefont {Chen}},\
	}\bibfield  {title} {\bibinfo {title} {Quantum circuit complexity of
			one-dimensional topological phases},\ }\href
	{https://doi.org/10.1103/PhysRevB.91.195143} {\bibfield  {journal} {\bibinfo
			{journal} {Phys. Rev. B}\ }\textbf {\bibinfo {volume} {91}},\ \bibinfo
		{pages} {195143} (\bibinfo {year} {2015})}\BibitemShut {NoStop}%
	\bibitem [{\citenamefont {Levin}(2020)}]{levin2020constraints}%
	\BibitemOpen
	\bibfield  {author} {\bibinfo {author} {\bibfnamefont {M.}~\bibnamefont
			{Levin}},\ }\bibfield  {title} {\bibinfo {title} {Constraints on order and
			disorder parameters in quantum spin chains},\ }\href
	{https://link.springer.com/article/10.1007/s00220-020-03802-4} {\bibfield
		{journal} {\bibinfo  {journal} {Communications in Mathematical Physics}\
		}\textbf {\bibinfo {volume} {378}},\ \bibinfo {pages} {1081} (\bibinfo {year}
		{2020})}\BibitemShut {NoStop}%
	\bibitem [{\citenamefont {Wen}(2012)}]{wen2012symmetry}%
	\BibitemOpen
	\bibfield  {author} {\bibinfo {author} {\bibfnamefont {X.-G.}\ \bibnamefont
			{Wen}},\ }\bibfield  {title} {\bibinfo {title} {Symmetry-protected
			topological phases in noninteracting fermion systems},\ }\href
	{https://doi.org/10.1103/PhysRevB.85.085103} {\bibfield  {journal} {\bibinfo
			{journal} {Phys. Rev. B}\ }\textbf {\bibinfo {volume} {85}},\ \bibinfo
		{pages} {085103} (\bibinfo {year} {2012})}\BibitemShut {NoStop}%
	\bibitem [{\citenamefont {Ryu}\ \emph {et~al.}(2010)\citenamefont {Ryu},
		\citenamefont {Schnyder}, \citenamefont {Furusaki},\ and\ \citenamefont
		{Ludwig}}]{ryu2010topological}%
	\BibitemOpen
	\bibfield  {author} {\bibinfo {author} {\bibfnamefont {S.}~\bibnamefont
			{Ryu}}, \bibinfo {author} {\bibfnamefont {A.~P.}\ \bibnamefont {Schnyder}},
		\bibinfo {author} {\bibfnamefont {A.}~\bibnamefont {Furusaki}},\ and\
		\bibinfo {author} {\bibfnamefont {A.~W.}\ \bibnamefont {Ludwig}},\ }\bibfield
	{title} {\bibinfo {title} {Topological insulators and superconductors:
			tenfold way and dimensional hierarchy},\ }\href
	{http://dx.doi.org/10.1088/1367-2630/12/6/065010} {\bibfield  {journal}
		{\bibinfo  {journal} {New Journal of Physics}\ }\textbf {\bibinfo {volume}
			{12}},\ \bibinfo {pages} {065010} (\bibinfo {year} {2010})}\BibitemShut
	{NoStop}%
	\bibitem [{\citenamefont {Jamio{\l}kowski}(1972)}]{jamiolkowski1972linear}%
	\BibitemOpen
	\bibfield  {author} {\bibinfo {author} {\bibfnamefont {A.}~\bibnamefont
			{Jamio{\l}kowski}},\ }\bibfield  {title} {\bibinfo {title} {Linear
			transformations which preserve trace and positive semidefiniteness of
			operators},\ }\href
	{https://www.sciencedirect.com/science/article/abs/pii/0034487772900110}
	{\bibfield  {journal} {\bibinfo  {journal} {Reports on Mathematical Physics}\
		}\textbf {\bibinfo {volume} {3}},\ \bibinfo {pages} {275} (\bibinfo {year}
		{1972})}\BibitemShut {NoStop}%
	\bibitem [{\citenamefont {Choi}(1975)}]{choi1975completely}%
	\BibitemOpen
	\bibfield  {author} {\bibinfo {author} {\bibfnamefont {M.-D.}\ \bibnamefont
			{Choi}},\ }\bibfield  {title} {\bibinfo {title} {Completely positive linear
			maps on complex matrices},\ }\href
	{https://www.sciencedirect.com/science/article/pii/0024379575900750}
	{\bibfield  {journal} {\bibinfo  {journal} {Linear algebra and its
				applications}\ }\textbf {\bibinfo {volume} {10}},\ \bibinfo {pages} {285}
		(\bibinfo {year} {1975})}\BibitemShut {NoStop}%
	\bibitem [{\citenamefont {Thouless}(1984)}]{thouless1984wannier}%
	\BibitemOpen
	\bibfield  {author} {\bibinfo {author} {\bibfnamefont {D.}~\bibnamefont
			{Thouless}},\ }\bibfield  {title} {\bibinfo {title} {Wannier functions for
			magnetic sub-bands},\ }\href
	{https://iopscience.iop.org/article/10.1088/0022-3719/17/12/003} {\bibfield
		{journal} {\bibinfo  {journal} {Journal of Physics C: Solid State Physics}\
		}\textbf {\bibinfo {volume} {17}},\ \bibinfo {pages} {L325} (\bibinfo {year}
		{1984})}\BibitemShut {NoStop}%
	\bibitem [{\citenamefont {Read}\ and\ \citenamefont
		{Green}(2000)}]{read2000paired}%
	\BibitemOpen
	\bibfield  {author} {\bibinfo {author} {\bibfnamefont {N.}~\bibnamefont
			{Read}}\ and\ \bibinfo {author} {\bibfnamefont {D.}~\bibnamefont {Green}},\
	}\bibfield  {title} {\bibinfo {title} {Paired states of fermions in two
			dimensions with breaking of parity and time-reversal symmetries and the
			fractional quantum hall effect},\ }\href
	{https://doi.org/10.1103/PhysRevB.61.10267} {\bibfield  {journal} {\bibinfo
			{journal} {Phys. Rev. B}\ }\textbf {\bibinfo {volume} {61}},\ \bibinfo
		{pages} {10267} (\bibinfo {year} {2000})}\BibitemShut {NoStop}%
	\bibitem [{\citenamefont {Schindler}\ \emph {et~al.}(2020)\citenamefont
		{Schindler}, \citenamefont {Bradlyn}, \citenamefont {Fischer},\ and\
		\citenamefont {Neupert}}]{schindler2020pairing}%
	\BibitemOpen
	\bibfield  {author} {\bibinfo {author} {\bibfnamefont {F.}~\bibnamefont
			{Schindler}}, \bibinfo {author} {\bibfnamefont {B.}~\bibnamefont {Bradlyn}},
		\bibinfo {author} {\bibfnamefont {M.~H.}\ \bibnamefont {Fischer}},\ and\
		\bibinfo {author} {\bibfnamefont {T.}~\bibnamefont {Neupert}},\ }\bibfield
	{title} {\bibinfo {title} {Pairing obstructions in topological
			superconductors},\ }\href {https://doi.org/10.1103/PhysRevLett.124.247001}
	{\bibfield  {journal} {\bibinfo  {journal} {Phys. Rev. Lett.}\ }\textbf
		{\bibinfo {volume} {124}},\ \bibinfo {pages} {247001} (\bibinfo {year}
		{2020})}\BibitemShut {NoStop}%
	\bibitem [{\citenamefont {Freedman}\ and\ \citenamefont
		{Hastings}(2013)}]{freedman2013quantum}%
	\BibitemOpen
	\bibfield  {author} {\bibinfo {author} {\bibfnamefont {M.~H.}\ \bibnamefont
			{Freedman}}\ and\ \bibinfo {author} {\bibfnamefont {M.~B.}\ \bibnamefont
			{Hastings}},\ }\bibfield  {title} {\bibinfo {title} {Quantum systems on non-$
			k $-hyperfinite complexes: A generalization of classical statistical
			mechanics on expander graphs},\ }\href {https://arxiv.org/abs/1301.1363}
	{\bibfield  {journal} {\bibinfo  {journal} {arXiv preprint arXiv:1301.1363}\
		} (\bibinfo {year} {2013})}\BibitemShut {NoStop}%
	\bibitem [{\citenamefont {Anshu}\ \emph {et~al.}(2022)\citenamefont {Anshu},
		\citenamefont {Breuckmann},\ and\ \citenamefont {Nirkhe}}]{anshu2022nlts}%
	\BibitemOpen
	\bibfield  {author} {\bibinfo {author} {\bibfnamefont {A.}~\bibnamefont
			{Anshu}}, \bibinfo {author} {\bibfnamefont {N.}~\bibnamefont {Breuckmann}},\
		and\ \bibinfo {author} {\bibfnamefont {C.}~\bibnamefont {Nirkhe}},\
	}\bibfield  {title} {\bibinfo {title} {Nlts hamiltonians from good quantum
			codes},\ }\href {https://dl.acm.org/doi/10.1145/3564246.3585114} {\bibfield
		{journal} {\bibinfo  {journal} {arXiv preprint arXiv:2206.13228}\ } (\bibinfo
		{year} {2022})}\BibitemShut {NoStop}%
	\bibitem [{\citenamefont {Leverrier}\ and\ \citenamefont
		{Z{\'e}mor}(2022)}]{leverrier2022quantum}%
	\BibitemOpen
	\bibfield  {author} {\bibinfo {author} {\bibfnamefont {A.}~\bibnamefont
			{Leverrier}}\ and\ \bibinfo {author} {\bibfnamefont {G.}~\bibnamefont
			{Z{\'e}mor}},\ }\bibfield  {title} {\bibinfo {title} {Quantum tanner codes},\
	}\href {https://doi.ieeecomputersociety.org/10.1109/FOCS54457.2022.00117}
	{\bibfield  {journal} {\bibinfo  {journal} {2022 IEEE 63rd Annual Symposium
				on Foundations of Computer Science (FOCS)}\ ,\ \bibinfo {pages} {872}}
		(\bibinfo {year} {2022})}\BibitemShut {NoStop}%
	\bibitem [{\citenamefont {Panteleev}\ and\ \citenamefont
		{Kalachev}(2022)}]{panteleev2022asymptotically}%
	\BibitemOpen
	\bibfield  {author} {\bibinfo {author} {\bibfnamefont {P.}~\bibnamefont
			{Panteleev}}\ and\ \bibinfo {author} {\bibfnamefont {G.}~\bibnamefont
			{Kalachev}},\ }\bibfield  {title} {\bibinfo {title} {Asymptotically good
			quantum and locally testable classical ldpc codes},\ }\href
	{https://dl.acm.org/doi/abs/10.1145/3519935.3520017} {\bibfield  {journal}
		{\bibinfo  {journal} {Proceedings of the 54th Annual ACM SIGACT Symposium on
				Theory of Computing}\ ,\ \bibinfo {pages} {375}} (\bibinfo {year}
		{2022})}\BibitemShut {NoStop}%
	\bibitem [{\citenamefont {Dinur}\ \emph {et~al.}(2023)\citenamefont {Dinur},
		\citenamefont {Hsieh}, \citenamefont {Lin},\ and\ \citenamefont
		{Vidick}}]{dinur2023good}%
	\BibitemOpen
	\bibfield  {author} {\bibinfo {author} {\bibfnamefont {I.}~\bibnamefont
			{Dinur}}, \bibinfo {author} {\bibfnamefont {M.-H.}\ \bibnamefont {Hsieh}},
		\bibinfo {author} {\bibfnamefont {T.-C.}\ \bibnamefont {Lin}},\ and\ \bibinfo
		{author} {\bibfnamefont {T.}~\bibnamefont {Vidick}},\ }\bibfield  {title}
	{\bibinfo {title} {Good quantum ldpc codes with linear time decoders},\
	}\href {https://dl.acm.org/doi/10.1145/3564246.3585101} {\bibfield  {journal}
		{\bibinfo  {journal} {Proceedings of the 55th Annual ACM Symposium on Theory
				of Computing}\ ,\ \bibinfo {pages} {905}} (\bibinfo {year}
		{2023})}\BibitemShut {NoStop}%
	\bibitem [{\citenamefont {Anshu}\ and\ \citenamefont
		{Nirkhe}(2020)}]{anshu2020circuit}%
	\BibitemOpen
	\bibfield  {author} {\bibinfo {author} {\bibfnamefont {A.}~\bibnamefont
			{Anshu}}\ and\ \bibinfo {author} {\bibfnamefont {C.}~\bibnamefont {Nirkhe}},\
	}\bibfield  {title} {\bibinfo {title} {Circuit lower bounds for low-energy
			states of quantum code hamiltonians},\ }\href
	{https://drops.dagstuhl.de/entities/document/10.4230/LIPIcs.ITCS.2022.6}
	{\bibfield  {journal} {\bibinfo  {journal} {arXiv preprint arXiv:2011.02044}\
		} (\bibinfo {year} {2020})}\BibitemShut {NoStop}%
	\bibitem [{\citenamefont {Lu}\ \emph {et~al.}(2020)\citenamefont {Lu},
		\citenamefont {Hsieh},\ and\ \citenamefont {Grover}}]{lu2020detecting}%
	\BibitemOpen
	\bibfield  {author} {\bibinfo {author} {\bibfnamefont {T.-C.}\ \bibnamefont
			{Lu}}, \bibinfo {author} {\bibfnamefont {T.~H.}\ \bibnamefont {Hsieh}},\ and\
		\bibinfo {author} {\bibfnamefont {T.}~\bibnamefont {Grover}},\ }\bibfield
	{title} {\bibinfo {title} {Detecting topological order at finite temperature
			using entanglement negativity},\ }\href
	{https://doi.org/10.1103/PhysRevLett.125.116801} {\bibfield  {journal}
		{\bibinfo  {journal} {Phys. Rev. Lett.}\ }\textbf {\bibinfo {volume} {125}},\
		\bibinfo {pages} {116801} (\bibinfo {year} {2020})}\BibitemShut {NoStop}%
	\bibitem [{\citenamefont {Roberts}\ \emph {et~al.}(2017)\citenamefont
		{Roberts}, \citenamefont {Yoshida}, \citenamefont {Kubica},\ and\
		\citenamefont {Bartlett}}]{roberts2017symmetry}%
	\BibitemOpen
	\bibfield  {author} {\bibinfo {author} {\bibfnamefont {S.}~\bibnamefont
			{Roberts}}, \bibinfo {author} {\bibfnamefont {B.}~\bibnamefont {Yoshida}},
		\bibinfo {author} {\bibfnamefont {A.}~\bibnamefont {Kubica}},\ and\ \bibinfo
		{author} {\bibfnamefont {S.~D.}\ \bibnamefont {Bartlett}},\ }\bibfield
	{title} {\bibinfo {title} {Symmetry-protected topological order at nonzero
			temperature},\ }\href {https://doi.org/10.1103/PhysRevA.96.022306} {\bibfield
		{journal} {\bibinfo  {journal} {Phys. Rev. A}\ }\textbf {\bibinfo {volume}
			{96}},\ \bibinfo {pages} {022306} (\bibinfo {year} {2017})}\BibitemShut
	{NoStop}%
	\bibitem [{\citenamefont {Terhal}\ \emph {et~al.}(2002)\citenamefont {Terhal},
		\citenamefont {Horodecki}, \citenamefont {Leung},\ and\ \citenamefont
		{DiVincenzo}}]{terhal2002entanglement}%
	\BibitemOpen
	\bibfield  {author} {\bibinfo {author} {\bibfnamefont {B.~M.}\ \bibnamefont
			{Terhal}}, \bibinfo {author} {\bibfnamefont {M.}~\bibnamefont {Horodecki}},
		\bibinfo {author} {\bibfnamefont {D.~W.}\ \bibnamefont {Leung}},\ and\
		\bibinfo {author} {\bibfnamefont {D.~P.}\ \bibnamefont {DiVincenzo}},\
	}\bibfield  {title} {\bibinfo {title} {The entanglement of purification},\
	}\href
	{https://pubs.aip.org/aip/jmp/article-abstract/43/9/4286/230896/The-entanglement-of-purification?redirectedFrom=fulltext}
	{\bibfield  {journal} {\bibinfo  {journal} {Journal of Mathematical Physics}\
		}\textbf {\bibinfo {volume} {43}},\ \bibinfo {pages} {4286} (\bibinfo {year}
		{2002})}\BibitemShut {NoStop}%
	\bibitem [{\citenamefont {White}(2009)}]{white2009minimally}%
	\BibitemOpen
	\bibfield  {author} {\bibinfo {author} {\bibfnamefont {S.~R.}\ \bibnamefont
			{White}},\ }\bibfield  {title} {\bibinfo {title} {Minimally entangled typical
			quantum states at finite temperature},\ }\href
	{https://doi.org/10.1103/PhysRevLett.102.190601} {\bibfield  {journal}
		{\bibinfo  {journal} {Phys. Rev. Lett.}\ }\textbf {\bibinfo {volume} {102}},\
		\bibinfo {pages} {190601} (\bibinfo {year} {2009})}\BibitemShut {NoStop}%
	\bibitem [{\citenamefont {Horodecki}\ \emph {et~al.}(2009)\citenamefont
		{Horodecki}, \citenamefont {Horodecki}, \citenamefont {Horodecki},\ and\
		\citenamefont {Horodecki}}]{horodecki2009quantum}%
	\BibitemOpen
	\bibfield  {author} {\bibinfo {author} {\bibfnamefont {R.}~\bibnamefont
			{Horodecki}}, \bibinfo {author} {\bibfnamefont {P.}~\bibnamefont
			{Horodecki}}, \bibinfo {author} {\bibfnamefont {M.}~\bibnamefont
			{Horodecki}},\ and\ \bibinfo {author} {\bibfnamefont {K.}~\bibnamefont
			{Horodecki}},\ }\bibfield  {title} {\bibinfo {title} {Quantum entanglement},\
	}\href {https://doi.org/10.1103/RevModPhys.81.865} {\bibfield  {journal}
		{\bibinfo  {journal} {Rev. Mod. Phys.}\ }\textbf {\bibinfo {volume} {81}},\
		\bibinfo {pages} {865} (\bibinfo {year} {2009})}\BibitemShut {NoStop}%
	\bibitem [{\citenamefont {Lu}\ and\ \citenamefont
		{Vijay}(2023)}]{lu2023characterizing}%
	\BibitemOpen
	\bibfield  {author} {\bibinfo {author} {\bibfnamefont {T.-C.}\ \bibnamefont
			{Lu}}\ and\ \bibinfo {author} {\bibfnamefont {S.}~\bibnamefont {Vijay}},\
	}\bibfield  {title} {\bibinfo {title} {Characterizing long-range entanglement
			in a mixed state through an emergent order on the entangling surface},\
	}\href {https://doi.org/10.1103/PhysRevResearch.5.033031} {\bibfield
		{journal} {\bibinfo  {journal} {Phys. Rev. Res.}\ }\textbf {\bibinfo {volume}
			{5}},\ \bibinfo {pages} {033031} (\bibinfo {year} {2023})}\BibitemShut
	{NoStop}%
	\bibitem [{\citenamefont {Dennis}\ \emph {et~al.}(2002)\citenamefont {Dennis},
		\citenamefont {Kitaev}, \citenamefont {Landahl},\ and\ \citenamefont
		{Preskill}}]{dennis2002}%
	\BibitemOpen
	\bibfield  {author} {\bibinfo {author} {\bibfnamefont {E.}~\bibnamefont
			{Dennis}}, \bibinfo {author} {\bibfnamefont {A.}~\bibnamefont {Kitaev}},
		\bibinfo {author} {\bibfnamefont {A.}~\bibnamefont {Landahl}},\ and\ \bibinfo
		{author} {\bibfnamefont {J.}~\bibnamefont {Preskill}},\ }\bibfield  {title}
	{\bibinfo {title} {Topological quantum memory},\ }\href
	{https://doi.org/10.1063/1.1499754} {\bibfield  {journal} {\bibinfo
			{journal} {Journal of Mathematical Physics}\ }\textbf {\bibinfo {volume}
			{43}},\ \bibinfo {pages} {4452} (\bibinfo {year} {2002})}\BibitemShut
	{NoStop}%
	\bibitem [{\citenamefont {Yoshida}(2011)}]{yoshida2011}%
	\BibitemOpen
	\bibfield  {author} {\bibinfo {author} {\bibfnamefont {B.}~\bibnamefont
			{Yoshida}},\ }\bibfield  {title} {\bibinfo {title} {Feasibility of
			self-correcting quantum memory and thermal stability of topological order},\
	}\href {https://doi.org/https://doi.org/10.1016/j.aop.2011.06.001} {\bibfield
		{journal} {\bibinfo  {journal} {Annals of Physics}\ }\textbf {\bibinfo
			{volume} {326}},\ \bibinfo {pages} {2566 } (\bibinfo {year}
		{2011})}\BibitemShut {NoStop}%
	\bibitem [{\citenamefont {Shor}(1996)}]{shor1996fault}%
	\BibitemOpen
	\bibfield  {author} {\bibinfo {author} {\bibfnamefont {P.~W.}\ \bibnamefont
			{Shor}},\ }\bibfield  {title} {\bibinfo {title} {Fault-tolerant quantum
			computation},\ }\href {https://ieeexplore.ieee.org/document/548464}
	{\bibfield  {journal} {\bibinfo  {journal} {Proceedings of 37th conference on
				foundations of computer science}\ ,\ \bibinfo {pages} {56}} (\bibinfo {year}
		{1996})}\BibitemShut {NoStop}%
	\bibitem [{\citenamefont {Aharonov}\ and\ \citenamefont
		{Ben-Or}(1997)}]{aharonov1997fault}%
	\BibitemOpen
	\bibfield  {author} {\bibinfo {author} {\bibfnamefont {D.}~\bibnamefont
			{Aharonov}}\ and\ \bibinfo {author} {\bibfnamefont {M.}~\bibnamefont
			{Ben-Or}},\ }\bibfield  {title} {\bibinfo {title} {Fault-tolerant quantum
			computation with constant error},\ }\href
	{https://arxiv.org/abs/quant-ph/9906129} {\bibfield  {journal} {\bibinfo
			{journal} {Proceedings of the twenty-ninth annual ACM symposium on Theory of
				computing}\ ,\ \bibinfo {pages} {176}} (\bibinfo {year} {1997})}\BibitemShut
	{NoStop}%
	\bibitem [{\citenamefont {Kitaev}(2003)}]{kitaev2003fault}%
	\BibitemOpen
	\bibfield  {author} {\bibinfo {author} {\bibfnamefont {A.~Y.}\ \bibnamefont
			{Kitaev}},\ }\bibfield  {title} {\bibinfo {title} {Fault-tolerant quantum
			computation by anyons},\ }\href
	{https://doi.org/10.1016%2Fs0003-4916%2802%2900018-0} {\bibfield  {journal}
		{\bibinfo  {journal} {Annals of Physics}\ }\textbf {\bibinfo {volume}
			{303}},\ \bibinfo {pages} {2} (\bibinfo {year} {2003})}\BibitemShut {NoStop}%
	\bibitem [{\citenamefont {Knill}\ \emph {et~al.}(1998)\citenamefont {Knill},
		\citenamefont {Laflamme},\ and\ \citenamefont {Zurek}}]{knill1998resilient}%
	\BibitemOpen
	\bibfield  {author} {\bibinfo {author} {\bibfnamefont {E.}~\bibnamefont
			{Knill}}, \bibinfo {author} {\bibfnamefont {R.}~\bibnamefont {Laflamme}},\
		and\ \bibinfo {author} {\bibfnamefont {W.~H.}\ \bibnamefont {Zurek}},\
	}\bibfield  {title} {\bibinfo {title} {Resilient quantum computation},\
	}\href {https://www.science.org/doi/10.1126/science.279.5349.342} {\bibfield
		{journal} {\bibinfo  {journal} {Science}\ }\textbf {\bibinfo {volume}
			{279}},\ \bibinfo {pages} {342} (\bibinfo {year} {1998})}\BibitemShut
	{NoStop}%
	\bibitem [{\citenamefont {Preskill}(1998)}]{preskill1998reliable}%
	\BibitemOpen
	\bibfield  {author} {\bibinfo {author} {\bibfnamefont {J.}~\bibnamefont
			{Preskill}},\ }\bibfield  {title} {\bibinfo {title} {Reliable quantum
			computers},\ }\href {https://doi.org/10.1098%2Frspa.1998.0167} {\bibfield
		{journal} {\bibinfo  {journal} {Proceedings of the Royal Society of London.
				Series A: Mathematical, Physical and Engineering Sciences}\ }\textbf
		{\bibinfo {volume} {454}},\ \bibinfo {pages} {385} (\bibinfo {year}
		{1998})}\BibitemShut {NoStop}%
	\bibitem [{\citenamefont {Terhal}(2015)}]{terhal2015quantum}%
	\BibitemOpen
	\bibfield  {author} {\bibinfo {author} {\bibfnamefont {B.~M.}\ \bibnamefont
			{Terhal}},\ }\bibfield  {title} {\bibinfo {title} {Quantum error correction
			for quantum memories},\ }\href {https://doi.org/10.1103/RevModPhys.87.307}
	{\bibfield  {journal} {\bibinfo  {journal} {Rev. Mod. Phys.}\ }\textbf
		{\bibinfo {volume} {87}},\ \bibinfo {pages} {307} (\bibinfo {year}
		{2015})}\BibitemShut {NoStop}%
	\bibitem [{\citenamefont {Wang}\ \emph {et~al.}(2003)\citenamefont {Wang},
		\citenamefont {Harrington},\ and\ \citenamefont
		{Preskill}}]{wang2003confinement}%
	\BibitemOpen
	\bibfield  {author} {\bibinfo {author} {\bibfnamefont {C.}~\bibnamefont
			{Wang}}, \bibinfo {author} {\bibfnamefont {J.}~\bibnamefont {Harrington}},\
		and\ \bibinfo {author} {\bibfnamefont {J.}~\bibnamefont {Preskill}},\
	}\bibfield  {title} {\bibinfo {title} {Confinement-higgs transition in a
			disordered gauge theory and the accuracy threshold for quantum memory},\
	}\href {https://doi.org/10.1016%2Fs0003-4916%2802%2900019-2} {\bibfield
		{journal} {\bibinfo  {journal} {Annals of Physics}\ }\textbf {\bibinfo
			{volume} {303}},\ \bibinfo {pages} {31} (\bibinfo {year} {2003})}\BibitemShut
	{NoStop}%
	\bibitem [{Note1()}]{Note1}%
	\BibitemOpen
	\bibinfo {note} {Here we ignore the non-contractible `charges' corresponding
		to $\DOTSB \prod@ \slimits@ _{\ell } x_e$ where $\ell $ is a non-contractible
		loop around the torus on which the system lives. This is because we will only
		be concerned with observables involving operators in the bulk of the system
		and such observables are insensitive to non-contractible
		charges.}\BibitemShut {Stop}%
	\bibitem [{\citenamefont {Honecker}\ \emph {et~al.}(2001)\citenamefont
		{Honecker}, \citenamefont {Picco},\ and\ \citenamefont
		{Pujol}}]{honecker2001universality}%
	\BibitemOpen
	\bibfield  {author} {\bibinfo {author} {\bibfnamefont {A.}~\bibnamefont
			{Honecker}}, \bibinfo {author} {\bibfnamefont {M.}~\bibnamefont {Picco}},\
		and\ \bibinfo {author} {\bibfnamefont {P.}~\bibnamefont {Pujol}},\ }\bibfield
	{title} {\bibinfo {title} {Universality class of the nishimori point in the
			2d $\ifmmode\pm\else\textpm\fi{}\mathit{J}$ random-bond ising model},\ }\href
	{https://doi.org/10.1103/PhysRevLett.87.047201} {\bibfield  {journal}
		{\bibinfo  {journal} {Phys. Rev. Lett.}\ }\textbf {\bibinfo {volume} {87}},\
		\bibinfo {pages} {047201} (\bibinfo {year} {2001})}\BibitemShut {NoStop}%
	\bibitem [{\citenamefont {Yoshida}(2016)}]{yoshida2016topological}%
	\BibitemOpen
	\bibfield  {author} {\bibinfo {author} {\bibfnamefont {B.}~\bibnamefont
			{Yoshida}},\ }\bibfield  {title} {\bibinfo {title} {Topological phases with
			generalized global symmetries},\ }\href
	{https://doi.org/10.1103/PhysRevB.93.155131} {\bibfield  {journal} {\bibinfo
			{journal} {Phys. Rev. B}\ }\textbf {\bibinfo {volume} {93}},\ \bibinfo
		{pages} {155131} (\bibinfo {year} {2016})}\BibitemShut {NoStop}%
	\bibitem [{\citenamefont {Kitaev}(2001)}]{kitaev2001unpaired}%
	\BibitemOpen
	\bibfield  {author} {\bibinfo {author} {\bibfnamefont {A.~Y.}\ \bibnamefont
			{Kitaev}},\ }\bibfield  {title} {\bibinfo {title} {Unpaired majorana fermions
			in quantum wires},\ }\href
	{https://iopscience.iop.org/article/10.1070/1063-7869/44/10S/S29} {\bibfield
		{journal} {\bibinfo  {journal} {Physics-uspekhi}\ }\textbf {\bibinfo {volume}
			{44}},\ \bibinfo {pages} {131} (\bibinfo {year} {2001})}\BibitemShut
	{NoStop}%
	\bibitem [{\citenamefont {Levin}\ and\ \citenamefont
		{Gu}(2012)}]{levin2012braiding}%
	\BibitemOpen
	\bibfield  {author} {\bibinfo {author} {\bibfnamefont {M.}~\bibnamefont
			{Levin}}\ and\ \bibinfo {author} {\bibfnamefont {Z.-C.}\ \bibnamefont {Gu}},\
	}\bibfield  {title} {\bibinfo {title} {Braiding statistics approach to
			symmetry-protected topological phases},\ }\href
	{https://doi.org/10.1103/PhysRevB.86.115109} {\bibfield  {journal} {\bibinfo
			{journal} {Phys. Rev. B}\ }\textbf {\bibinfo {volume} {86}},\ \bibinfo
		{pages} {115109} (\bibinfo {year} {2012})}\BibitemShut {NoStop}%
	\bibitem [{\citenamefont {Klein~Kvorning}\ \emph {et~al.}(2020)\citenamefont
		{Klein~Kvorning}, \citenamefont {Sp\aa{}nsl\"att}, \citenamefont {Chan},\
		and\ \citenamefont {Ryu}}]{kvorning2020nonlocal}%
	\BibitemOpen
	\bibfield  {author} {\bibinfo {author} {\bibfnamefont {T.}~\bibnamefont
			{Klein~Kvorning}}, \bibinfo {author} {\bibfnamefont {C.}~\bibnamefont
			{Sp\aa{}nsl\"att}}, \bibinfo {author} {\bibfnamefont {A.~P.~O.}\ \bibnamefont
			{Chan}},\ and\ \bibinfo {author} {\bibfnamefont {S.}~\bibnamefont {Ryu}},\
	}\bibfield  {title} {\bibinfo {title} {Nonlocal order parameters for states
			with topological electromagnetic response},\ }\href
	{https://doi.org/10.1103/PhysRevB.101.205101} {\bibfield  {journal} {\bibinfo
			{journal} {Phys. Rev. B}\ }\textbf {\bibinfo {volume} {101}},\ \bibinfo
		{pages} {205101} (\bibinfo {year} {2020})}\BibitemShut {NoStop}%
	\bibitem [{\citenamefont {Girvin}\ and\ \citenamefont
		{MacDonald}(1987)}]{girvin1987off}%
	\BibitemOpen
	\bibfield  {author} {\bibinfo {author} {\bibfnamefont {S.~M.}\ \bibnamefont
			{Girvin}}\ and\ \bibinfo {author} {\bibfnamefont {A.~H.}\ \bibnamefont
			{MacDonald}},\ }\bibfield  {title} {\bibinfo {title} {Off-diagonal long-range
			order, oblique confinement, and the fractional quantum hall effect},\ }\href
	{https://doi.org/10.1103/PhysRevLett.58.1252} {\bibfield  {journal} {\bibinfo
			{journal} {Phys. Rev. Lett.}\ }\textbf {\bibinfo {volume} {58}},\ \bibinfo
		{pages} {1252} (\bibinfo {year} {1987})}\BibitemShut {NoStop}%
	\bibitem [{\citenamefont {Grover}\ \emph {et~al.}(2014)\citenamefont {Grover},
		\citenamefont {Sheng},\ and\ \citenamefont
		{Vishwanath}}]{grover2014emergent}%
	\BibitemOpen
	\bibfield  {author} {\bibinfo {author} {\bibfnamefont {T.}~\bibnamefont
			{Grover}}, \bibinfo {author} {\bibfnamefont {D.}~\bibnamefont {Sheng}},\ and\
		\bibinfo {author} {\bibfnamefont {A.}~\bibnamefont {Vishwanath}},\ }\bibfield
	{title} {\bibinfo {title} {Emergent space-time supersymmetry at the boundary
			of a topological phase},\ }\href {http://dx.doi.org/10.1126/science.1248253}
	{\bibfield  {journal} {\bibinfo  {journal} {Science}\ }\textbf {\bibinfo
			{volume} {344}},\ \bibinfo {pages} {280} (\bibinfo {year}
		{2014})}\BibitemShut {NoStop}%
	\bibitem [{\citenamefont {Rahmani}\ \emph {et~al.}(2015)\citenamefont
		{Rahmani}, \citenamefont {Zhu}, \citenamefont {Franz},\ and\ \citenamefont
		{Affleck}}]{rahmani2015emergent}%
	\BibitemOpen
	\bibfield  {author} {\bibinfo {author} {\bibfnamefont {A.}~\bibnamefont
			{Rahmani}}, \bibinfo {author} {\bibfnamefont {X.}~\bibnamefont {Zhu}},
		\bibinfo {author} {\bibfnamefont {M.}~\bibnamefont {Franz}},\ and\ \bibinfo
		{author} {\bibfnamefont {I.}~\bibnamefont {Affleck}},\ }\bibfield  {title}
	{\bibinfo {title} {Emergent supersymmetry from strongly interacting majorana
			zero modes},\ }\href {https://doi.org/10.1103/PhysRevLett.115.166401}
	{\bibfield  {journal} {\bibinfo  {journal} {Phys. Rev. Lett.}\ }\textbf
		{\bibinfo {volume} {115}},\ \bibinfo {pages} {166401} (\bibinfo {year}
		{2015})}\BibitemShut {NoStop}%
	\bibitem [{\citenamefont {Li}\ and\ \citenamefont
		{Haldane}(2008)}]{li2008entanglement}%
	\BibitemOpen
	\bibfield  {author} {\bibinfo {author} {\bibfnamefont {H.}~\bibnamefont
			{Li}}\ and\ \bibinfo {author} {\bibfnamefont {F.~D.~M.}\ \bibnamefont
			{Haldane}},\ }\bibfield  {title} {\bibinfo {title} {Entanglement spectrum as
			a generalization of entanglement entropy: Identification of topological order
			in non-abelian fractional quantum hall effect states},\ }\href@noop {}
	{\bibfield  {journal} {\bibinfo  {journal} {Physical review letters}\
		}\textbf {\bibinfo {volume} {101}},\ \bibinfo {pages} {010504} (\bibinfo
		{year} {2008})}\BibitemShut {NoStop}%
	\bibitem [{\citenamefont {Kitaev}\ and\ \citenamefont
		{Preskill}(2006)}]{Kitaev06_1}%
	\BibitemOpen
	\bibfield  {author} {\bibinfo {author} {\bibfnamefont {A.}~\bibnamefont
			{Kitaev}}\ and\ \bibinfo {author} {\bibfnamefont {J.}~\bibnamefont
			{Preskill}},\ }\bibfield  {title} {\bibinfo {title} {Topological entanglement
			entropy},\ }\href {https://doi.org/10.1103/PhysRevLett.96.110404} {\bibfield
		{journal} {\bibinfo  {journal} {Phys. Rev. Lett.}\ }\textbf {\bibinfo
			{volume} {96}},\ \bibinfo {pages} {110404} (\bibinfo {year}
		{2006})}\BibitemShut {NoStop}%
	\bibitem [{\citenamefont {Kim}\ \emph {et~al.}(2022{\natexlab{a}})\citenamefont
		{Kim}, \citenamefont {Shi}, \citenamefont {Kato},\ and\ \citenamefont
		{Albert}}]{kim2022chiral}%
	\BibitemOpen
	\bibfield  {author} {\bibinfo {author} {\bibfnamefont {I.~H.}\ \bibnamefont
			{Kim}}, \bibinfo {author} {\bibfnamefont {B.}~\bibnamefont {Shi}}, \bibinfo
		{author} {\bibfnamefont {K.}~\bibnamefont {Kato}},\ and\ \bibinfo {author}
		{\bibfnamefont {V.~V.}\ \bibnamefont {Albert}},\ }\bibfield  {title}
	{\bibinfo {title} {Chiral central charge from a single bulk wave function},\
	}\href {https://doi.org/10.1103/PhysRevLett.128.176402} {\bibfield  {journal}
		{\bibinfo  {journal} {Phys. Rev. Lett.}\ }\textbf {\bibinfo {volume} {128}},\
		\bibinfo {pages} {176402} (\bibinfo {year} {2022}{\natexlab{a}})}\BibitemShut
	{NoStop}%
	\bibitem [{\citenamefont {Kim}\ \emph {et~al.}(2022{\natexlab{b}})\citenamefont
		{Kim}, \citenamefont {Shi}, \citenamefont {Kato},\ and\ \citenamefont
		{Albert}}]{kim2022modular}%
	\BibitemOpen
	\bibfield  {author} {\bibinfo {author} {\bibfnamefont {I.~H.}\ \bibnamefont
			{Kim}}, \bibinfo {author} {\bibfnamefont {B.}~\bibnamefont {Shi}}, \bibinfo
		{author} {\bibfnamefont {K.}~\bibnamefont {Kato}},\ and\ \bibinfo {author}
		{\bibfnamefont {V.~V.}\ \bibnamefont {Albert}},\ }\bibfield  {title}
	{\bibinfo {title} {Modular commutator in gapped quantum many-body systems},\
	}\href {https://doi.org/10.1103/PhysRevB.106.075147} {\bibfield  {journal}
		{\bibinfo  {journal} {Phys. Rev. B}\ }\textbf {\bibinfo {volume} {106}},\
		\bibinfo {pages} {075147} (\bibinfo {year} {2022}{\natexlab{b}})}\BibitemShut
	{NoStop}%
	\bibitem [{\citenamefont {Zou}\ \emph {et~al.}(2022)\citenamefont {Zou},
		\citenamefont {Shi}, \citenamefont {Sorce}, \citenamefont {Lim},\ and\
		\citenamefont {Kim}}]{zou2022modular}%
	\BibitemOpen
	\bibfield  {author} {\bibinfo {author} {\bibfnamefont {Y.}~\bibnamefont
			{Zou}}, \bibinfo {author} {\bibfnamefont {B.}~\bibnamefont {Shi}}, \bibinfo
		{author} {\bibfnamefont {J.}~\bibnamefont {Sorce}}, \bibinfo {author}
		{\bibfnamefont {I.~T.}\ \bibnamefont {Lim}},\ and\ \bibinfo {author}
		{\bibfnamefont {I.~H.}\ \bibnamefont {Kim}},\ }\bibfield  {title} {\bibinfo
		{title} {Modular commutators in conformal field theory},\ }\href
	{https://doi.org/10.1103/PhysRevLett.129.260402} {\bibfield  {journal}
		{\bibinfo  {journal} {Phys. Rev. Lett.}\ }\textbf {\bibinfo {volume} {129}},\
		\bibinfo {pages} {260402} (\bibinfo {year} {2022})}\BibitemShut {NoStop}%
	\bibitem [{\citenamefont {Fan}(2022)}]{fan2022entanglement}%
	\BibitemOpen
	\bibfield  {author} {\bibinfo {author} {\bibfnamefont {R.}~\bibnamefont
			{Fan}},\ }\bibfield  {title} {\bibinfo {title} {From entanglement generated
			dynamics to the gravitational anomaly and chiral central charge},\ }\href
	{https://doi.org/10.1103/PhysRevLett.129.260403} {\bibfield  {journal}
		{\bibinfo  {journal} {Phys. Rev. Lett.}\ }\textbf {\bibinfo {volume} {129}},\
		\bibinfo {pages} {260403} (\bibinfo {year} {2022})}\BibitemShut {NoStop}%
	\bibitem [{\citenamefont {Zou}\ \emph {et~al.}(2023)\citenamefont {Zou},
		\citenamefont {Sang},\ and\ \citenamefont {Hsieh}}]{zou2023channeling}%
	\BibitemOpen
	\bibfield  {author} {\bibinfo {author} {\bibfnamefont {Y.}~\bibnamefont
			{Zou}}, \bibinfo {author} {\bibfnamefont {S.}~\bibnamefont {Sang}},\ and\
		\bibinfo {author} {\bibfnamefont {T.~H.}\ \bibnamefont {Hsieh}},\ }\bibfield
	{title} {\bibinfo {title} {Channeling quantum criticality},\ }\href
	{https://doi.org/10.1103/PhysRevLett.130.250403} {\bibfield  {journal}
		{\bibinfo  {journal} {Phys. Rev. Lett.}\ }\textbf {\bibinfo {volume} {130}},\
		\bibinfo {pages} {250403} (\bibinfo {year} {2023})}\BibitemShut {NoStop}%
	\bibitem [{\citenamefont {Xue}\ \emph {et~al.}(2024)\citenamefont {Xue},
		\citenamefont {Lee},\ and\ \citenamefont {Bao}}]{xue2024tensor}%
	\BibitemOpen
	\bibfield  {author} {\bibinfo {author} {\bibfnamefont {H.}~\bibnamefont
			{Xue}}, \bibinfo {author} {\bibfnamefont {J.~Y.}\ \bibnamefont {Lee}},\ and\
		\bibinfo {author} {\bibfnamefont {Y.}~\bibnamefont {Bao}},\ }\bibfield
	{title} {\bibinfo {title} {Tensor network formulation of symmetry protected
			topological phases in mixed states},\ }\href@noop {} {\bibfield  {journal}
		{\bibinfo  {journal} {arXiv preprint arXiv:2403.17069}\ } (\bibinfo {year}
		{2024})}\BibitemShut {NoStop}%
	\bibitem [{\citenamefont {Ma}\ and\ \citenamefont
		{Turzillo}(2024)}]{ma2024symmetry}%
	\BibitemOpen
	\bibfield  {author} {\bibinfo {author} {\bibfnamefont {R.}~\bibnamefont
			{Ma}}\ and\ \bibinfo {author} {\bibfnamefont {A.}~\bibnamefont {Turzillo}},\
	}\bibfield  {title} {\bibinfo {title} {Symmetry protected topological phases
			of mixed states in the doubled space},\ }\href@noop {} {\bibfield  {journal}
		{\bibinfo  {journal} {arXiv preprint arXiv:2403.13280}\ } (\bibinfo {year}
		{2024})}\BibitemShut {NoStop}%
	\bibitem [{\citenamefont {Guo}\ \emph {et~al.}(2024)\citenamefont {Guo},
		\citenamefont {Zhang}, \citenamefont {Yang},\ and\ \citenamefont
		{Bi}}]{guo2024locally}%
	\BibitemOpen
	\bibfield  {author} {\bibinfo {author} {\bibfnamefont {Y.}~\bibnamefont
			{Guo}}, \bibinfo {author} {\bibfnamefont {J.-H.}\ \bibnamefont {Zhang}},
		\bibinfo {author} {\bibfnamefont {S.}~\bibnamefont {Yang}},\ and\ \bibinfo
		{author} {\bibfnamefont {Z.}~\bibnamefont {Bi}},\ }\bibfield  {title}
	{\bibinfo {title} {Locally purified density operators for symmetry-protected
			topological phases in mixed states},\ }\href@noop {} {\bibfield  {journal}
		{\bibinfo  {journal} {arXiv preprint arXiv:2403.16978}\ } (\bibinfo {year}
		{2024})}\BibitemShut {NoStop}%
	\bibitem [{\citenamefont {Li}\ and\ \citenamefont
		{Mong}(2024)}]{li2024replica}%
	\BibitemOpen
	\bibfield  {author} {\bibinfo {author} {\bibfnamefont {Z.}~\bibnamefont
			{Li}}\ and\ \bibinfo {author} {\bibfnamefont {R.~S.}\ \bibnamefont {Mong}},\
	}\bibfield  {title} {\bibinfo {title} {Replica topological order in quantum
			mixed states and quantum error correction},\ }\href@noop {} {\bibfield
		{journal} {\bibinfo  {journal} {arXiv preprint arXiv:2402.09516}\ } (\bibinfo
		{year} {2024})}\BibitemShut {NoStop}%
	\bibitem [{Note2()}]{Note2}%
	\BibitemOpen
	\bibinfo {note} {We note that the C-J isomorphism discussed here is a bit
		different from the original C-J isomorphism between channels $\protect
		\mathcal {E}[\cdot ]$ and operators $\protect \mathcal {E}\otimes I[|\Phi
		\rangle \langle \Phi |]$ introduced in Ref.\protect \citep
		{jamiolkowski1972linear, choi1975completely}, and is along the lines of
		super-operator formalism in Refs.\cite {schmutz1978real,prosen2008third}. We
		use C-J isomorphism as a mnemonic to transform bra(ket) to ket(bra) spaces
		using maximally-entangled states. See App.\ref {sec:CJ_fermions} for more
		discussion.}\BibitemShut {Stop}%
	\bibitem [{\citenamefont {Ashida}\ \emph {et~al.}(2023)\citenamefont {Ashida},
		\citenamefont {Furukawa},\ and\ \citenamefont {Oshikawa}}]{ashida2023system}%
	\BibitemOpen
	\bibfield  {author} {\bibinfo {author} {\bibfnamefont {Y.}~\bibnamefont
			{Ashida}}, \bibinfo {author} {\bibfnamefont {S.}~\bibnamefont {Furukawa}},\
		and\ \bibinfo {author} {\bibfnamefont {M.}~\bibnamefont {Oshikawa}},\
	}\bibfield  {title} {\bibinfo {title} {System-environment entanglement phase
			transitions},\ }\href {https://arxiv.org/abs/2311.16343} {\bibfield
		{journal} {\bibinfo  {journal} {arXiv preprint arXiv:2311.16343}\ } (\bibinfo
		{year} {2023})}\BibitemShut {NoStop}%
	\bibitem [{\citenamefont {Peschel}\ and\ \citenamefont
		{Chung}(2011)}]{peschel2011relation}%
	\BibitemOpen
	\bibfield  {author} {\bibinfo {author} {\bibfnamefont {I.}~\bibnamefont
			{Peschel}}\ and\ \bibinfo {author} {\bibfnamefont {M.-C.}\ \bibnamefont
			{Chung}},\ }\bibfield  {title} {\bibinfo {title} {On the relation between
			entanglement and subsystem hamiltonians},\ }\href
	{http://dx.doi.org/10.1209/0295-5075/96/50006} {\bibfield  {journal}
		{\bibinfo  {journal} {Europhysics Letters}\ }\textbf {\bibinfo {volume}
			{96}},\ \bibinfo {pages} {50006} (\bibinfo {year} {2011})}\BibitemShut
	{NoStop}%
	\bibitem [{\citenamefont {Qi}\ \emph {et~al.}(2012)\citenamefont {Qi},
		\citenamefont {Katsura},\ and\ \citenamefont {Ludwig}}]{qi2012general}%
	\BibitemOpen
	\bibfield  {author} {\bibinfo {author} {\bibfnamefont {X.-L.}\ \bibnamefont
			{Qi}}, \bibinfo {author} {\bibfnamefont {H.}~\bibnamefont {Katsura}},\ and\
		\bibinfo {author} {\bibfnamefont {A.~W.~W.}\ \bibnamefont {Ludwig}},\
	}\bibfield  {title} {\bibinfo {title} {General relationship between the
			entanglement spectrum and the edge state spectrum of topological quantum
			states},\ }\href {https://doi.org/10.1103/PhysRevLett.108.196402} {\bibfield
		{journal} {\bibinfo  {journal} {Phys. Rev. Lett.}\ }\textbf {\bibinfo
			{volume} {108}},\ \bibinfo {pages} {196402} (\bibinfo {year}
		{2012})}\BibitemShut {NoStop}%
	\bibitem [{\citenamefont {Raussendorf}\ \emph {et~al.}(2005)\citenamefont
		{Raussendorf}, \citenamefont {Bravyi},\ and\ \citenamefont
		{Harrington}}]{raussendorf2005long}%
	\BibitemOpen
	\bibfield  {author} {\bibinfo {author} {\bibfnamefont {R.}~\bibnamefont
			{Raussendorf}}, \bibinfo {author} {\bibfnamefont {S.}~\bibnamefont
			{Bravyi}},\ and\ \bibinfo {author} {\bibfnamefont {J.}~\bibnamefont
			{Harrington}},\ }\bibfield  {title} {\bibinfo {title} {Long-range quantum
			entanglement in noisy cluster states},\ }\href
	{https://doi.org/10.1103/PhysRevA.71.062313} {\bibfield  {journal} {\bibinfo
			{journal} {Phys. Rev. A}\ }\textbf {\bibinfo {volume} {71}},\ \bibinfo
		{pages} {062313} (\bibinfo {year} {2005})}\BibitemShut {NoStop}%
	\bibitem [{\citenamefont {Aguado}\ \emph {et~al.}(2008)\citenamefont {Aguado},
		\citenamefont {Brennen}, \citenamefont {Verstraete},\ and\ \citenamefont
		{Cirac}}]{aguado2008creation}%
	\BibitemOpen
	\bibfield  {author} {\bibinfo {author} {\bibfnamefont {M.}~\bibnamefont
			{Aguado}}, \bibinfo {author} {\bibfnamefont {G.~K.}\ \bibnamefont {Brennen}},
		\bibinfo {author} {\bibfnamefont {F.}~\bibnamefont {Verstraete}},\ and\
		\bibinfo {author} {\bibfnamefont {J.~I.}\ \bibnamefont {Cirac}},\ }\bibfield
	{title} {\bibinfo {title} {Creation, manipulation, and detection of abelian
			and non-abelian anyons in optical lattices},\ }\href
	{https://doi.org/10.1103/PhysRevLett.101.260501} {\bibfield  {journal}
		{\bibinfo  {journal} {Phys. Rev. Lett.}\ }\textbf {\bibinfo {volume} {101}},\
		\bibinfo {pages} {260501} (\bibinfo {year} {2008})}\BibitemShut {NoStop}%
	\bibitem [{\citenamefont {Verresen}\ \emph {et~al.}(2021)\citenamefont
		{Verresen}, \citenamefont {Tantivasadakarn},\ and\ \citenamefont
		{Vishwanath}}]{verresen2021efficiently}%
	\BibitemOpen
	\bibfield  {author} {\bibinfo {author} {\bibfnamefont {R.}~\bibnamefont
			{Verresen}}, \bibinfo {author} {\bibfnamefont {N.}~\bibnamefont
			{Tantivasadakarn}},\ and\ \bibinfo {author} {\bibfnamefont {A.}~\bibnamefont
			{Vishwanath}},\ }\bibfield  {title} {\bibinfo {title} {Efficiently preparing
			schr$\backslash$" odinger's cat, fractons and non-abelian topological order
			in quantum devices},\ }\href {https://arxiv.org/abs/2112.03061} {\bibfield
		{journal} {\bibinfo  {journal} {arXiv preprint arXiv:2112.03061}\ } (\bibinfo
		{year} {2021})}\BibitemShut {NoStop}%
	\bibitem [{\citenamefont {Tantivasadakarn}\ \emph {et~al.}(2021)\citenamefont
		{Tantivasadakarn}, \citenamefont {Thorngren}, \citenamefont {Vishwanath},\
		and\ \citenamefont {Verresen}}]{tantivasadakarn2021long}%
	\BibitemOpen
	\bibfield  {author} {\bibinfo {author} {\bibfnamefont {N.}~\bibnamefont
			{Tantivasadakarn}}, \bibinfo {author} {\bibfnamefont {R.}~\bibnamefont
			{Thorngren}}, \bibinfo {author} {\bibfnamefont {A.}~\bibnamefont
			{Vishwanath}},\ and\ \bibinfo {author} {\bibfnamefont {R.}~\bibnamefont
			{Verresen}},\ }\bibfield  {title} {\bibinfo {title} {Long-range entanglement
			from measuring symmetry-protected topological phases},\ }\href
	{https://arxiv.org/abs/2112.01519} {\bibfield  {journal} {\bibinfo  {journal}
			{arXiv preprint arXiv:2112.01519}\ } (\bibinfo {year} {2021})}\BibitemShut
	{NoStop}%
	\bibitem [{\citenamefont {Vijay}\ \emph {et~al.}(2016)\citenamefont {Vijay},
		\citenamefont {Haah},\ and\ \citenamefont {Fu}}]{vijay2016fracton}%
	\BibitemOpen
	\bibfield  {author} {\bibinfo {author} {\bibfnamefont {S.}~\bibnamefont
			{Vijay}}, \bibinfo {author} {\bibfnamefont {J.}~\bibnamefont {Haah}},\ and\
		\bibinfo {author} {\bibfnamefont {L.}~\bibnamefont {Fu}},\ }\bibfield
	{title} {\bibinfo {title} {Fracton topological order, generalized lattice
			gauge theory, and duality},\ }\href
	{https://doi.org/10.1103/PhysRevB.94.235157} {\bibfield  {journal} {\bibinfo
			{journal} {Phys. Rev. B}\ }\textbf {\bibinfo {volume} {94}},\ \bibinfo
		{pages} {235157} (\bibinfo {year} {2016})}\BibitemShut {NoStop}%
	\bibitem [{\citenamefont {Cornfeld}\ \emph {et~al.}(2018)\citenamefont
		{Cornfeld}, \citenamefont {Goldstein},\ and\ \citenamefont
		{Sela}}]{cornfeld2018imbalance}%
	\BibitemOpen
	\bibfield  {author} {\bibinfo {author} {\bibfnamefont {E.}~\bibnamefont
			{Cornfeld}}, \bibinfo {author} {\bibfnamefont {M.}~\bibnamefont
			{Goldstein}},\ and\ \bibinfo {author} {\bibfnamefont {E.}~\bibnamefont
			{Sela}},\ }\bibfield  {title} {\bibinfo {title} {Imbalance entanglement:
			Symmetry decomposition of negativity},\ }\href
	{https://doi.org/10.1103/PhysRevA.98.032302} {\bibfield  {journal} {\bibinfo
			{journal} {Phys. Rev. A}\ }\textbf {\bibinfo {volume} {98}},\ \bibinfo
		{pages} {032302} (\bibinfo {year} {2018})}\BibitemShut {NoStop}%
	\bibitem [{\citenamefont {Elben}\ \emph {et~al.}(2020)\citenamefont {Elben},
		\citenamefont {Kueng}, \citenamefont {Huang}, \citenamefont {van Bijnen},
		\citenamefont {Kokail}, \citenamefont {Dalmonte}, \citenamefont {Calabrese},
		\citenamefont {Kraus}, \citenamefont {Preskill}, \citenamefont {Zoller},\
		and\ \citenamefont {Vermersch}}]{elben2020mixed}%
	\BibitemOpen
	\bibfield  {author} {\bibinfo {author} {\bibfnamefont {A.}~\bibnamefont
			{Elben}}, \bibinfo {author} {\bibfnamefont {R.}~\bibnamefont {Kueng}},
		\bibinfo {author} {\bibfnamefont {H.-Y.~R.}\ \bibnamefont {Huang}}, \bibinfo
		{author} {\bibfnamefont {R.}~\bibnamefont {van Bijnen}}, \bibinfo {author}
		{\bibfnamefont {C.}~\bibnamefont {Kokail}}, \bibinfo {author} {\bibfnamefont
			{M.}~\bibnamefont {Dalmonte}}, \bibinfo {author} {\bibfnamefont
			{P.}~\bibnamefont {Calabrese}}, \bibinfo {author} {\bibfnamefont
			{B.}~\bibnamefont {Kraus}}, \bibinfo {author} {\bibfnamefont
			{J.}~\bibnamefont {Preskill}}, \bibinfo {author} {\bibfnamefont
			{P.}~\bibnamefont {Zoller}},\ and\ \bibinfo {author} {\bibfnamefont
			{B.}~\bibnamefont {Vermersch}},\ }\bibfield  {title} {\bibinfo {title}
		{Mixed-state entanglement from local randomized measurements},\ }\href
	{https://doi.org/10.1103/PhysRevLett.125.200501} {\bibfield  {journal}
		{\bibinfo  {journal} {Phys. Rev. Lett.}\ }\textbf {\bibinfo {volume} {125}},\
		\bibinfo {pages} {200501} (\bibinfo {year} {2020})}\BibitemShut {NoStop}%
	\bibitem [{\citenamefont {Brand{\~a}o}\ and\ \citenamefont
		{Kastoryano}(2019)}]{brandao2019finite}%
	\BibitemOpen
	\bibfield  {author} {\bibinfo {author} {\bibfnamefont {F.~G.}\ \bibnamefont
			{Brand{\~a}o}}\ and\ \bibinfo {author} {\bibfnamefont {M.~J.}\ \bibnamefont
			{Kastoryano}},\ }\bibfield  {title} {\bibinfo {title} {Finite correlation
			length implies efficient preparation of quantum thermal states},\ }\href
	{https://link.springer.com/article/10.1007/s00220-018-3150-8} {\bibfield
		{journal} {\bibinfo  {journal} {Communications in Mathematical Physics}\
		}\textbf {\bibinfo {volume} {365}},\ \bibinfo {pages} {1} (\bibinfo {year}
		{2019})}\BibitemShut {NoStop}%
	\bibitem [{\citenamefont {Barahona}(1982)}]{barahona1982on}%
	\BibitemOpen
	\bibfield  {author} {\bibinfo {author} {\bibfnamefont {F.}~\bibnamefont
			{Barahona}},\ }\bibfield  {title} {\bibinfo {title} {On the computational
			complexity of ising spin glass models},\ }\href
	{https://doi.org/10.1088/0305-4470/15/10/028} {\bibfield  {journal} {\bibinfo
			{journal} {Journal of Physics A: Mathematical and General}\ }\textbf
		{\bibinfo {volume} {15}},\ \bibinfo {pages} {3241} (\bibinfo {year}
		{1982})}\BibitemShut {NoStop}%
	\bibitem [{\citenamefont {Lucas}(2014)}]{lucas2014ising}%
	\BibitemOpen
	\bibfield  {author} {\bibinfo {author} {\bibfnamefont {A.}~\bibnamefont
			{Lucas}},\ }\bibfield  {title} {\bibinfo {title} {Ising formulations of many
			np problems},\ }\href {https://doi.org/10.3389/fphy.2014.00005} {\bibfield
		{journal} {\bibinfo  {journal} {Frontiers in physics}\ }\textbf {\bibinfo
			{volume} {2}},\ \bibinfo {pages} {74887} (\bibinfo {year}
		{2014})}\BibitemShut {NoStop}%
	\bibitem [{\citenamefont {Piroli}\ \emph {et~al.}(2021)\citenamefont {Piroli},
		\citenamefont {Styliaris},\ and\ \citenamefont {Cirac}}]{piroli2021quantum}%
	\BibitemOpen
	\bibfield  {author} {\bibinfo {author} {\bibfnamefont {L.}~\bibnamefont
			{Piroli}}, \bibinfo {author} {\bibfnamefont {G.}~\bibnamefont {Styliaris}},\
		and\ \bibinfo {author} {\bibfnamefont {J.~I.}\ \bibnamefont {Cirac}},\
	}\bibfield  {title} {\bibinfo {title} {Quantum circuits assisted by local
			operations and classical communication: Transformations and phases of
			matter},\ }\href {https://doi.org/10.1103/PhysRevLett.127.220503} {\bibfield
		{journal} {\bibinfo  {journal} {Phys. Rev. Lett.}\ }\textbf {\bibinfo
			{volume} {127}},\ \bibinfo {pages} {220503} (\bibinfo {year}
		{2021})}\BibitemShut {NoStop}%
	\bibitem [{\citenamefont {Chen}\ \emph {et~al.}(2023)\citenamefont {Chen},
		\citenamefont {Zhu}, \citenamefont {Verresen}, \citenamefont {Seif},
		\citenamefont {Ba{\"u}mer}, \citenamefont {Layden}, \citenamefont
		{Tantivasadakarn}, \citenamefont {Zhu}, \citenamefont {Sheldon},
		\citenamefont {Vishwanath} \emph {et~al.}}]{chen2023realizing}%
	\BibitemOpen
	\bibfield  {author} {\bibinfo {author} {\bibfnamefont {E.~H.}\ \bibnamefont
			{Chen}}, \bibinfo {author} {\bibfnamefont {G.-Y.}\ \bibnamefont {Zhu}},
		\bibinfo {author} {\bibfnamefont {R.}~\bibnamefont {Verresen}}, \bibinfo
		{author} {\bibfnamefont {A.}~\bibnamefont {Seif}}, \bibinfo {author}
		{\bibfnamefont {E.}~\bibnamefont {Ba{\"u}mer}}, \bibinfo {author}
		{\bibfnamefont {D.}~\bibnamefont {Layden}}, \bibinfo {author} {\bibfnamefont
			{N.}~\bibnamefont {Tantivasadakarn}}, \bibinfo {author} {\bibfnamefont
			{G.}~\bibnamefont {Zhu}}, \bibinfo {author} {\bibfnamefont {S.}~\bibnamefont
			{Sheldon}}, \bibinfo {author} {\bibfnamefont {A.}~\bibnamefont {Vishwanath}},
		\emph {et~al.},\ }\bibfield  {title} {\bibinfo {title} {Realizing the
			nishimori transition across the error threshold for constant-depth quantum
			circuits},\ }\href {https://arxiv.org/abs/2309.02863} {\bibfield  {journal}
		{\bibinfo  {journal} {arXiv preprint arXiv:2309.02863}\ } (\bibinfo {year}
		{2023})}\BibitemShut {NoStop}%
	\bibitem [{\citenamefont {Schmutz}(1978)}]{schmutz1978real}%
	\BibitemOpen
	\bibfield  {author} {\bibinfo {author} {\bibfnamefont {M.}~\bibnamefont
			{Schmutz}},\ }\bibfield  {title} {\bibinfo {title} {Real-time green's
			functions in many body problems},\ }\href
	{https://link.springer.com/article/10.1007/BF01323673} {\bibfield  {journal}
		{\bibinfo  {journal} {Zeitschrift f{\"u}r Physik B Condensed Matter}\
		}\textbf {\bibinfo {volume} {30}},\ \bibinfo {pages} {97} (\bibinfo {year}
		{1978})}\BibitemShut {NoStop}%
	\bibitem [{\citenamefont {Prosen}(2008)}]{prosen2008third}%
	\BibitemOpen
	\bibfield  {author} {\bibinfo {author} {\bibfnamefont {T.}~\bibnamefont
			{Prosen}},\ }\bibfield  {title} {\bibinfo {title} {Third quantization: a
			general method to solve master equations for quadratic open fermi systems},\
	}\href {https://doi.org/10.1088/1367-2630/10/4/043026} {\bibfield  {journal}
		{\bibinfo  {journal} {New Journal of Physics}\ }\textbf {\bibinfo {volume}
			{10}},\ \bibinfo {pages} {043026} (\bibinfo {year} {2008})}\BibitemShut
	{NoStop}%
	\bibitem [{\citenamefont {Shapourian}\ \emph {et~al.}(2017)\citenamefont
		{Shapourian}, \citenamefont {Shiozaki},\ and\ \citenamefont
		{Ryu}}]{shapourian2017partial}%
	\BibitemOpen
	\bibfield  {author} {\bibinfo {author} {\bibfnamefont {H.}~\bibnamefont
			{Shapourian}}, \bibinfo {author} {\bibfnamefont {K.}~\bibnamefont
			{Shiozaki}},\ and\ \bibinfo {author} {\bibfnamefont {S.}~\bibnamefont
			{Ryu}},\ }\bibfield  {title} {\bibinfo {title} {Partial time-reversal
			transformation and entanglement negativity in fermionic systems},\ }\href
	{https://doi.org/10.1103/PhysRevB.95.165101} {\bibfield  {journal} {\bibinfo
			{journal} {Phys. Rev. B}\ }\textbf {\bibinfo {volume} {95}},\ \bibinfo
		{pages} {165101} (\bibinfo {year} {2017})}\BibitemShut {NoStop}%
\end{thebibliography}
%


\appendix

\section{Details of string order parameter for 1d cluster state}

\label{sec:appendix_1dcluster}
This section provides details of evaluating the string order parameter for 1d cluster state with respect to each $\rho_{Q_a, Q_b}$, i.e., $\tr(\rho_{Q_a, Q_b}S_{a/b})/\tr(\rho_{Q_a, Q_b})$.
We first compute the denominator in this expression, namely, the trace of $\rho_{Q_a, Q_b}$ (= probability of sector with charge $(Q_a, Q_b)$) by inserting the complete basis $\{ |x_\mathbf{a,b}\rangle \}$ and $\{ |z_\mathbf{a,b}\rangle \}$: 
\begin{equation}
	\label{Eq:tr_rho_ab}
	\begin{aligned}
		\tr(\rho_{Q_a, Q_b} ) 
		& \propto \sum_{x_\mathbf{a,b}, z_\mathbf{a,b}} \langle x_\mathbf{a,b}|\rho_{Q_a}|z_\mathbf{a,b}\rangle \langle z_\mathbf{a,b} |\rho_{Q_b} |x_\mathbf{a,b} \rangle \\
		& \propto \sum_{x_\mathbf{a,b} \in Q_{a,b}, z_\mathbf{a,b}} \langle x_\mathbf{a,b}|\rho_{a}|z_\mathbf{a,b}\rangle \langle z_\mathbf{a,b} |\rho_{b} |x_\mathbf{a,b} \rangle,
	\end{aligned}
\end{equation}
where $\sum_{x_\mathbf{a,b} \in Q_{a,b}}$ denotes summation over all  possible $x_\mathbf{a,b}$ in the  $Q_{a}$ and $Q_{b}$ sectors, i.e., $\prod_j (x_{a,j}) = (-1)^{Q_a}$ and $\prod_j (x_{b,j}) = (-1)^{Q_b}$.
Now, 
\begin{equation}
	\begin{aligned}
		\langle x_{\mathbf{a,b}}|\rho_{a}|z_\mathbf{a,b}\rangle 
		& \propto e^{-\beta_{a} \sum_j z_{b,j-1}x_{a,j}z_{b,j+1}} \langle x_{\mathbf{a,b}}|z_{\mathbf{a,b}} \rangle \\
		& \propto  e^{-\beta_{a} \sum_j z_{b,j-1}x_{a,j}z_{b,j+1}}.
	\end{aligned}
\end{equation}
Similarly, $ \langle z_\mathbf{a,b} |\rho_{b} |x_\mathbf{a,b} \rangle \propto e^{-\beta_b \sum_{j} {z_{a,j} x_{b,j} z_{a,j+1}}}$.
It follows that $\tr(\rho_{Q_a, Q_b}) \propto  \sum_{x_{\mathbf{a}} \in Q_a} \mathcal{Z}_{\text{1D Ising}, x_\mathbf{a}}
\sum_{x_{\mathbf{b}} \in Q_b}  \mathcal{Z}_{\text{1D Ising}, x_\mathbf{b}}$, where $\mathcal{Z}_{\text{1D Ising}, x_\mathbf{a}} = \sum_{z_{\mathbf{b}}}e^{\beta_{a} \sum_j x_{a,j} z_{b,j-1} z_{b,j}} $ is the partition function of 1D Ising model with the Ising interaction determined by $x_\mathbf{a}$ (the expression for $\mathcal{Z}_{\text{1D Ising}, x_\mathbf{b}}$ is analogously obtained by interchanging $a$ and $b$).
For a system with periodic boundary condition, one can parameterize all $x_\mathbf{a} \in Q_{a}$ by performing the transformation $x_{a, j} \rightarrow x_{a,j} s_{b,j-1} s_{b,j},\ s_{b,j} = \pm 1$ from any $x_{\mathbf{a}}$ that belongs to $Q_a$.
Besides, one can easily verify $\mathcal{Z}_{\text{1D Ising}, x_\mathbf{a}}$ is invariant under the transformation by changing the dummy variables $z_{b,j-1} \rightarrow s_{b,j-1}z_{b,j-1}$. 
In other words, $\mathcal{Z}_{\text{1D Ising}, x_\mathbf{a}} = \mathcal{Z}_{\text{1D Ising}, Q_a}$ only depends on the charge $Q_a$.
Therefore, 
\begin{equation}
	\tr(\rho_{Q_a,Q_b}) \propto \mathcal{Z}_{\text{1D Ising}, Q_{a}} \mathcal{Z}_{\text{1D Ising}, Q_{b}},
\end{equation}
which is the product of the partition functions of the 1D Ising model in the $Q_a$ and $Q_b$ sectors.
The evaluation of the numerator of the string order can be done in a similar way.
The only term that doesn't cancel out with the the denominator is associated with $\langle x_{\mathbf{a,b}}|\rho S_a(j,k)|z_\mathbf{a,b}\rangle $, which can be evaluated as:
\begin{equation}
	\begin{aligned}
		& \langle x_{\mathbf{a,b}}|\rho S_a(j,k)|z_\mathbf{a,b}\rangle \\
		& \propto e^{-\beta_{a} \sum_j z_{b,j-1}x_{a,j}z_{b,j+1}} z_{b,j-1} (\prod_{l = j}^k x_{a,l}) z_{b,k} \langle x_{\mathbf{a,b}}|z_{\mathbf{a,b}} \rangle \\
		& \propto (\prod_{l = j}^k x_{a,l})  e^{-\beta_{a} \sum_j z_{b,j-1}x_{a,j}z_{b,j+1}} z_{b,j-1} z_{b,k}.
	\end{aligned}
\end{equation}
It follows that 
\begin{equation}
	\label{Eq:string_order_rhoab}
	\frac{ \tr(\rho_{Q_a, Q_b} S_a(j,k))}{ \tr( \rho_{Q_a, Q_b})} =  (\prod_{l = j}^k x_{a,l})  \langle z_{j-1} z_k \rangle_{\text{1D Ising}, x_\mathbf{a}}\Big|_{x_\mathbf{a} \in Q_a},
\end{equation}
where $ \langle z_{j-1} z_k \rangle_{\text{1D Ising}, x_{\mathbf{a}} \in Q_{a}}$ is the spin-spin correlation function of the 1D Ising model with the Ising interaction determined by any $x_\mathbf{a}$ belonging to the $Q_a$ sector.
Note that $(\prod_{l = j}^k x_{a,l})  \langle z_{j-1} z_k \rangle_{\text{1D Ising}, x_\mathbf{a}\in Q_a}$ is invariant under the transformation $x_{a, j} \rightarrow x_{a,j} s_{b,j-1} s_{b,j},\ s_{b,j} = \pm 1$, and thus Eq.\eqref{Eq:string_order_rhoab} is independent of the choice for any $x_{\mathbf{a}} \in Q_a$. 
For example, one can choose $x_{a, j} = 1,\ \forall j$ if $Q_a = 0$. On the other hand, if $Q_a = 1$, one can choose $x_{a, j} = 1,\ \forall j\neq N$ and $x_{a, N} = -1$.
It follows that the string order of $\rho_{Q_a,Q_b}$ for both $Q_a = 0,1$ can be mapped to $\langle z_{j-1} z_k \rangle_{\text{1D Ising}}$, the spin-spin correlation function of the 1D ferromagnetic Ising model.
The results for $S_b$ is similar.
Since $\langle z_{j-1} z_k \rangle_{\text{1D Ising}}$ decays exponentially with $|j-k|$ for any $\beta < \infty$, we conclude that $\rho_{Q_a,Q_b}$ has no string order as long as $p_a, p_b > 0$.

\section{Details of calculations for chiral fermions subjected to decoherence}

\subsection{Covariance matrix under channel linear in fermion operators}
 \label{sec:covariance}
We are interested in subjecting the ground state $\rho_0 = |\psi_0\rangle \langle \psi_0|$ of a Gaussian fermionic Hamiltonian $H$ to the composition of the following single-majorana channel on all sites:
\begin{equation}
	\label{Eq:app_single_maj_channel}
	\mathcal{E}_j[\rho] = (1-p)\rho+p\gamma_j \rho \gamma_j.
\end{equation}
The goal of this section is to show that the resulting covariance matrix is $\mathcal{E}(M) = (1-2p)^2 M$ and the resulting density matrix $\rho \propto e^{- i \beta \sum_j \xi_{2j-1} \xi_{2j}} $, where $\tanh \beta = (1-2p)^2$ and $\xi_j = (O^T)_{jk} \gamma_k$ are the majorana operators that block diagonalize the original Hamiltonian $H$.
We note that one can also pair up two majorana fermions to get regular fermions through $\alpha_j = (\xi_{2j-1} + i \xi_{2j})/2$, and the density matrix take the form $\rho \propto e^{-2\beta \sum_j \alpha^\dagger_j \alpha_j}$ mentioned in Sec.\ref{sec:chiral_free}.

To proceed, we note that Eq.\eqref{Eq:app_single_maj_channel} will map a Gaussion state to a Gaussian state.
A Gaussian fermionic state $\rho$, whether pure or mixed, is fully specified by the covariance matrix $M$, defined as
\begin{equation}
	\label{Eq:covariance_matrix_def}
	M_{j k} = -i \tr(\rho (\gamma_j \gamma_k - \delta_{jk})).
\end{equation}
Therefore, to determine the evolution of the density matrix under the channel, it suffices to determine how the covariance matrix evolves, which we denote as $\mathcal{E}(M)$.
Using $\gamma_l (\gamma_j \gamma_k) \gamma_l = (-1)^{\delta_{j l} + \delta_{k l}} \gamma_j \gamma_k$, the element of the resulting covariance matrix $[\mathcal{E}_l(M)]_{jk}$ can be easily computed:
\begin{equation}
	\begin{aligned}
		[\mathcal{E}_l(M)]_{jk} & = -i \tr(\mathcal{E}_l[\rho] (\gamma_j \gamma_k - \delta_{jk})) \\
		& = (1-p) M_{jk} + (-1)^{\delta_{j l} + \delta_{k l}} p M_{jk} \\
		& = [(1-p)+(-1)^{\delta_{j l} + \delta_{k l}} p] M_{jk} \\
		& = \begin{cases}
			M_{jk} & \text{for}\ j \neq l\  \text{and}\ k \neq l, \\
			(1-2p)M_{jk}  & \text{for}\ j = l\ \text{or}\  k = l.\\
		\end{cases}
	\end{aligned}
\end{equation}
It follows that the composition of the channel $\mathcal{E}_l$ on all sites gives
\begin{equation}
	\label{Eq:all_maj_channel}
	\mathcal{E}(M) = (1-2p)^2 M.
\end{equation}

To see how one can use Eq.\eqref{Eq:all_maj_channel} to deduce the resulting decohered mixed state, let us first explicitly write down the relation between a general density matrix $\rho$ and its covariance matrix $M$. Let us write the density matrix as $\rho \propto e^{-H_\rho}$. Since $\rho$ is Gaussian, $H_{\rho}$ can be written as a sum of Majorana bilinears:
\begin{equation}
	H_\rho = \frac{i}{2} \sum_{j, k =1 }^{2N} \gamma_j K_{jk} \gamma_k,
\end{equation}
where $K$ is a $2N \times 2N$ antisymmetric matrix and we denote the number of Majorana modes as $2N$. To see how the matrix $M$ is related to the matrix $K$, we begin by block diagonalizing $K$:
\begin{equation}
	K = O (K_d \otimes (iY)) O^T,
\end{equation}
where $K_d$ is $N \times N$ diagonal matrix and $Y$ is the Pauli-Y matrix $(= [0 \,\,-i;\, i\,\, 0])$.
Denoting ${\xi}_j = (O^T)_{jk} \gamma_k$, the density matrix then takes the following form:
\begin{equation}
	\label{Eq:gaussian_dm}
	\rho \propto \prod_j [I-\tanh ((K_d)_{j,j}) (i {\xi}_{2j-1} {\xi}_{2j})].
\end{equation}
Using $-i \tr(\rho \xi_{2j-1} \xi_{2j}) = \tanh((K_d)_{j,j})$ and the relation between $\xi_j$ and $\gamma_j$, the covariance matrix can then be obtained as 
\begin{equation}
	\label{Eq:covariance_matrix}
	M = O (\tanh(K_d) \otimes (iY)) O^T.
\end{equation}

Now, let's determine the matrix $O$ and the relation between $\tanh(K_d)$ and $p$ using the property of the initial pure state $\rho_0$ and Eq.\eqref{Eq:all_maj_channel} .
Since $\rho_0 = |\psi_0\rangle \langle \psi_0|$ is the ground state of Hamiltonian $H$, the matrix $O$ at $p = 0$ is precisely the one that diagonalizes $H$. 
Besides, using $\rho_0^2 = \rho_0$, one finds $\tanh(K_d) = I$  when $p = 0$, and thus $M_0 = O(I \otimes (iY))O^T$.
Eq.\eqref{Eq:all_maj_channel} then gives $M(p) = O[(1-2p)^2I  \otimes (iY)] O^T$.
This implies that $O$ remains unchanged and $\tanh (K_d)_{j,j} = (1-2p)^2 $ is independent of $j$. 
Therefore, the entanglement Hamiltonian takes the form
\begin{equation}
	H_\rho(p) = i \beta \sum_ j \xi_{2j-1} \xi_{2j},
\end{equation}
where $\tanh \beta = (1-2p)^2$ and $\xi_j = (O^T)_{jk} \gamma_k$ are the majorana operators that block diagonalize the original Hamiltonian $H$.

\subsection{Double-state formalism for fermions}
 \label{sec:CJ_fermions}
\begin{figure}
	\centering
	\includegraphics[width=\linewidth]{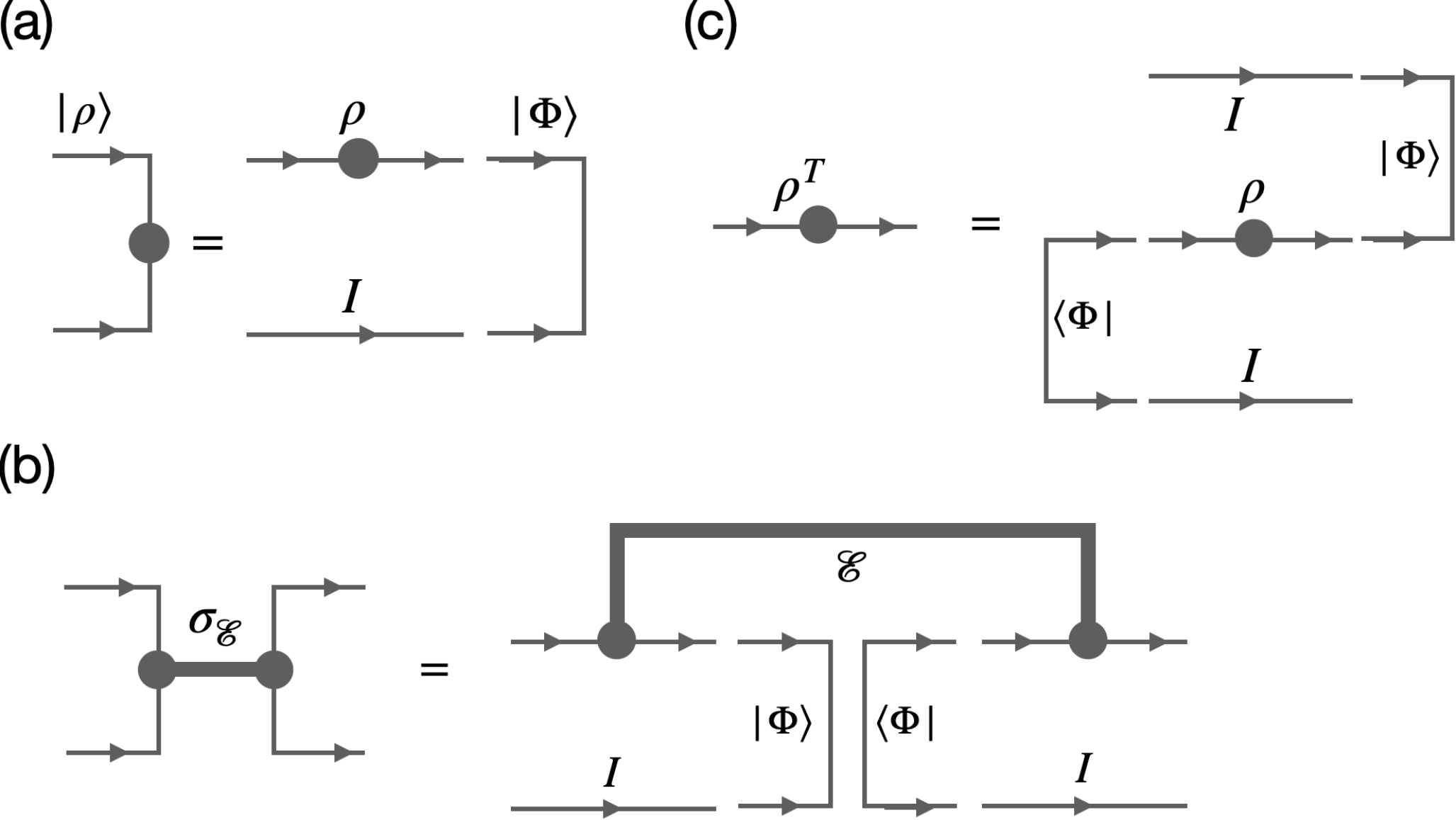}
	\caption{Tensor network representations of C-J isomorphisn for (a) $|\rho\rangle = \rho \otimes I |\Phi\rangle$, (b) $\sigma_\mathcal{E} = \mathcal{E} \otimes I [|\Phi\rangle \langle \Phi|]$, and (c) $ \langle m' |\rho^T|m\rangle = (\langle m'| \otimes \langle \Phi|) I \otimes \rho \otimes I (|\Phi\rangle \otimes |m\rangle)$.} 
	\label{Fig:cjam}
\end{figure}

In this section, we derive the double-state formalism for fermions.
As a warmup, we will first derive how the bosonic density matrices and channels are mapped to pure states and operators, respectively, under 
\begin{equation}
	\label{Eq:app_double_state}
	|\rho\rangle = \rho \otimes I |\Phi\rangle
\end{equation}
We will later derive the similar correspondence for fermions using Grassmann algebra.

For a general bosonic density matrix $\rho = \sum_{p,q} \rho^p_q |p\rangle \langle q|$, one can  compute $|\rho\rangle$ using  Eq.\eqref{Eq:app_double_state}:
\begin{equation}
	\label{Eq:cjam_bosons}
	\begin{aligned}
		|\rho\rangle & = \sum_{p,q}  \rho^p_q |p\rangle \langle q| \big( \sum_r |r\rangle \otimes |r\rangle \big)\\
		& = \sum_{p,q, r}  \rho^p_q \delta^{q}_{r}|p\rangle  \otimes |r\rangle = \sum_{p,q}  \rho^p_q |p, q\rangle,
	\end{aligned}
\end{equation}
which is tantamount to flipping the bra vector of $\rho$ in the original Hilbert space $\mathcal{H}$ to the ket vector in $\bar{\mathcal{H}}$.
This intuition can be visualized by expressing Eq.\eqref{Eq:app_double_state} using tensor network [see Fig.\ref{Fig:cjam}(a)].
In this sense, one can regard the maximally entangled states as a tool to transform the bra(ket) to ket(bra) spaces, and such a trick  is called Choi–Jamio{\l}kowski (C-J) isomorphism.
We note that C-J isomorphism was originally used to map quantum channels $\mathcal{E}$ (superoperators) to quantum states $\sigma$ (density matrices) \cite{jamiolkowski1972linear, choi1975completely}:
\begin{equation}
	\label{Eq:cjam_original}
	\sigma_\mathcal{E} =  \mathcal{E} \otimes I [|\Phi\rangle \langle \Phi|],
\end{equation}
which can be visualized in Fig.\ref{Fig:cjam}(b).
However, while both Eq.\eqref{Eq:cjam_original} and Eq.\eqref{Eq:app_double_state} map quantum channels to operators, the operators they map to are different in general.
Specifically, consider the channel $\mathcal{E}[\cdot] = \sum_{\alpha} K_\alpha (\cdot) K^\dagger_\alpha $ and denote its corresponding operator Eq.\eqref{Eq:app_double_state} maps to as $\mathcal{N}$, i.e., $|\mathcal{E}[\rho] \rangle = \mathcal{N} |\rho\rangle$.
The operator $\mathcal{N}$ can be obtain by the direct evaluation of  $|\mathcal{E}[\rho]\rangle$ for any $\rho$:
\begin{equation}
	\begin{aligned}
		\label{Eq:double_state_kraus}
		|\mathcal{E}[\rho]\rangle  & = \sum_{\alpha} K_\alpha \rho K^\dagger_\alpha \otimes I |\Phi\rangle \\
		& = \sum_{\alpha,} K_\alpha
		\big(\sum_q |q\rangle \langle q| \big) \rho \big(\sum_r |r\rangle \langle r| \big) K^\dagger_\alpha \big( \sum_p |p\rangle \otimes |p\rangle \big) \\
		& = \sum_{\alpha} \sum_{p,q,r} K_\alpha|q\rangle \langle r|K^\dagger_\alpha| p \rangle \langle q|\rho| r\rangle  |p\rangle \\
		& = \sum_{\alpha} \sum_{q,r} K_\alpha|q\rangle \big( \sum_p|p\rangle \langle p| \big) \bar{K}| r \rangle \langle q|\rho| r\rangle   \\
		& = \sum_{\alpha} \sum_{q,r} K_\alpha \otimes \bar{K}_\alpha | q, r \rangle \langle q, r|\rho \rangle   \\
		& = \sum_{\alpha} K_\alpha \otimes \bar{K}_\alpha|\rho\rangle.
	\end{aligned}
\end{equation}
Therefore, one finds
\begin{equation}
	\mathcal{N}_\mathcal{E} = \sum_{\alpha} K_\alpha \otimes \bar{K}_\alpha.
\end{equation}
On the other hand, the corresponding operator $\sigma_\mathcal{E}$ for the channel $\mathcal{E}$ is evaluated using Eq.\eqref{Eq:cjam_original}, and one finds
\begin{equation}
	\sigma_\mathcal{E} = \sum_\alpha |K_\alpha\rangle \langle K_\alpha|,
\end{equation}
where $|K_\alpha \rangle = K_\alpha \otimes I |\Phi\rangle$.
It is then obvious that $\mathcal{N}_\mathcal{E}$ and $\sigma_\mathcal{E}$ are different.
For example, if $\mathcal{E}[\cdot] = K(\cdot)K^\dagger$ only consists of a single Kraus operator $K$, $\sigma_\mathcal{E}$ is necessarily a projector while $N_\mathcal{E}$ is not in general.

Now we turn to the C-J map for fermions, i.e., the analog of Eq.\eqref{Eq:double_state_kraus} for fermions. Note that the derivation of $\mathcal{N}_\mathcal{E}$ in Eq.\eqref{Eq:double_state_kraus} required the insertion of a complete basis.
For fermions, this can be achieved by using Grassmann algebra.
To build intuition, we consider the mixed state density matrix $\rho$ with a single fermionic mode with creation/annihilation operators $\mathbf{c}^{\dagger}/\mathbf{c}$ (which act on the Hilbert space $\mathcal{H}$ in our notation), i.e., $\rho=\rho(\mathbf{c},\mathbf{c}^\dagger)$.
The maximally entangled state in the double Hilbert space for fermions can then be defined as 
\begin{equation}
	|\Phi\rangle \equiv (I + e^{i \theta}\bold{c}^\dagger \bold{d}^\dagger)|00\rangle.
\end{equation}
Here $\mathbf{d}^\dagger$ denotes the fermionic creation operators in the Hilbert space $\mathcal{\overline{H}}$,  $|00\rangle$ is the vacuum defined by $\bold{c} |00 \rangle = \bold{d}|00\rangle =0$, and $\theta \in [0,2\pi)$ is an arbitrary phase that we will set to zero for convenience. To derive a compact form for $|\rho\rangle$, we make use of the coherent state $|c,d\rangle = e^{-c \bold c^\dagger } e^{-d \bold d^\dagger }|00\rangle$ where $c$ and $d$ are Grassmann numbers.
The maximally entangled state in the coherent state basis can be easily computed:
\begin{equation}
	\braket{\bar{c} \bar{d}}{\Phi} = \langle \bar{c} \bar{d}|I + \bold{c}^\dagger \bold{d}^\dagger |00\rangle = (1+\bar{c} \bar{d}) \braket{\bar{c} \bar{d}}{00} = e^{\bar{c} \bar{d}}.
\end{equation}
Similarly, we can compute $|\rho\rangle$ in the coherent state basis:
\begin{equation}
	\label{eqn:rho_choi}
	\begin{aligned}
		\langle \bar{c} \bar{d}|\rho \rangle & = \int \mathcal{D} \bar{\alpha} \mathcal{D} \alpha \langle \bar{c}|\rho|\alpha\rangle e^{-\bar{\alpha} \alpha} \langle \bar{\alpha} \bar{d}|\Phi\rangle\\
		& = \int \mathcal{D} \bar{\alpha} e^{\bar{\alpha}(\bar{d} - \alpha)} \mathcal{D} \alpha \langle
		\bar{c}|\rho
		|\alpha \rangle = \int \mathcal{D} \alpha (\alpha - \bar{d}) \langle \bar{c}|\rho|\alpha\rangle \\
		& = \langle \bar{c}|\rho|\bar{d}\rangle.
	\end{aligned}
\end{equation}

In the final line, we use the fact that $(\alpha - \bar{d}) = \delta(\bar{d} - \alpha
)$. Therefore, we arrive at the following conclusion: Given the density matrix in the coherent state basis $\langle \bar{c}|\rho|c\rangle$, the corresponding double state in the coherent state basis $\langle \bar{c} \bar{d}|\rho \rangle$ can be simply obtained by substituting $c \rightarrow \bar{d}$.
For example, if $\rho_0 = |\Psi_0\rangle \langle \Psi_0|$ is the density matrix of the pure state, then 
\begin{equation}
	\begin{aligned}
		\langle \bar{c} \bar{d}|\rho_0 \rangle & = \langle \bar{c}|\Psi_0 \rangle \langle \Psi_0|\bar{d}\rangle 
		= \langle \bar{c}|\Psi_0 \rangle \langle \bar{d|}\Psi_0\rangle^* \\
		& = \langle \bar{c} \bar{d}|\Psi_0, \Psi^*_0 \rangle,
	\end{aligned}
\end{equation}
which is consistent with our intuition on bosonic fields.
We emphasize that the left-hand side of Eq.\eqref{eqn:rho_choi} is defined in the double Hilbert space spanned by the Fock basis $\{ |00\rangle, \mathbf{c}^\dagger|00\rangle,  \mathbf{d}^\dagger|00\rangle,  \mathbf{c}^\dagger \mathbf{d}^\dagger|00\rangle \}$. On the other hand, the right hand side of Eq.\eqref{eqn:rho_choi} is defined in the original Hilbert space spanned by $\{ |0\rangle, \mathbf{c}^\dagger|0\rangle \}$.

We are now ready to work out the corresponding operator $\mathcal{N}_\mathcal{E}$ for the channel $\mathcal{E}[\cdot] = \sum_\alpha K_\alpha (\cdot) K_\alpha^\dagger$ under Eq.\eqref{Eq:app_double_state}.
As mentioned in the main text, since Eq.\eqref{Eq:app_double_state} is linear, one can consider each $K_\alpha (\cdot) K_\alpha^\dagger$ individually.
Using $|K_\alpha \rho K^\dagger_\alpha \rangle = (K_\alpha \rho K^\dagger_\alpha) \otimes I |\Phi\rangle = K_\alpha (\rho K_\alpha^\dagger \otimes I |\Phi\rangle) = K_\alpha |\rho K^\dagger_\alpha\rangle$, one finds $K_\alpha$ is unchanged under Eq.\eqref{Eq:double_state}.
Besides, since one can always write $K^\dagger_\alpha$ as a function of $\mathbf{c}$ and $\mathbf{c}^\dagger$, it suffices to consider how to express $|\rho \mathbf{c}\rangle$ and  $|\rho \mathbf{c}^\dagger\rangle$ as an operator applying to $|\rho\rangle$.
Now, using Eq.\eqref{eqn:rho_choi} for $|\rho \mathbf{c} \rangle$, we find
\begin{equation}
	\begin{aligned}
		\langle \bar{c} \bar{d}|\rho \mathbf{c} \otimes I |\Phi\rangle & = \langle \bar{c}|\rho \mathbf{c}|\bar{d}\rangle 
		= \langle\bar{c}|\rho|\bar{d}\rangle \bar{d} \\
		& =\bar{d}\langle \bar{c} \bar{d}|\rho \rangle =  \langle \bar{c} \bar{d}|\mathbf{d}^\dagger|\rho\rangle,
	\end{aligned}
\end{equation}
{In the second line, we use the fact that $\rho$ preserves the fermionic parity for $\rho$. }
It is then obvious that $|\rho \mathbf{c}\rangle = \mathbf{d}^\dagger|\rho\rangle$.

Similarly, using Eq.\eqref{eqn:rho_choi} for $\rho \mathbf{c}^\dagger$, we get $\langle \bar{c} \bar{d}|\rho \mathbf{c}^\dagger \otimes I |\Phi \rangle = \langle \bar{c}|\rho \mathbf{c}^\dagger |\bar{d}\rangle$. However, the evaluation of $ \langle \bar{c}|\rho \mathbf{c}^\dagger |\bar{d}\rangle$ is not as straightforward as $\langle \bar{c}|\rho \mathbf{c}|\bar{d}\rangle $ since  $ \langle \bar{c}|\rho \mathbf{c}^\dagger |\bar{d}\rangle$  is not normally ordered. To proceed, we insert the identity between $\rho$ and $\mathbf{c}^\dagger$:
\begin{equation}
	\begin{aligned}
		\langle \bar{c}|\rho I \mathbf{c}^\dagger |\bar{d}\rangle & = 
		\int \mathcal{D} \bar{\alpha} \mathcal{D} \alpha \langle \bar{c}|\rho|\alpha \rangle e^{-\bar{a} \alpha} \langle \bar{\alpha}| \mathbf{c}^\dagger |\bar{d}\rangle \\
		& =\int  \mathcal{D} \bar{\alpha} \mathcal{D} \alpha \ \bar{\alpha}  e^{\bar{\alpha} (\bar{d} - \alpha)} \langle \bar{c}|\rho|\alpha \rangle.
	\end{aligned}
\end{equation}
In the second line, we use the fact that $\langle \bar{\alpha}| \mathbf{c}^\dagger |\bar{d}\rangle = \bar{\alpha} \langle \bar{\alpha}|\bar{d} \rangle = \bar{\alpha} e^{\bar{\alpha}\bar{d}}$.
Now we change the variable $\alpha (\bar{\alpha})$ as $\bar{\beta} (-\beta)$ so that we can substitute $\langle \bar{c} \bar{\beta}|\rho \rangle$ for $\langle \bar{c}|\rho|\alpha \rangle$:
\begin{equation}
	\begin{aligned}
		\langle \bar{c}|\rho \mathbf{c}^\dagger |\bar{d}\rangle & = \int \mathcal{D}(-\beta) \mathcal{D} \bar{\beta} (-\beta) e^{-\beta (\bar{d} - \bar{\beta})} \langle \bar{c} \bar{\beta}|\rho \rangle \\
		& =\int \mathcal{D} \beta \mathcal{D} \bar{\beta}(\beta e^{\bar{d} \beta}) e^{-\bar{\beta} \beta} \langle \bar{c} \bar{\beta}|\rho \rangle \\
		& = \int \mathcal{D} \beta \mathcal{D} \bar{\beta} \langle \bar{d}|\mathbf{d}|\beta\rangle e^{-\bar{\beta} \beta} \langle \bar{c} \bar{\beta}|\rho \rangle \\
		& =  \int  \mathcal{D} \bar{\beta} \mathcal{D}\beta  \langle \bar{d}|\mathbf{d}|\beta\rangle e^{-\bar{\beta} \beta} \langle \bar{c} \bar{\beta}|\rho \rangle\\
		& = \langle \bar{c} \bar{d}|-\mathbf{d}|\rho\rangle.
	\end{aligned}
\end{equation}
In the fourth line, the minus sign is attributed to the exchange of $\mathcal{D} \beta$ and $\mathcal{D} \bar{\beta}$. 
It follows that $|\rho \mathbf{c}^\dagger\rangle =  -\mathbf{d}|\rho \rangle$.

Interestingly, treating C-J map as a general way to transform the bra(ket) space to ket(bra) space leads to other useful applications for fermionic problems.
For example, the fermionic transpose can be naturally defined using C-J map, and we find that this definition is consistent with the fermionic time-reversal, which was proposed in Ref.\citep{shapourian2017partial} to resolve the issue that the conventional definition of the fermionic-transpose fails to capture the entanglement negativity due to the formation of the edge Majorana fermions.
Specifically, the fermionic transpose can be defined as follows:
\begin{equation}
	\label{Eq:fermion_transpose}
	\langle m' |\rho^T|m\rangle =  (\langle m'| \otimes \langle \Phi|) I \otimes \rho \otimes I (|\Phi\rangle \otimes |m\rangle),
\end{equation}
where $\{ |m\rangle \}$ is an arbitrary complete basis for fermions. 
One may show that this definition makes sense by expressing Eq.\eqref{Eq:fermion_transpose} in terms of tensor network (see Fig.\ref{Fig:cjam}(c)).
We now show that this definition coincides with the ones proposed in Ref.\citep{shapourian2017partial} using fermionic time-reversal.
Noting that $\langle \bar{c} \bar{d}|\Phi\rangle = e^{\bar{c} \bar{d}}$ and $\langle \Phi|c d \rangle = e^{-c d}$, one can evaluate $\rho$ in the coherent state basis:
\begin{equation}
	\begin{aligned}
		\langle \bar{c}|\rho^T|c \rangle & = \int \mathcal{D} \bar{\alpha} \mathcal{D} \alpha \mathcal{D} \bar{\beta} \mathcal{D} \beta
		\langle \Phi|\alpha c\rangle e^{-\bar{\alpha} \alpha} \langle \bar{\alpha} \rho | \beta \rangle e^{-\bar{\beta} \beta} e^{\bar{c} \bar{\beta}} \\
		& = \int \mathcal{D} \bar{\alpha} \mathcal{D} \alpha \mathcal{D} \bar{\beta} \mathcal{D} \beta e^{\alpha (\bar{\alpha} - c)} \langle \bar{\alpha} \rho|\beta \rangle e^{-\bar{\beta}(\beta + \bar{c}) }\\
		& = \langle c|\rho|-\bar{c}\rangle.
	\end{aligned}
\end{equation}
Therefore, one can obtain $\langle \bar{c}|\rho^T|c \rangle$ by simply substituting $c \rightarrow - \bar{c}, \bar{c} \rightarrow c$ in $\langle \bar{c}|\rho |c\rangle$.

\end{document}